\documentclass[11pt]{article}

\usepackage[T1]{fontenc}
\usepackage{amssymb, amsmath, amsthm, graphicx}
\usepackage{thmtools,thm-restate}
\makeatletter
\@ifundefined{newcounteralias}{}{%
  \renewcommand\thmt@autorefsetup{%
    \@xa\def\csname\thmt@envname autorefname\@xa\endcsname
      \@xa{\thmt@thmname}%
  }%
}
\makeatother
\usepackage[shortlabels]{enumitem}
\usepackage[dvipsnames]{xcolor}
\usepackage{color,soul}
\usepackage{url}
\usepackage{float}
\usepackage{fullpage}
\usepackage[colorlinks,linkcolor=magenta,hypertexnames=false]{hyperref}
\usepackage{cleveref}
\usepackage{braket}
\usepackage{subfiles}

\usepackage{graphicx,tikz}
\usetikzlibrary{arrows.meta,calc,decorations.pathreplacing}

\usepackage[framemethod=TikZ]{mdframed}

\newmdenv[
  linewidth=0.4pt,
  roundcorner=3pt,
  backgroundcolor=gray!3, 
  innertopmargin=8pt,
  innerbottommargin=8pt,
  innerleftmargin=10pt,
  innerrightmargin=10pt,
  skipabove=8pt,
  skipbelow=8pt
]{protocolframe}



\definecolor{refred}{RGB}{218,49,34}
\definecolor{citeblue}{RGB}{46,30,244}

\hypersetup{
    colorlinks=true,
    linkcolor=refred,
    citecolor=citeblue,
    urlcolor=citeblue
}

\usepackage[lambda,adversary,advantage,asymptotics,sets,landau,probability,operators]{cryptocode}
\usepackage[T1]{fontenc}

\newcommand*{\Z}{\mathbb{Z}}

\newcommand{\xor}{\oplus}

\newtheorem{theorem}{Theorem}[section]
\newtheorem{definition}[theorem]{Definition}

\newtheorem{lemma}[theorem]{Lemma}
\newtheorem{corollary}[theorem]{Corollary}
\newtheorem{proposition}[theorem]{Proposition}

\newtheorem{question}{Open Question}

\newcommand*{\keepstatementanchor}[1]{%
  \AtBeginEnvironment{#1}{\Needspace{3\baselineskip}}%
}

\forcsvlist{\keepstatementanchor}{theorem,definition,remark,lemma,corollary,proposition,claim,observation,fact,assumption,question}

\AtBeginEnvironment{restatable}{\Needspace{3\baselineskip}}

\newcommand{\heading}[6]{
  \renewcommand{\thepage}{\arabic{page}}
  \noindent
  \begin{center}
  \framebox{
    \vbox{
      \hbox to 5.78in {\textbf{#2} \hfill #3 }
      \vspace{4mm}
      \hbox to 5.78in {{\Large \hfill #6  \hfill} }
      \vspace{2mm}
      \hbox to 5.78in {{\bf #4} \hfill #5}
    }
  }
  \end{center}
  \vspace*{4mm}
}

\newcounter{problemnum}
\newcommand{\theproblem}{Problem \arabic{problemnum}}

\makeatletter
\newcommand{\problem}[1][{}]{\item \let\@currentlabel=\theproblem \textbf{#1}}
\makeatother

\newcounter{problempartnum}[problemnum]

\def\sol{1}

\ifnum\sol=1
\newcommand{\solution}[1]{
  \noindent\newline
  {
     {\bf \underline{Solution}}\newline
        #1
    
  }
  \bigskip
}
\else
\newcommand{\solution}[1]{}
\fi 

\newcommand{\class}[1]{\ensuremath{\mathbf{#1}}}

\newcommand{\PSPACE}{\class{PSPACE}}
\newcommand{\NC}{\class{NC}}

\newcommand{\IP}{\class{IP}}

\newcommand{\IPP}{\class{IPP}}
\newcommand{\BQP}{\class{BQP}}

\newcommand{\QIP}{\class{QIP}}
\newcommand{\QMAP}{\class{QMAP}}

\newcommand{\QCIPP}{\class{QCIPP}}
\newcommand{\QPT}{\class{QPT}}

\newcommand{\sfQIP}{\mathsf{QIP}}

\newcommand{\sfQIPP}{\mathsf{QIPP}}
\newcommand{\sfQCIPP}{\mathsf{QCIPP}}

\newcommand{\crs}{\mathsf{crs}}
\newcommand{\Acc}{\mathsf{Acc}}

\newcommand{\N}{\mathbb{N}}
\newcommand{\F}{\mathbb{F}}
\newcommand{\R}{\mathbb{R}}
\newcommand{\C}{\mathbb{C}}

\newcommand{\bbE}{\mathbb{E}}

\newcommand{\bbI}{\mathbb{I}}

\newcommand{\sfD}{\mathsf{D}}

\newcommand{\sfL}{\mathsf{L}}

\DeclareMathOperator{\poly}{poly}
\DeclareMathOperator{\polylog}{polylog}

\DeclareMathOperator{\negl}{negl}

\newcommand{\tr}{\operatorname{tr}}
\let\epsilon\varepsilon

\newcommand{\pval}{\ensuremath{{\mathsf{PVAL}}}}

\newcommand{\calA}{\mathcal{A}}

\newcommand{\calH}{\mathcal{H}}
\newcommand{\calI}{\mathcal{I}}

\newcommand{\calK}{\mathcal{K}}
\newcommand{\calL}{\mathcal{L}}
\newcommand{\calM}{\mathcal{M}}
\newcommand{\calN}{\mathcal{N}}

\newcommand{\calP}{\mathcal{P}}
\newcommand{\calQ}{\mathcal{Q}}
\newcommand{\calR}{\mathcal{R}}
\newcommand{\calS}{\mathcal{S}}

\newcommand{\calU}{\mathcal{U}}

\newcommand{\calW}{\mathcal{W}}
\newcommand{\calX}{\mathcal{X}}
\newcommand{\calY}{\mathcal{Y}}
\newcommand{\calZ}{\mathcal{Z}}

\newcommand{\Parity}{\mathrm{Parity}}
\newcommand{\MOD}{\mathrm{MOD}}
\newcommand{\softO}{\widetilde O}
\newcommand{\dist}{\mathrm{d}}

\title{Quantum Interactive Proofs of Proximity}
\author{
Kenan Wood\thanks{Georgia Institute of Technology. Email: \href{mailto:kwood67@gatech.edu}{\texttt{kwood67@gatech.edu}}.}
}
\date{}

\begin{document}

\maketitle

\begin{abstract}
We initiate the systematic study of quantum interactive proofs of proximity (QIPPs). These are quantum interactive proof systems with sublinear total communication, where the verifier is required to accept inputs in a property $\Pi$ and reject inputs $\epsilon$-far from $\Pi$, while only querying the input a sublinear number of times. This work builds on \textit{classical} interactive proofs of proximity (IPPs) and quantum \textit{noninteractive} proofs of proximity (QMAPs), which have been previously studied. Our main contributions are as follows.

\begin{enumerate}
    \item We show that all logspace-uniform $\NC$ properties---those computable with logspace uniform circuits of polynomial size and polylogarithmic depth---admit QIPPs with query, communication and time complexity all $\softO(n^{2/5})$. This improves on the best known classical total complexity of \(\softO(\sqrt n)\), which is optimal in the worst case for polynomial-time verifiers under the PRG assumption of Kalai and Rothblum (CRYPTO, 2015). This resolves an open question of Dall’Agnol, Gur, Roy Moulik, and Thaler (Quantum, 2022) in the affirmative.

    \item Assuming the existence of a post-quantum subexponentially secure PRG computable in logspace-uniform $\NC$, we prove that, for every nondecreasing, logspace-computable query bound $q(n)\le n^{1/2-\Omega(1)}$, there exists a property $\Pi$ in logspace-uniform $\NC$ and a constant $\epsilon>0$ such that no polynomial-time $\epsilon$-QIPP for $\Pi$ can achieve query complexity $q(n)$ and communication complexity $c(n)$ satisfying $q(n)+c(n)=o(n^{1/3})$.

    \item We establish a bi-directional connection between QIPPs for logspace-uniform $\NC$ and query-model QIPs for parity. A QIP for $k$-bit parity with total complexity $k^{2/3-\Omega(1)}$ yields QIPPs for every logspace-uniform $\NC$ property with total complexity $n^{2/5-\Omega(1)}$. Conversely, achieving total complexity $o(n^{2/5})$ within the standard $\pval$-based framework requires improving the known query--communication tradeoff for parity.
\end{enumerate}

Beyond logspace-uniform \(\NC\), we prove that, for every constant \(\beta\in(0,1)\), almost every property \(\Pi\subseteq\{0,1\}^n\) of cardinality \(2^{\beta n}\) requires linear total QIPP complexity at a constant proximity parameter depending only on \(\beta\). This lower bound is unconditional and tight up to constant factors.

We further give general transformations yielding quantum query speedups for $\polylog(n)$-round IPPs and three-round compression of public-coin QIPPs, and exponentially separate QIPPs from both round-unrestricted IPPs and QMAPs.

\end{abstract}

\cleardoublepage

\tableofcontents

\cleardoublepage


\section{Introduction}

Property testing is a fundamental model of sublinear computation~\cite{rubinfeld-sudan:robust-polynomials,
goldreich-goldwasser-ron:property-testing}. A tester is given query access
to a large object and must solve the approximate decision problem of membership in a property
$\Pi$: it should accept inputs in $\Pi$ and reject inputs that are far from $\Pi$, while making
only a small number of queries. Building on the large literature on classical property testing~\cite{rubinfeld-sudan:robust-polynomials,goldreich-goldwasser-ron:property-testing},
proofs of proximity study the analogous verification problem, where a sublinear-time verifier is
aided by an untrusted prover. These include, among others, Merlin--Arthur proofs of proximity (MAP)~\cite{GR-map}, interactive proofs of proximity (IPP)~\cite{RVW13}, and probabilistically checkable proofs of proximity (PCPP)~\cite{BGHSV-PCPP}.

Quantum query algorithms provide a second way to strengthen property testers, allowing the
verifier to query the input in coherent superposition~\cite{BFNR08}. Quantum property testing has exhibited
strong separations from classical testing~\cite{BFNR08,MW-qpt-survey}. Simultaneously, quantum proof systems such as quantum Merlin--Arthur proofs (QMA, the quantum analogue of NP) and
quantum interactive proofs (QIP) display rich structural phenomena with no known classical analogue~\cite{VW-quantum-proofs,KW-QIP3,KKMV}. These two directions suggest a
natural question:

\[
\textit{What is the power of quantum interactive proofs for property testing?}
\]

\subsection{Quantum interactive proofs of proximity}
This work initiates the systematic study of quantum interactive proofs of proximity
(QIPPs). 
Here, a property is a family $\Pi=(\Pi_n)_{n\in\N}$,
where $\Pi_n\subseteq\{0,1\}^n$; we suppress the subscript $n$
when the input length is clear.
A QIPP for $\Pi$ is a quantum interactive proof system in the property testing setting:
the verifier has coherent oracle access to an input string $x \in \{0,1\}^n$ via the unitary $U_x \ket{i}\ket{b} = \ket{i}\ket{b \xor x_i}$ (for $i \in [n]$ and $b \in \{0,1\}$), interacts with an untrusted quantum
prover via quantum channels, and must distinguish $x\in \Pi$ from inputs that are $\varepsilon$-far from $\Pi$ in normalized Hamming distance. Completeness is the condition that for inputs in $\Pi$, some prover strategy must make the verifier accept with probability at least $2/3$, while soundness is that for inputs $\epsilon$-far from $\Pi$, no prover strategy can make the verifier accept with probability greater than $1/3$. 
Unless otherwise specified, we assume that all proofs of proximity have completeness $2/3$ and soundness $1/3$, but these can be amplified by repetition.
When a proximity parameter $\epsilon$ is left unspecified in informal language, one may take $\epsilon$ to be some arbitrarily small constant not depending on $n$.

If the communication of a QIPP is entirely classical, we call it a QCIPP. A \emph{non-interactive} QIPP in which the prover sends a sublinear-sized state as the only proof is called a QMAP (for quantum MA proof of proximity)~\cite{dallagnol-gur-roymoulik-thaler:qmap}.
\emph{Public-coin} protocols form an important class of IPPs, QCIPPs, and QIPPs. In a public-coin protocol, the verifier's messages consist only of random coin tosses while the prover's messages are unrestricted.
In these protocols, the verifier's oracle queries can be deferred until after the communication has finished.

The relevant resources are the verifier's \emph{query complexity} (number of calls to the verifier's oracle), the \emph{communication complexity} (total number of qubits transmitted), the
\emph{round complexity} (the number of rounds of communication), and the \emph{time complexity} of the verifier. In nontrivial protocols,
the query and communication complexities are sublinear in the input length. The sum of query and communication complexities will be called the \emph{total complexity}.

\subsection{Results}

Our main result is a public-coin QCIPP for every
logspace-uniform $\NC$ property with query, communication,
and verifier time complexity $\softO(n^{2/5})$.
We complement this construction with a conditional lower
bound and reductions relating further improvements to
query-model QIPs for $\Parity$.
We also prove a tight unconditional lower bound for random
properties, give general transformations for proofs of
proximity, and establish exponential separations from
classical IPPs and quantum noninteractive proofs.

\subsubsection{Logspace-uniform NC admits QCIPPs with \texorpdfstring{$o(\sqrt{n})$}{o(sqrt(n))} complexity}\label{Subsec:intro-nc-qcipp}

Since the conception of IPPs of Rothblum, Vadhan, and Wigderson~\cite{RVW13}, there has been considerable interest in IPP algorithms for logspace-uniform $\NC$, the class of properties computable by logspace-uniform circuits of polynomial size and polylogarithmic depth~\cite{RVW13,RR20}, and related classes like $\mathbf{SC}$~\cite{RRR16:constant-round-delegating-computation}. Moreover, IPPs for logspace-uniform $\NC$ have been valuable components of doubly efficient interactive proofs and batch verification of unambiguous versions of $\mathbf{NP}$~\cite{RR20,RRR16:constant-round-delegating-computation,RRR18:batch-verification-up} and $\mathbf{IP}$ languages~\cite{BGHK:batch-UIP}. Thus, the study of IPPs for this class of properties appears to be especially important. In particular, the study of \emph{quantum advantage} for this class of IPPs may have more significant implications beyond the immediate property testing applications.


The best known IPP for logspace-uniform $\NC$ is due to Rothblum and Rothblum~\cite{RR20,RR20-revised}, achieving query and communication complexity with $q + c = \softO(\sqrt{n})$ and $q \cdot c = \softO(n)$, improving the protocol of~\cite{RVW13} that achieves $q+c = n^{\frac{1}{2}+o(1)}$ and $q \cdot c = n^{1+o(1)}$. Furthermore, Kalai and Rothblum~\cite{KR-arguments-of-proximity} show a matching lower bound: if subexponentially secure, logspace-uniform $\NC$-computable PRGs exist, then no (polynomial-time) IPP for all of logspace-uniform $\NC$ can have query complexity $q$ and communication complexity $c$ satisfying $q \cdot c = o(n)$.
A natural question then emerges, first posed by Dall'Agnol, Gur, Roy Moulik, and Thaler~\cite{dallagnol-gur-roymoulik-thaler:qmap}:

\begin{center}
    \itshape
    Does logspace-uniform $\NC$ admit QIPPs with query complexity $q$ and communication\\ complexity $c$ such that $q + c = o(\sqrt{n})$?
\end{center}

We answer this question in the affirmative by constructing a QCIPP protocol for any logspace-uniform $\NC$ property with query, communication, and time complexity $\softO(n^{2/5})$. 

\begin{theorem}[Theorem~\ref{thm:nc-in-qcipp},
    informally stated]
\label{thm:nc-qcipp-informal-main}
\label{thm:nc-qcipp-informal-2}
Every property in logspace-uniform $\NC$ admits a
public-coin $n^{-3/5}$-QCIPP with query, communication,
and time complexity $\softO(n^{2/5})$.

Moreover, for every constant $\delta\in(0,1/2]$, every
such property admits a public-coin QCIPP with query
complexity $\softO(n^{\delta/2})$ and communication
and time complexity $\softO(n^{1-\delta})$.

All of these protocols use $\polylog(n)$ rounds.
\end{theorem}


Our construction exploits a structural feature of the
Rothblum--Rothblum protocol~\cite{RR20-revised}: its remaining checks
can be expressed as linear identities whose coefficients
and input locations are efficiently computable from a
succinct transcript.
This lets us apply the block-verification idea
of~\cite{dallagnol-gur-roymoulik-thaler:qmap} to reduce
the query complexity.
Achieving the stated time bound requires a further
ingredient: an amplitude-amplification procedure that
amortizes the cost of accessing the prover's classical
data across all amplification iterations.
We explain these ingredients in
Section~\ref{Subsec:technical-nc-qcipp}.

\subsubsection{A lower bound on QIPP for logspace-uniform NC}


Our second main result gives a query--communication lower
bound under a cryptographic assumption:

\begin{theorem}[Theorem~\ref{thm:main_lower_bound}, informally stated]
\label{thm:main_lower_bound_informal}
Assume that there exists a PRG
that is computable in logspace-uniform $\NC$ and secure against $2^{o(\lambda)}$-time quantum
distinguishers, for a security parameter $\lambda$. Let $q : \N \to \N$ be a nondecreasing logspace-uniformly computable function such that $q(n)\le n^{1/2-\Omega(1)}$. Then there is a property
$\Pi$ in logspace-uniform $\NC$ and a constant $\epsilon>0$ such that no polynomial-time $\epsilon$-QIPP for $\Pi$
with query complexity $q(n)$ and communication complexity $c(n)$ can satisfy
\[
    q(n)^2 c(n) = o(n).
\]
In particular, such an $\epsilon$-QIPP cannot satisfy $q(n) + c(n) = o(n^{1/3})$.
\end{theorem}

We adapt the lower-bound framework of Kalai and
Rothblum~\cite{KR-arguments-of-proximity}.
Classically, completeness is preserved by arguing that
the verifier is unlikely to query a hidden modified seed
block.
Coherent queries invalidate this argument, and applying
a quantum hybrid bound requires controlling the
input-dependent prover strategy.
We arrange the reduction so that the hidden block index
is independent of the valid input and its honest prover
strategy.
Fixing that strategy then turns the entire interaction
into a fixed query algorithm, allowing us to bound the
perturbation by $O(q/\sqrt m)$ for $m$ seed blocks.
Taking $m=\Theta(q^2)$ preserves average-case completeness;
the short-communication QIP simulation
of~\cite{BSW-QIP-short-messages} then yields the claimed
tradeoff. The technical overview contains further details.


Theorem~\ref{thm:nc-qcipp-informal-2} gives public-coin QCIPPs
for every logspace-uniform $\NC$ property with query complexity
$q(n)=\softO(n^{\delta/2})$ and communication complexity
$c(n)=\softO(n^{1-\delta})$, for every constant
$\delta\in(0,1/2]$.
These parameters satisfy $q(n)^2c(n)=\softO(n)$, matching
the conditional lower bound up to polylogarithmic factors
in this regime, where $c\ge\Omega(\sqrt{n})$.

The optimal total complexity remains open.
Theorem~\ref{thm:nc-qcipp-informal-main} achieves
$q+c=\softO(n^{2/5})$, whereas
Theorem~\ref{thm:main_lower_bound_informal} rules out
$q+c=o(n^{1/3})$ under its cryptographic assumption.
Thus, the matching tradeoff above does not establish
optimality of our protocol at its balanced parameters.




 \subsubsection{QIPPs for NC and the QIP complexity of Parity}
\label{Subsec:intro-nc-parity}

Our results leave a gap in the total complexity of QIPPs for
logspace-uniform $\NC$: our conditional lower bound rules out
$q+c=o(n^{1/3})$, while our best construction achieves only
$q+c=\softO(n^{2/5})$. We show that improving this construction
is closely connected to improving quantum interactive proofs
for the elementary problem of computing parity.

For inputs of length $k$, define
\[
    \Parity_k := \left\{x\in\{0,1\}^k :
    \bigoplus_{i=1}^k x_i=0\right\}.
\]
Throughout this subsection, $k$ denotes the input length for
$\Parity$ and $n$ denotes the input length for a
logspace-uniform $\NC$ property.
The QIPs we discuss are in the query model and verify exact
membership, with soundness required on every input outside
the property. The following noninteractive protocol is known.

\begin{proposition}[\cite{dallagnol-gur-roymoulik-thaler:qmap}]\label{prop:intro-parity-qmap}
    For every integer $1\le c\le k$, $\Parity_k$ admits a QMA proof
    with a classical proof of length $O(c)$ and query complexity
    $O(k/\sqrt{c})$. The protocol verifies exact membership
    in $\Parity_k$.
    In particular, choosing $c=k^{2/3}$ gives total complexity
    $O(k^{2/3})$.
\end{proposition}

Can quantum interaction improve this noninteractive tradeoff?
We relate this question to the complexity of QIPPs for the
entire class of logspace-uniform $\NC$ properties.
First, improved QIPs for $\Parity$ yield improved QIPPs
for every property in this class.

\begin{theorem}[Theorem~\ref{thm:qip-parity-implies-qipp-nc}, informally stated]\label{thm:intro-parity-to-nc}
    If $\Parity$ has a QIP with total complexity
    $k^{2/3-\Omega(1)}$, then every logspace-uniform $\NC$
    property has an $n^{-3/5}$-QIPP with total complexity
    $n^{2/5-\Omega(1)}$.
\end{theorem}

Conversely, reducing total complexity to $o(n^{2/5})$ at the
proximity parameter $n^{-3/5}$ achieved by
Theorem~\ref{thm:nc-qcipp-informal-main} would improve the known
query--communication tradeoff for $\Parity$.

\begin{theorem}[Theorem~\ref{thm:qipp-nc-implies-qip-parity}, informally stated]\label{thm:intro-nc-to-parity}
    There exists a property $\Pi$ in logspace-uniform $\NC$
    such that an $n^{-3/5}$-QIPP for $\Pi$ with total complexity
    $o(n^{2/5})$ implies a QIP for $\Parity_k$ with query
    complexity $o(k^{4/5})$ and communication complexity
    $o(k^{2/5})$.
\end{theorem}

The resulting parameters satisfy $q=o(k/\sqrt{c})$,
improving the query--communication tradeoff achieved by
Proposition~\ref{prop:intro-parity-qmap} at an unbalanced
parameter pair.
They do not necessarily improve the best known
$O(k^{2/3})$ total complexity for $\Parity$.

Although this converse uses a vanishing proximity parameter,
it also constrains constant-proximity improvements within
the following $\pval$-based framework~\cite{RVW13,RR20}.
Consider protocols that first spend $\Omega(\eta n)$
communication reducing verification to an $\eta$-proximity
claim for $\pval$, and then verify that claim while
preserving the $\eta$-proximity guarantee.
Total complexity $o(n^{2/5})$ therefore forces
$\eta=o(n^{-3/5})$, so the resulting protocol is also an
$n^{-3/5}$-QIPP.
By Theorem~\ref{thm:intro-nc-to-parity}, improving our bound
within this framework requires improving the known QIP
query--communication tradeoff for $\Parity$, even when only
constant proximity is required.




\subsubsection{Random properties require linear total complexity}
\label{Subsec:intro-random-properties}

We now turn from logspace-uniform $\NC$ to arbitrary
properties and ask whether QIPPs can always achieve
sublinear total complexity.
For every fixed $\beta\in(0,1)$, we show that almost every
property of cardinality $2^{\beta n}$ requires linear
total complexity at a suitable constant proximity
parameter, even with unrestricted verifier time and
arbitrarily many rounds.


\begin{theorem}[Theorem~\ref{thm:qipp-random-lower-bound},
    informally stated]\label{thm:intro-random-lower-bound}
    For every constant $\beta\in(0,1)$, there exist constants
    $\epsilon,\alpha>0$ such that the following holds for all
    sufficiently large $n$. Let $\Pi\subseteq\{0,1\}^n$ be sampled
    uniformly among properties of cardinality $2^{\beta n}$.
    Then with probability at least $1-2^{-2^{\Omega(n)}}$, every
    $\epsilon$-QIPP for $\Pi$ with query complexity $q$ and
    communication complexity $c$ satisfies
    \(
        q+c\ge\alpha n.
    \)
\end{theorem}

This matches the trivial $n$-query upper bound up to
constant factors.
The main technical challenge is obtaining a finite
approximation of an interactive verifier that preserves
acceptance probabilities uniformly over inputs and
quantum prover strategies.
We describe this approximation in the technical overview.

\subsubsection{General transformations}
\label{Subsec:intro-general-transformations}

Combining the Goldwasser--Sipser
transformation~\cite{goldwasser-sipser:private-public-coins}
with amplitude amplification, we transform any classical
$\epsilon$-IPP with query complexity $q$, communication
complexity $c\ge\log n$, and $\polylog(n)$ rounds into a
public-coin $\epsilon$-QCIPP with query complexity
$\softO(\sqrt q)$ and communication complexity
$\softO(c+q)$.
This applies to private-coin protocols with adaptive
queries; see Theorem~\ref{thm:ipp-to-qcipp} for the general
dependence on the number of rounds.
When $q=\softO(c)$, the transformation gives a quadratic
query improvement while preserving communication up to
polylogarithmic factors.

Adapting the round reduction
of~\cite{KKMV}, we also transform any public-coin
$r$-round $\epsilon$-QIPP into a three-round
$\epsilon$-QIPP with $\poly(r)$ overhead in query,
communication, and time complexity
(Theorem~\ref{thm:pub-coin-round-reduction}).
The two transformations compose.
Round reduction also gives a three-round version of our
protocol for logspace-uniform $\NC$, with query,
communication, and time complexity $\softO(n^{2/5})$
(Corollary~\ref{cor:nc-in-qipp-3}).

\subsubsection{Separations}

We adapt the Forrelation lower bound of Bouland
et al.~\cite{BQP-IP-separation-26} to property testing.
For a suitable constant proximity parameter, the resulting
property admits a constant-query quantum tester, whereas
every classical IPP requires $n^{\Omega(1)}$ total
complexity, regardless of its number of rounds
(Theorem~\ref{thm:separation}).

Combining this consequence with the known permutation-testing
separation~\cite{GLR-MA-AM-proximity,sherstov-thaler:vanishing-error-qma}
gives a property admitting a two-round QCIPP with
$O(\log n)$ query and communication complexity at constant
proximity, while every classical IPP or QMAP for the same
proximity parameter requires $n^{\Omega(1)}$ total complexity
(Corollary~\ref{cor:qipp-vs-ipp-cup-qmap}).

\subsection{Technical overview}\label{Sec:technical-overview}
This section contains a technical overview of the high-level ideas of the proofs of our main results. We begin by explaining how we achieve our $\softO(n^{2/5})$ total complexity QCIPP for logspace-uniform $\NC$, and how we tighten the time to also $\softO(n^{2/5})$. Then we describe the key ideas behind the connection to Parity. The proof idea of our cryptographic lower bound in Theorem~\ref{thm:main_lower_bound_informal} follows, including the technical challenges and our solution. We conclude with the new techniques developed for the unconditional random-property lower bound.

\subsubsection{Constructing QCIPPs for NC with \texorpdfstring{$o(\sqrt{n})$}{o(sqrt(n))} total complexity}
\label{Subsec:technical-nc-qcipp}

\paragraph{The classical starting point.}
We build on the IPP of Rothblum and Rothblum~\cite{RR20-revised}, following
the framework introduced by Rothblum, Vadhan, and
Wigderson~\cite{RVW13}. The protocol first uses parallel executions
of GKR~\cite{goldwasser-kalai-rothblum:delegating-computation} to
reduce proximity to a logspace-uniform $\NC$ property to proximity
to a polynomial-evaluation property, denoted $\pval$. Such a
property specifies prescribed evaluations of the input's low-degree
extension. This reduction requires no input queries: the verifier's
remaining task is to check proximity to the resulting $\pval$
instance.

The central ingredient of~\cite{RR20-revised} is a recursive
distance-amplification procedure based on \cite{ben-sasson-kopparty-saraf:distance-to-code}. Each step folds the current input, initialized to $x$,
into a smaller derived input while amplifying its distance from the
corresponding $\pval$ instance. Eventually, the folded input $y$ is
small enough for the prover to send the entire purported table $y'$, which is equal to $y$ in the honest case.
The verifier checks that the received table $y'$ satisfies the final $\pval$ instance,
then compares $y$ with $y'$ at few uniformly sampled locations. For an
$\epsilon$-far input, with constant probability the verifier has
already rejected or $y$ and $y'$ disagree on an inverse-polylogarithmic
fraction of locations. Thus only $\polylog(n)$ such comparisons are
needed. Each query to $y$, however, unfolds into $O(1/\epsilon)$
queries to the original input. This yields classical query complexity
$\softO(1/\epsilon)$ and communication complexity
$\softO(\epsilon n)$, balancing at total complexity
$\softO(\sqrt n)$. 

\paragraph{Accelerating the expensive virtual queries.}
This description exposes the obstacle to a direct quantum speedup:
amplitude amplification over the final comparisons saves only
polylogarithmic factors, since most of the cost lies in evaluating
one location of $y$. Alternatively, the prover could supply all the
needed input bits, which the verifier checks using amplitude
amplification. This improves queries quadratically but gives
communication $\softO(\epsilon n+1/\epsilon)$, whose optimized
bound is still $\softO(\sqrt n)$. Although this idea does not resolve the question of obtaining $o(\sqrt{n})$ total complexity for logspace-uniform $\NC$, the approach is far more general. Indeed, our IPP-to-QCIPP transformation in Section~\ref{Subsec:intro-general-transformations} first uses the Goldwasser--Sipser transformation~\cite{goldwasser-sipser:private-public-coins} to obtain a public-coin IPP. The prover then supplies the oracle answers needed to simulate the verifier, and the verifier uses amplitude amplification to check their consistency with the input.

To exploit the linearity of the recursive folding, we need
an explicit representation that can be evaluated coherently
and efficiently.
We show that each location $k$ of the final folded input
can be written as
\[
    y(k)=\sum_{i\in[p]}\alpha_k(i)x[q_k(i)],
\]
where $p=O(1/\epsilon)$, the coefficients
$\alpha_k(i)$ belong to a characteristic-$2$ field $\F$
of size $\poly(n)$, and $q_k(i)\in[n]$.
Crucially, both $\alpha_k(i)$ and $q_k(i)$ are computable
in $\polylog(n)$ time from $k$, $i$, and a
$\polylog(n)$-length portion of the transcript.
Thus, a term of the sum can be evaluated coherently using
one input query and $\polylog(n)$ gates.
This representation lets us apply block verification
directly to the linear identity $y(k)=y'(k)$, without
asking the prover to supply all $p$ input values.


Equipartition $[p]$ into $b$ blocks $B_1,\ldots,B_b$ of nearly the same size and ask the prover
for the partial sums $s_j = \sum_{i \in B_j} \alpha_k(i) x[q_k(i)]$, over every block $B_j$, for every $j \in [b]$. Upon receiving claimed sums $s_1', \dots, s_b'$, the verifier checks
$\sum_{j\in [b]} s_j' = y'(k)$. If this check passes but the original claim is
false, some block $B_j$ must satisfy
\[
    s_j'\ne\sum_{i\in B_j}\alpha_k(i)x[q_k(i)].
\]
Checking one block takes $O(\lceil p/b\rceil)$ queries, so amplitude
amplification finds an incorrect block with $O(p/\sqrt b)$ queries
when $b\le p$. This adapts the block-verification idea used in the
known Parity protocol and other QMAPs for decomposable properties~\cite{dallagnol-gur-roymoulik-thaler:qmap} to
the linear claims produced by the classical IPP. Taking
$b=\min\{p,\lceil\epsilon n\rceil\}$ keeps the additional
communication within $\softO(\epsilon n)$ and gives query complexity $\softO \left( \frac{1}{\epsilon^{3 / 2}\sqrt n} +\frac{1}{\sqrt\epsilon}\right)$ and communication complexity $\softO(\epsilon n)$. Choosing $\epsilon=n^{-3/5}$ balances query and communication
complexity at $\softO(n^{2/5})$.

\paragraph{Matching the time complexity.}
The query improvement alone does not give the desired
running time.
A straightforward coherent lookup of the $b$ claimed
partial sums costs $\softO(b)$ gates per amplification
iteration.
Over $O(\sqrt b)$ iterations, this introduces
$\softO(b^{3/2})$ overhead.
At the balanced parameters $b=\Theta(n^{2/5})$,
this gives only a $\softO(n^{3/5})$ time bound.

We avoid this repeated loading cost with a general
amplitude-amplification lemma for classical advice.
Given a classical table of $M$ rows of $d$ bits and a
reversible predicate that uses $Q$ input queries and $T$
gates to check a row, our procedure finds a successful
row with probability at least $2/3$, whenever one exists,
using
\[
    O(Q\sqrt M)
    \quad\text{queries and}\quad
    \softO\bigl(M(d+\log M)+T\sqrt M\bigr)
    \quad\text{gates}.
\]
The gate bound includes the cost of preparing the
table-dependent quantum states.

The construction combines parallel lookup~\cite{beals}
and sample-based reflections~\cite{kimmel} with fixed-point
amplitude amplification~\cite{ylc}.
For a table $(d_1,\ldots,d_M)$, we prepare $O(M)$
independent copies of
\[
    \ket{\psi}
    =\frac{1}{\sqrt M}\sum_{h=1}^M\ket{h,d_h}
\]
together, using parallel lookup.
Fresh copies then implement approximate reflections
about $\ket{\psi}$, with the errors chosen so that their
accumulation remains bounded.
Preparing the copies together amortizes the table-loading
cost across all amplification iterations.

In our application, each table row contains one claimed
partial sum.
The coefficients and input locations are computed from
the succinct transcript.
Taking $M=b$, $d=O(\log n)$,
$Q=O(\lceil p/b\rceil)$, and
$T=\softO(\lceil p/b\rceil)$ gives time
$\softO(b+p/\sqrt b)$ for the block search.
Consequently, the verifier's total time is
$\softO(q+c)$, giving $\softO(n^{2/5})$ time
at the balanced parameters.

\paragraph{The Parity connection.}
Over a field $\F$ of characteristic 2, each linear claim
$\sum_{i \in [p]} \alpha_k(i) x[q_k(i)] = y'(k)$ is equivalent to
$\log|\F|=O(\log n)$ binary parity constraints.
Each query to these constraints can be simulated with at most one
query to $x$. Substituting QIPs for Parity into the verification
procedure therefore gives the forward reduction, Theorem~\ref{thm:intro-parity-to-nc}.

For the converse, encode a Parity input blockwise using a systematic,
constant-rate, constant-distance code computable in logspace-uniform
$\NC$. Let $E$ denote this encoding and set
$\Pi=\{E(x):x\in\Parity_k\}$. Choose the number of blocks
$\ell=\Theta(1/\epsilon)$ so that encoded odd-parity inputs are
$\epsilon$-far from $\Pi$. Each encoded query can be simulated by
reading one block of $O(\epsilon k)$ input bits. Simulating an
$\epsilon$-QIPP for $\Pi$ thus gives a QIP for Parity with an
$O(\epsilon k)$ query overhead and unchanged communication. This is the idea behind Theorem~\ref{thm:intro-nc-to-parity}.

\subsubsection{A lower bound for NC}
\label{Subsec:technical-nc-lower-bound}

We adapt the lower-bound framework of Kalai and
Rothblum~\cite{KR-arguments-of-proximity}. Let
$G:\{0,1\}^{\lambda}\to\{0,1\}^{2\lambda}$ be a subexponentially
secure post-quantum PRG computable in logspace-uniform $\NC$, and
write $\Pi_G=\operatorname{Im}(G)$. The proof has two steps.
First, a sufficiently efficient QIPP for a property
$\Lambda$ in logspace-uniform $\NC$ yields a low-communication QIP for $\Pi_G$, with
shared randomness and average-case completeness. Second, simulating
this QIP yields a subexponential-time quantum algorithm distinguishing
$G(U_\lambda)$ from $U_{2\lambda}$, contradicting security.
Here $U_t$ denotes the uniform distribution on $\{0,1\}^t$;
every PRG output belongs to $\Pi_G$, whereas
$\Pr[U_{2\lambda}\in\Pi_G]\le 2^{-\lambda}$.

\paragraph{From proximity proofs to small-communication QIPs.}
Fix $q=q(n)\le n^{1/2-\Omega(1)}$ as in
Theorem~\ref{thm:main_lower_bound_informal}. Choose
$m=\Theta(q^2)$ with a sufficiently large constant and
$\lambda=\Theta(n/m)$. Let $C$ be a constant-rate,
constant-distance code with encoding computable in logspace-uniform
$\NC$~\cite{Spielman1996LinearTime,GuruswamiKabanets2008Hardness}. Writing $\mathbf r=(r_1,\ldots,r_m)$ and
$\mathbf s=(s_1,\ldots,s_m)$, define
\[
    \Lambda_n=
    \left\{\bigl(C(\mathbf r),\mathbf s\bigr):
    \mathbf s\in(\{0,1\}^{\lambda})^m,\quad
    r_i=G(s_i)\text{ for all }i\in[m]
    \right\}.
\]
Then $\Lambda$ is in logspace-uniform $\NC$. Suppose $(P,V)$ is an $\epsilon$-QIPP
for $\Lambda$ with a polynomial-time verifier, query complexity $q$,
and communication $c$ satisfying $q^2c=o(n)$, for a sufficiently
small constant $\epsilon>0$.

On a challenge $r\in\{0,1\}^{2\lambda}$, the reduced protocol
samples a common random string
$R=(s_1,\ldots,s_m,I)$ consisting of independent uniform seeds and
$I\in[m]$. Set $r_I=r$ and $r_j=G(s_j)$ for $j\ne I$.
The reduced verifier $V'$ simulates $V$ with oracle
\[
    X=\bigl(C(r_1,\ldots,r_m),s_1,\ldots,s_m\bigr).
\]
If $r\notin\Pi_G$, every valid codeword in $\Lambda$ encodes a
different message. Code distance therefore gives
$\Delta(X,\Lambda)>\epsilon$, implying soundness at most $1/3$
against every prover, for every $R$.

\paragraph{Quantum average-case completeness.}
When $r\in\Pi_G$, the honest prover replaces $s_I$ by a uniform
preimage of $r$, obtaining $X^*\in\Lambda$, and simulates
$P(X^*)$. The verifier still uses $X$, so completeness does not
follow directly. Classically, one conditions on never querying the
corrupted block; coherent queries require a state-distance bound
instead. For $r\sim G(U_\lambda)$, however, $X^*$ is
uniform in $\Lambda_n$ and independent of $I$; $X$ is obtained by
replacing its $I$th seed block by an independent uniform seed.

To compare these executions, fix $x\in\Lambda_n$ and its entire
honest prover strategy $P_x$. One of the key insights is that once the prover is fixed, the interaction $(P_x, V^z)$ can be viewed as a single query algorithm with oracle $z$, since $P_x$ uses fixed quantum channels. Thus the hybrid method for query lower bounds can be applied.
Let $\ket{\psi_x(z)}$ be the final purified state of $(P_x,V^z)$,
and let $x^{(i,s)}$ replace the $i$th seed block of $x$ by $s$.
The hybrid method~\cite{BBBV-strengths-weaknesses} then gives
\[
    \mathbb E_{I\leftarrow[m],\,S\leftarrow U_\lambda}
    \left\|\ket{\psi_x(x)}-
    \ket{\psi_x(x^{(I,S)})}\right\|_2
    \le \frac{2q}{\sqrt m}.
\]
Taking $m=\Theta(q^2)$ preserves average-case
completeness at least $1/2$. This replaces the classical
$O(q/m)$ estimate, and is the source of the quadratic dependence
on $q$ in our lower bound.

\paragraph{From small communication to a PRG distinguisher.}
Beigi, Shor, and Watrous~\cite{BSW-QIP-short-messages} supply the
second step.
\begin{proposition}[Beigi--Shor--Watrous]
\label{prop:BSW_small_comm_QIP_simulation}
    A quantum interactive proof for a promise problem on $N$-bit
    inputs, with a polynomial-time verifier, communication $c$,
    and a constant completeness--soundness gap, can be simulated
    by a quantum algorithm in time $2^{O(c)}\poly(N)$.
\end{proposition}

Our reduced protocol has only average-case completeness,
so we apply Proposition~\ref{prop:BSW_small_comm_QIP_simulation}
to a promise problem on pairs $(r,R)$, where $r$ is the
challenge and $R$ is the shared randomness.
Let $\omega(r,R)$ denote the reduced verifier's maximum
acceptance probability.
Average-case completeness at least $1/2$ implies that
$\omega(r,R)\ge2/5$ on at least a $1/6$ fraction of pairs
sampled with $r\sim G(U_\lambda)$.
In contrast, soundness gives $\omega(r,R)\le1/3$ for
every $r\notin\Pi_G$ and every $R$.

Define YES instances by $\omega(r,R)\ge2/5$ and NO
instances by $\omega(r,R)\le1/3$.
Simulating this promise problem with sufficiently small
constant error, and sampling $R$, gives a PRG
distinguisher running in time $2^{O(c)}\poly(n)$:
PRG samples produce YES instances with constant
probability, whereas uniform samples produce NO instances
except with probability at most $2^{-\lambda}$.

Finally, $\lambda=\Theta(n/q^2)$ and $q^2c=o(n)$ give
$c=o(\lambda)$.
The restriction $q\le n^{1/2-\Omega(1)}$ ensures
$n=\poly(\lambda)$, so the distinguisher runs in time
$2^{O(c)}\poly(n)=2^{o(\lambda)}$, contradicting the
assumed security of $G$.


\subsubsection{Techniques for a linear lower bound on random properties}

Let us first describe a simpler technique to show that a random property does not admit an IPP with sublinear total complexity, which is a strictly weaker task than for QIPPs. The idea is to bound the number of possible sublinear verifier strategies, after making a restriction on the randomness complexity that does not alter the maximal acceptance probabilities by much. If the number of such verifiers is substantially smaller than the number of properties of a specific cardinality, a random property is unlikely to admit a sublinear IPP. This approach was used by Gur and Rothblum~\cite{GR-map} to show an analogous lower bound for \emph{noninteractive} proofs of proximity.

Suppose that $V$ is an IPP verifier that makes $q$ queries and expects $c \ge \log n$ total communication. As~\cite{RVW13} points out, we may assume without loss of generality that $V$ uses at most $R = c + O(1)$ random bits, along the lines of the random-tape sampling of~\cite{GoldreichSheffet2010}. The key observation, allowing us to extend the approach of~\cite{GR-map} to IPPs instead of being limited to MAPs, is the following. Instead of viewing a verifier given an explicit proof string as a decision tree on its own, we view the entire interactive verifier strategy as a decision tree having additional designated inputs and outputs for communication, randomness, input queries, and final decision. With this view, the verifier $V$ can be represented as a decision tree of depth at most $R + q + c$. Since each node has $O(1)$ possible labels for non-query nodes and $n$ labels for query nodes, we can bound the number of verifier strategies by $2^{2^{R+q+c} \log n} = 2^{2^{O(q+c)} \log n}$. Thus if $q+c$ is sublinear, we use the property-counting analysis of~\cite{GR-map} to obtain only a negligible probability that a random property of a specific cardinality admits an $\epsilon$-IPP with a sufficiently small constant $\epsilon > 0$.

Extending this idea to \emph{quantum} communication and queries uses a similar underlying idea: count the entire verifier description under a specific restriction that does not alter expressiveness. There is a continuum of possible sequences of quantum channels, so a direct counting approach does not work. The key idea of our solution is to discretize a \emph{single} operator with a sufficiently small matrix representation that captures all of the verifier's quantum channels. The rounding method follows the $\delta$-netting approach previously used for quantum property testers~\cite{BFNR08} and exact-decision QMA proofs~\cite{RazShpilka2004}.

To start, we represent a quantum interactive verifier strategy $V$ with $q$ queries and $c$ communication as an ensemble of acceptance operators $\{\rho^x_1\}_{x \in \{0,1\}^n}$, using the strategy representation of quantum games due to Gutoski and Watrous~\cite{GutoskiWatrous2007QuantumGames}. The maximum acceptance probability of $V$ on input $x$ can then be described as $\omega_V(x) = \max_{P \in \calP} \braket{\rho_1^x, P}$, for a compact feasible set $\calP$ of prover representations. The size of $\rho_1^x$ is governed by $c$.

Then, we use the polynomial method of~\cite{BealsEtAlPolynomials} to show that every matrix entry of $\rho^x_1$ is a degree at most $2q$ multilinear polynomial in $x$. Hence, for sublinear $q$, each entry of $\rho^x_1$ has a bounded number $M = \sum_{j=0}^{2q}\binom nj$ of nonzero Fourier coefficients, all bounded by a constant magnitude. A key step in our proof is to show that rounding the Fourier coefficients to a $\delta$-net with $\delta = \Theta((2^{2c}M)^{-1})$ preserves the acceptance error uniformly over
\emph{every} input and compatible prover strategy, hence also after
maximization.

Finally, thresholding a rounded description gives a set $S$ satisfying
$\Pi\subseteq S\subseteq N_\epsilon(\Pi)$ for every property it
verifies, where $N_\epsilon(\Pi)$ is the $\epsilon$-neighborhood of
$\Pi$. A Hamming-ball bound limits how many properties of cardinality
$2^{\beta n}$ can share such a set. For $q+c\le\alpha n$, the
description count is small enough when
$2\alpha+H(2\alpha)<\beta$ and $\beta+H(\epsilon)<1$, giving the
linear lower bound. 

\subsection{Related work}


IPPs have been studied under a range of security requirements
and input-access models. These include zero-knowledge
IPPs~\cite{zk-IPP}, IPPs with proof-oblivious
queries~\cite{GRS-proof-oblivious-queries}, sample-based IPPs
where the verifier only receives random labeled samples
from the input~\cite{GR-sample-based}, and distribution-free
IPPs where distance is measured with respect to an unknown
distribution~\cite{AGRR-distribution-free}.
Other related directions include proofs of proximity for distribution testing~\cite{CG-distribution-testing}, locally verifiable codes and their applications to 3-round IPPs for CSPs~\cite{GG-ulvc}, and doubly-sublinear IPPs, in which both the verifier and honest prover have highly restricted query access~\cite{AGR-dsipp}.

Quantum property testing was first introduced by Buhrman, Fortnow, Newman, and Röhrig \cite{BFNR08}; their work provides the conceptual starting point for our lower bound against random properties. Raz and Shpilka~\cite{RazShpilka2004} extend this lower bound idea to treat exact-decision QMA protocols. Montanaro and de Wolf wrote an excellent survey of quantum property testing \cite{MW-qpt-survey}. Dall'Agnol, Gur, Roy Moulik, and Thaler introduced quantum proofs of proximity, or QMAPs, where the verifier receives a non-interactive quantum proof and makes quantum queries~\cite{dallagnol-gur-roymoulik-thaler:qmap}. Query lower bounds for QMA proofs have been studied in several works~\cite{BunJuvekarKing26,Aaronson2012PTP,sherstov-thaler:vanishing-error-qma}. 

Our round-collapse theorem builds on the theory of quantum interactive proofs. This theory includes the celebrated $\QIP = \PSPACE$ theorem of Jain, Ji, Upadhyay, and Watrous~\cite{JJUW-QIP-PSPACE}, the parallelization and amplification results of Kitaev and Watrous~\cite{KW-QIP3}, the constant-round QIP results of Watrous~\cite{Watrous-QIP4}, and the round-reduction of Kempe, Kobayashi, Matsumoto, and Vidick~\cite{KKMV}; see also the survey of Vidick and Watrous on quantum proofs~\cite{VW-quantum-proofs}. Recently, Sun and Vidick introduce \emph{quantum interactive oracle proofs} \cite{SunVidickQIOP26,SunVidickLocal26} as a strengthening of a quantum PCP and a classical interactive oracle proof (IOP): the prover can interactively send many-qubit quantum states as a proof that the verifier can query. Notably, this differs from our model, since the verifier still has full access to the input and need not preserve locality of the input.

Finally, our separation of QIPPs from IPPs and QMAPs is based on Forrelation~\cite{aaronson-ambainis:forrelation}, which has been extensively studied. Our lower bound on IPPs is based on the recent work of Bouland, Huang, Natarajan, Shalit, and Tal~\cite{BQP-IP-separation-26}, who gave an oracle relative to which $\BQP \not\subseteq \IP$ by showing that a promise version of Forrelation does not have an interactive proof in the query model with sufficiently small query and communication complexity, independent of the number of rounds. Prior works gave lower bounds on Forrelation for $\mathsf{AC}^0$ circuits~\cite{raz-tal:bqp-ph,tal:tight-fourier-ac0,bansal-sinha:k-forrelation}, which can imply constant-round IPP lower bounds when combined with techniques of~\cite{RVW13}. 

\subsection{Open questions}\label{Sec:open-questions}
The results within this work present many natural questions and directions for future work, so we discuss some of these here. 

Our main open question is whether the total complexity
$\softO(n^{2/5})$ achieved by
Theorem~\ref{thm:nc-qcipp-informal-main} is optimal.
The conditional lower bound of
Theorem~\ref{thm:main_lower_bound_informal} rules out
polynomial-time QIPPs of total complexity $o(n^{1/3})$
for all logspace-uniform $\NC$ properties.
Closing this gap would determine the optimal exponent
for the total complexity of QIPPs for this class.


\begin{question}\label{qst:qipp-nc}
    What is the infimum of $\delta \in (0, 1)$ such that all logspace-uniform $\NC$ properties admit a QIPP with query and communication complexity $\softO(n^{\delta})$?
\end{question}

We conjecture that the $\softO(n^{2/5})$ total complexity achieved by our method is optimal, but a resolution of this question in either direction would be extremely valuable in better understanding QIPPs.

Recall the bi-directional connection between QIPPs for logspace-uniform $\NC$ and QIPs for Parity. Theorem~\ref{thm:intro-parity-to-nc} implies that an $\Omega(n^{2/5})$ \emph{lower bound} on the total complexity of QIPPs for logspace-uniform $\NC$ would imply that Parity does not have an exact-decision QIP with total complexity $k^{2/3-\Omega(1)}$. The converse, Theorem~\ref{thm:intro-nc-to-parity}, also gives reason to believe that QIPs for Parity may capture the power of QIPPs for logspace-uniform $\NC$, since the reduction in the direction of Theorem~\ref{thm:intro-nc-to-parity} holds for the small proximity parameters that the standard $\pval$-based constructions achieve. Thus, toward answering Question~\ref{qst:qipp-nc}, we ask the following open question.

\begin{question}
    Does there exist some communication bound $c = c(k)$ satisfying $1 \le c(k) \le k$ such that Parity admits a QIP with query complexity $o(k/\sqrt{c})$ and communication complexity $O(c)$?
\end{question}



Given that all public-coin $r$-round QIPPs reduce to only 3 rounds with only $\poly(r)$ overhead, one may ask if we are able to prove a more complete hierarchy theorem for QIPPs. Known results are able to exponentially separate QMAPs from quantum property testers~\cite{dallagnol-gur-roymoulik-thaler:qmap} and 2-round QIPPs (more precisely, 2-round IPPs) from QMAPs~\cite{GLR-MA-AM-proximity,sherstov-thaler:vanishing-error-qma}. However, the relationship between 2-round QIPPs and 3-round QIPPs remains elusive, mirroring the mystery that persists with respect to placing 2-round quantum interactive proofs in surrounding complexity classes~\cite{VW-quantum-proofs}.


\begin{question}
    Does there exist a constant $\epsilon > 0$ such that some property admits a 3-round $\epsilon$-QIPP with query and communication complexity $O(\log(n))$, but does not admit a 2-round $\epsilon$-QIPP with query and communication complexity both $n^{o(1)}$?
\end{question}


\subsection{Organization}

The rest of this work is organized as follows. Section~\ref{Sec:preliminaries} gives the necessary background. Sections~\ref{Sec:nc-aa-speedup}--\ref{Sec:qipp-nc-equiv-qip-parity} study QIPPs for logspace-uniform $\NC$: Section~\ref{Sec:nc-aa-speedup} constructs our improved QCIPPs, Section~\ref{Sec:QIPP_lower_bound} proves a lower bound under a cryptographic assumption, and Section~\ref{Sec:qipp-nc-equiv-qip-parity} establishes the bidirectional connection to query-model QIPs for parity. Section~\ref{Sec:QIPP-random-lower-bound} then proves a tight unconditional lower bound on the total QIPP complexity of random properties. Sections~\ref{Sec:ipp-aa-speedup} and~\ref{Sec:QIPP_3_rounds} give general transformations for proofs of proximity: Section~\ref{Sec:ipp-aa-speedup} establishes a generic quantum query speedup for classical IPPs, while Section~\ref{Sec:QIPP_3_rounds} shows that public-coin QIPPs can be compressed to three rounds. Finally, Section~\ref{Sec:qipp-ipp-qmap-separation} gives an exponential separation of QCIPPs from both round-unrestricted IPPs and QMAPs.

\section{Preliminaries}\label{Sec:preliminaries}
In this section, we establish basic probability and asymptotic notation, and define the necessary notions related to quantum computing, classical and quantum interactive proofs of proximity.

\subsection{Notation}

Throughout this work, we denote the set $\{1, \dots, n\}$ for a positive integer $n$ by $[n]$. 

For an event $A$ in a probability space, we denote the probability of $A$ by $\Pr[A]$. If $X$ is a random variable, its expectation is denoted $\expect{X}$. We denote by $a \gets X$ or $a \sim X$ an element $a$ that is sampled from a distribution or random variable $X$.

Let $f : \N \to \N$ and $g : \N \to \N$. We write $f(n) = \polylog(n)$ if there exists a constant $k \in \N$ such that $f(n) = O(\log^k(n))$. We write $f(n) = \softO(g(n))$ if $f(n) = g(n) \cdot \polylog(n)$.\footnote{It is also common to use $f(n) = \softO(g(n))$ to mean that $f(n) = g(n) \cdot \polylog(g(n))$. Using $\softO$ to hide factors of $\polylog(n)$ is more convenient for this work.}

\subsection{Quantum computing}

\paragraph{Linear algebra.}
Let $\calH, \calK$ be finite-dimensional complex Hilbert spaces. The set of linear operators from $\calH$ into $\calK$ is denoted by $\sfL(\calH, \calK)$; the set of linear operators on $\calH$ is denoted $\sfL(\calH)$. The identity operator on $\calH$ is denoted $\mathbf{1}_{\calH}$. We say that a Hermitian operator $\rho \in \sfL(\calH)$ is \emph{positive semidefinite}, denoted by $\rho \ge 0$, if its eigenvalues are all nonnegative. 

A \emph{super-operator} is simply a linear operator of the form $\Phi : \sfL(\calH) \to \sfL(\calK)$. A super-operator $\Phi$ is \emph{positive} if for all $\rho \in \sfL(\calH)$ with $\rho \ge 0$, we have $\Phi(\rho) \ge 0$. The super-operator $\Phi$ is \emph{completely positive} if for every Hilbert space $\calM$, the map $\Phi \otimes \mathbf{1}_{\calM} : \sfL(\calH) \otimes \sfL(\calM) \to \sfL(\calK) \otimes \sfL(\calM)$ is a positive super-operator. The super-operator $\Phi$ is \emph{trace-preserving} if $\tr(\Phi(\rho)) = \tr(\rho)$ for all $\rho \in \sfL(\calH)$, where $\tr : \sfL(\calH) \to \C$ denotes the trace. The \emph{partial trace} super-operator $\tr_\calK : \sfL(\calH \otimes \calK) \to \sfL(\calH)$ on $\calK$ is defined as the unique linear map such that for all $\rho \in \sfL(\calH)$ and $\sigma \in \sfL(\calK)$, we have $\tr_{\calK}(\rho \otimes \sigma) = \tr(\sigma) \rho$.

\paragraph{Quantum states.}
A \emph{pure state} on $\calH$ is a unit vector $\ket{\psi} \in \calH$, represented with the usual Dirac notation. For a pure state $\ket{\psi}$, we define the dual to $\ket{\psi}$ as a linear functional $\bra{\psi} : \calH \to \C$ with $\bra{\psi}(\ket{\phi}) = \braket{\psi \mid \phi}$, where $\braket{\psi \mid \phi}$ denotes the inner product of $\psi$ and $\phi$ given by $\calH$. A \emph{mixed state} is an arbitrary distribution over pure states, say $\{p_k, \ket{\psi_k}\}_{k}$, represented as an operator $\rho = \sum_{k} p_k \ket{\psi_k} \bra{\psi_k}$. Mixed states on $\calH$ are precisely characterized by \emph{density operators} $\rho \in \mathsf{L}(\calH)$, which are those operators that are positive semidefinite with unit trace. The set of all density operators on $\calH$ is denoted $\sfD(\calH)$.

\paragraph{Quantum channels.}
All computation on states can be done via a combination of \emph{unitary transformations} and \emph{measurements}. A linear operator $U \in \mathsf{L}(\calH)$ is called \emph{unitary} if $U$ preserves inner products:
$
\braket{U\phi \mid U \psi} = \braket{\phi \mid \psi}.
$
Equivalently, we require that $U^\dagger U = I$, where $U^\dagger$ denotes the \emph{adjoint} of $U$. ($U^\dagger$ is also called the \emph{conjugate transpose} of $U$.)
Applying a unitary $U$ to a pure state $\ket{\psi}$ results in the state $U \ket{\psi}$; when applied to a mixed state $\rho$, we obtain the mixed state $U \rho U^\dagger$. 

Measuring a pure state $\ket{\psi}$ with respect to mutually
orthogonal projectors $\{P_1,\dots,P_m\}$ satisfying
$\sum_{k=1}^m P_k = I$ produces outcome $k$ with probability
$p_k = \braket{\psi \mid P_k \mid \psi}$.
Conditioned on an outcome $k$ with $p_k > 0$, the post-measurement
state is $P_k\ket{\psi}/\sqrt{p_k}$.
If the outcome is discarded, the resulting density operator is
$\sum_{k=1}^m P_k\ket{\psi}\bra{\psi}P_k$.

Both unitary evolution and measurements are special cases of quantum channels. A \emph{quantum channel} from $\calH$ to $\calK$ is given by a super-operator $\Phi : \sfL(\calH) \to \sfL(\calK)$ that is completely-positive and trace-preserving. When a quantum channel $\Phi$ is applied to a state $\rho \in \sfD(\calH)$, the resulting state is $\Phi(\rho)$.

For a unitary $U$ on $\calH$, the corresponding quantum channel is $\rho \mapsto U \rho U^\dagger$. 
For a fixed ancilla state $\sigma \in \sfD(\calK)$,
the channel that appends this ancilla is
$\Phi_\sigma : \sfL(\calH) \to \sfL(\calH \otimes \calK)$,
defined by $\Phi_\sigma(\rho) = \rho \otimes \sigma$.
Partial tracing also provides a convenient quantum channel: the partial trace map $\tr_{\calK} : \sfL(\calH \otimes \calK) \to \sfL(\calH)$ is a quantum channel that ``forgets'' everything in the register corresponding to $\calK$. By the Stinespring dilation theorem, any quantum channel can be decomposed as a composition of three channels corresponding to adding ancillas, unitaries, and partial tracing~\cite{Stinespring1955,NielsenChuang2010}.



\paragraph{Computation.}
A \emph{qubit} is a state in $\C^2$. We write the \emph{computational basis} of $\C^2$ by $\{\ket{0}, \ket{1}\}$, so that pure states can be written as $\alpha \ket{0} + \beta \ket{1}$ where $|\alpha|^2 + |\beta|^2 = 1$. Without loss of generality, computation is always done over the Hilbert space on $m$-qubits, $\calH = (\C^2)^{\otimes m}$. 

A \emph{quantum circuit} with explicit input $x \in \{0,1\}^n$ over $m$ qubits has initial state $\ket{x} \otimes \ket{\mathbf{0}} = \bigotimes_{k=1}^n \ket{x_k} \otimes \ket{\mathbf{0}}$, and applies unitary transformations on a bounded constant number of qubits at a time. The unitaries come from a small set of \emph{universal} gates for quantum computation, the details of which are unimportant. The circuit may include measurement gates at the end.

For the oracle algorithms that we work with in property testing, quantum circuits instead have the initial state $\ket{\mathbf{0}}$ and access to an \emph{oracle gate} $U_x$ that maps the basis states as $\ket{i}\ket{b} \mapsto \ket{i}\ket{b \xor x_i}$, for an input string $x$. Such an oracle gate not only allows one to query $x_i$ by passing the state $\ket{i}\ket{0}$ and getting $\ket{i}\ket{x_i}$, but one can query $x$ on a superposition of indexes as well.

Lastly, a \emph{uniform quantum algorithm} is a uniformly generated family of quantum circuits.

\subsection{Classical and quantum interactive protocols in the query model}\label{Subsec:quantum_interactive_proofs}

In this section, we define both classical and quantum interactive protocols in the query model. Our definition captures the standard notion of classical and quantum interactive proofs as well, since such a proof without loss of generality makes all $n$ queries at the beginning of the protocol.

\subsubsection{Classical interactive protocols} First, we define a randomized algorithm with oracle access. A $q$-query randomized algorithm on implicit length $n$ is a randomized algorithm $A$ having oracle access to some $x \in \{0,1\}^n$ and explicit access to some fixed-length string $w$ such that $A$ makes at most $q$ point queries $i \mapsto x_i$ to $x$. The output of $A$ on oracle input $x$ and explicit input $w$, with random tape $r \in \{0,1\}^*$ is denoted $A^x(w; r)$. The random variable described by selecting $r$ uniformly at random and outputting $A^x(w; r)$ is denoted $A^x(w)$. For a function $q : \N \to \N$, a \emph{$q$-query randomized algorithm} is a function $A$ that maps each $n \in \N$ to a $q(n)$-query randomized algorithm on implicit input length $n$. The map $A$ is said to have query complexity $q$. If $A$ is also uniformly computable in time $t$, then $A$ is said to have \emph{time complexity $t$}.

An \emph{interactive protocol} in the query model is a pair of functions $(P, V)$ that given $n \in \N$, $P(n)$ outputs functions $(P_1, \dots, P_m)$ and $V(n)$ outputs randomized query algorithms $(V_0, V_1, \dots, V_m)$, all taking oracle inputs of length $n$.
Define the output of $V$ in the interaction $(P, V)$ on input $x \in \{0,1\}^n$ as follows. Initialize $\gamma$ as an empty list.
Compute $\gamma_{V,0} \gets V_0^x(\gamma)$ and set $\gamma := (\gamma_{V,0})$. For all $i = 1, \dots, m$, compute $\gamma_{P,i} \gets P_i(x, \gamma)$ and append $\gamma_{P,i}$ to $\gamma$, then compute $\gamma_{V,i} \gets V_i^x(\gamma)$ and append $\gamma_{V,i}$ to $\gamma$. The list $\gamma$ is called the \emph{transcript} of the interaction. The \emph{output} of the interaction is $\gamma_{V,m}$, which must be in $\{0,1\}$, where $1$ indicates $\mathsf{accept}$ and $0$ indicates $\mathsf{reject}$. The random variable that gives the output of this interaction is denoted $\braket{P(n, x), V^x(n)}$ or just $\braket{P(x), V^x}$.

The function $V$ is called an \emph{interactive verifier} and $P$ is called an \emph{interactive prover}. If $P$ is a function that always sends and receives messages of the expected length (given by the fixed length of inputs expected by each randomized query algorithm), then $P$ is said to be a \emph{prover compatible with $V$}. 
For $x \in \{0,1\}^n$, let
\[
\omega_V(x) :=
\max_{P^*} \Pr[\braket{P^*(x), V^x} = 1],
\]
where the maximum ranges over all provers $P^*$ compatible with $V$.

The \emph{round complexity} of $(P, V)$ is $2m$ if $V_0$ does not output the empty string (which would indicate an empty message), and is $2m-1$ otherwise.
The \emph{query complexity} of $(P,V)$ is the sum of the query complexities of each $V_0, \dots, V_m$, and is a function of the input length $n$. (Note that we always assume that $P$ is all-powerful throughout this work, so its query complexity is not of interest to us.) The \emph{communication complexity} of $(P, V)$ is the sum over the length of the outputs of each $V_0^x(\gamma), P_1(x,\gamma), V_1^x(\gamma), \dots, P_m(x,\gamma)$ where $\gamma$ is the evolving transcript described above. The \emph{total complexity} of $(P, V)$ is the sum of its query and communication complexity. If $V$ can be computed uniformly in time $t : \N \to \N$, then $V$ has \emph{time complexity} $t$.
We assume that every round (except potentially the first round) transmits a nonzero number of bits. Thus, the round complexity is always bounded by the communication complexity.

\subsubsection{Quantum interactive protocols}
The definition of a classical interactive protocol given above crucially relies on the information sent in each round being copied into a local copy of the transcript. When quantum information is exchanged, the no-cloning theorem prohibits copying of states. We fix this issue by defining quantum interactive protocols so that both parties have access to a private register in which to store state between rounds. Our definition is inspired by Gutoski and Watrous~\cite{GutoskiWatrous2007QuantumGames} and Beigi, Shor, and Watrous~\cite{BSW-QIP-short-messages}.

First, let us define a quantum query channel. Recall that for $x \in \{0,1\}^n$, the unitary $U_x$ denotes the query unitary for $x$, which acts on $\C^n \otimes \C^2$; denote $\calI = \C^n \otimes \C^2$. The channel associated with $U_x$ is denoted $\calU_x$, acting on $\sfL(\calI)$ as $\rho \mapsto U_x \rho U_x^{\dagger}$. Let $\calH$ and $\calK$ be Hilbert spaces. An ensemble of channels $\{\Phi^x\}_{x \in \{0,1\}^n}$ is a \emph{$q$-query quantum channel ensemble from $\sfL(\calH)$ to $\sfL(\calK)$} if there exist Hilbert spaces $\calM_1, \dots, \calM_q$ and channels
\begin{itemize}
    \item $\Psi_1 : \sfL(\calH) \to \sfL(\calI) \otimes \sfL(\calM_1)$,
    \item $\Psi_i : \sfL(\calI) \otimes \sfL(\calM_{i-1}) \to \sfL(\calI) \otimes \sfL(\calM_i)$ for $2 \le i \le q$, and
    \item $\Psi_{q+1} : \sfL(\calI) \otimes \sfL(\calM_q) \to \sfL(\calK)$
\end{itemize}
such that for all $x \in \{0,1\}^n$,
\[
\Phi^x = \Psi_{q+1} \circ (\calU_x \otimes \mathbf{1}_{\sfL(\calM_q)}) \circ \Psi_q \circ \cdots \circ \Psi_2 \circ (\calU_x \otimes \mathbf{1}_{\sfL(\calM_1)}) \circ \Psi_1.
\]

Let $\{\calH_n\}_{n \in \N}$ and $\{\calK_n\}_{n \in \N}$ be ensembles of Hilbert spaces. For a function $q : \N \to \N$,
a \emph{$q$-query quantum channel from $\{\calH_n\}_{n \in \N}$ to $\{\calK_n\}_{n \in \N}$} is a function $Q$ that maps $n \in \N$ to a $q(n)$-query quantum channel ensemble $\{\Phi_n^x\}_{x \in \{0,1\}^n}$ from $\sfL(\calH_n)$ to $\sfL(\calK_n)$.

Suppose that $\calH_n = (\C^2)^{\otimes k_n}$ and $\calK_n = (\C^2)^{\otimes \ell_n}$ for some $k_n, \ell_n \in \N$, so that channels from $\calH_n$ to $\calK_n$ may be described by quantum circuits.
If there exists a uniform algorithm that, given $n$, generates in time $t(n)$ a description of the $q(n)$-query quantum circuit that implements $Q(n) = \{\Phi^x_n\}_{x \in \{0,1\}^n}$ in terms of the oracle $\calU_x$, then we say that $Q$ has \emph{time complexity} $t$. 

A \emph{quantum interactive protocol} in the query model is a pair of functions $(P, V)$ that, given $n \in \N$, $P(n)$ outputs ensembles of quantum channels $P_1 = \{\Psi_1^x\}_{x \in \{0,1\}^n}, \dots, P_m = \{\Psi_m^x\}_{x \in \{0,1\}^n}$ and $V(n)$ outputs quantum query channels $V_0 = \{\Phi_0^x\}_{x \in \{0,1\}^n}, \dots, V_m = \{\Phi_m^x\}_{x \in \{0,1\}^n}$. Let $\calA = \C^2$ with basis $\{\ket{0}, \ket{1}\}$. There must exist Hilbert spaces $\calX_1, \dots, \calX_{m}$ and $\calZ_1, \dots, \calZ_m$ and $\calY_1, \dots, \calY_{m}$ and $\calW_1, \dots, \calW_{m}$ such that
\begin{itemize}
    \item $V_0 : \sfL(\C) \to \sfL(\calX_1 \otimes \calZ_1)$;
    \item for $1 \le i \le m-1$, we have $V_i : \sfL(\calY_i \otimes \calZ_i) \to \sfL(\calX_{i+1}\otimes \calZ_{i+1})$;
    \item $V_m : \sfL(\calY_m \otimes \calZ_m) \to \sfL(\calA)$;
\end{itemize}
and also
\begin{itemize}
    \item $P_1 : \sfL(\calX_1) \to \sfL(\calY_1 \otimes \calW_1)$;
    \item for $2 \le i \le m-1$, we have $P_i : \sfL(\calX_i \otimes \calW_{i-1}) \to \sfL(\calY_{i}\otimes \calW_{i})$;
    \item $P_m : \sfL(\calX_m \otimes \calW_{m-1}) \to \sfL(\calY_{m})$.
\end{itemize}

The outcome qubit from the interaction $(P, V)$ on input $x \in \{0,1\}^n$ is computed as follows. The verifier computes a state in $\sfD(\calX_1 \otimes \calZ_1)$ from $\Phi^x_0$, and sends the register for $\calX_1$ to the prover while keeping the register in $\calZ_1$; the prover applies the channel $\Psi^x_1$ on the received state to obtain a state in $\sfD(\calY_1 \otimes \calW_1)$, and sends the register in $\sfD(\calY_1)$ to the verifier; the verifier applies $\Phi^x_1$ on the received state in $\sfD(\calY_1)$ and its leftover state in $\sfD(\calZ_1)$, and sends the part in $\sfD(\calX_2)$ to the prover. This exchange goes on until the final verifier channel obtains a state in $\sfD(\calY_m)$ from the prover, and applies $\Phi^x_m$ to this register and its leftover register $\sfD(\calZ_m)$ to obtain a qubit in $\sfD(\calA)$. The protocol outputs $1$ if a computational basis measurement outcome of the output qubit $\rho^x \in \sfD(\calA)$ is $\ket{1}$, which happens with probability $\tr(\ket{1}\bra{1} \rho) = \braket{1 \mid \rho \mid 1}$. Succinctly, the output qubit $\rho^x$ can be described by
\[
\rho^x
=
\Phi_m^x
\circ
\left(\Psi_m^x \otimes \mathbf{1}_{\sfL(\calZ_m)}\right)
\circ
\left(\Phi_{m-1}^x \otimes \mathbf{1}_{\sfL(\calW_{m-1})}\right)
\circ \cdots \circ
\left(\Phi_1^x \otimes \mathbf{1}_{\sfL(\calW_1)}\right)
\circ
\left(\Psi_1^x \otimes \mathbf{1}_{\sfL(\calZ_1)}\right)
\circ
\Phi_0^x(\mathbf{1}_{\C}),
\]
where $\mathbf{1}_\C$ is the unique density operator on $\C$. The outcome of a measurement of $\rho^x$ in the computational basis is denoted $\braket{P(n, x), V^x(n)}$ or just $\braket{P(x), V^x}$.

As with classical interactive protocols, $V$ is called a \emph{quantum interactive verifier} and $P$ is called a \emph{quantum interactive prover}. Furthermore, any prover $P^*$ that can be described in the above form is \emph{compatible with $V$}. Like for classical verifiers, if $V$ is a quantum interactive verifier, we denote by $\omega_V(x)$ the maximum value of $\Pr[\braket{P^*(x), V^x}=1]$ over all quantum interactive provers $P^*$ compatible with $V$.

The complexity definitions for quantum protocols are similar to those of classical protocols, but there are subtle differences.
The \emph{round complexity} of $(P, V)$ is $2m$ if $\dim \calX_1 > 1$ and is $2m-1$ otherwise.
The \emph{query complexity} of $(P,V)$ is the sum of the query complexities of each $V_0, \dots, V_m$, and is a function of the input length $n$. The \emph{communication complexity} of $(P, V)$ is
\[
\sum_{i=1}^m \ceil{\log \dim \calX_i} + \ceil{\log \dim \calY_i},
\]
since a state in a space of dimension $d$ can be described with at most $\ceil{\log d}$ qubits. The \emph{total complexity} of $(P, V)$ is the sum of its query and communication complexity.
If $V$ can be computed uniformly in time $t : \N \to \N$, then $V$ has \emph{time complexity} $t$. We assume that every round (except possibly the first) transmits a nonzero number of qubits, so that round complexity is always bounded by communication.

\paragraph{Quantum--classical interactive protocols.} 
A common subclass of quantum interactive protocols we study in this work is the class of \emph{quantum--classical interactive protocols}. These are interactive protocols in which all communication is classical, but the next-message query algorithms are allowed to be quantum algorithms that can implement quantum query channels, as described above. In particular, although the communication is constrained to classical strings, the next-message algorithms may query $U_x$ on states, rather than being constrained to point-queries of $x$. One may more formally define this model as a quantum interactive protocol, where the verifier always immediately measures in the computational basis any received states and measures states before sending them. Since this collapses any incoming or outgoing states to classical strings, we effectively simulate classical communication channels this way.

\paragraph{Public-coin interactive protocols.}
Another subclass of quantum (and classical) interactive protocols we study significantly in this work are those that are \emph{public-coin}. A public-coin protocol is one where the verifier's messages are classical and consist solely of $k_i(n)$ uniform bits for its $i$th message, where the length $k_i(n)$ is a function of only $n$. For classical protocols, we sometimes require that the verifier's query locations and decision are deterministic functions of the transcript and the input. Such classical protocols are called \emph{public-coin with deterministic decision}. A quantum public-coin protocol still allows the verifier to receive arbitrary states from the prover. Such protocols are conveniently \emph{query oblivious}, meaning that all queries to $U_x$ without loss of generality occur after all communication is complete. We make use of this structural property.

\subsection{Interactive proofs of proximity}

With the notion of classical and quantum interactive protocols established, we are now ready to define the corresponding notions of proofs of proximity for a given property.
We are primarily concerned with properties of binary strings, i.e. languages, but one may have analogous definitions for properties consisting of more general functions. See~\cite{zk-IPP} for more details on general properties.

A \emph{property} is an ensemble $\Pi = (\Pi_n)_{n \in \N}$ where $\Pi_n \subseteq \{0,1\}^n$.
For two strings $x, y \in \{0,1\}^n$, we denote their Hamming distance by $\dist(x, y) := |\{i \in [n] : x_i \ne y_i\}|$ and their normalized Hamming distance by
$
\Delta(x,y) := \frac{\dist(x, y)}{n}.
$
For $x \in \{0,1\}^n$ and a property $\Pi$, we define the distance between $x$ and $\Pi$ by $\Delta(x, \Pi) \coloneqq \min_{y \in \Pi_n} \Delta(x,y)$. If $\Delta(x, \Pi) > \epsilon$, then we say that $x$ is \emph{$\epsilon$-far} from $\Pi$; otherwise, $x$ is said to be \emph{$\epsilon$-close} to $\Pi$. Abusing notation slightly, we often use $\Pi$ to also refer to the set $\bigcup_{n \in \N} \Pi_n$.

Define logspace-uniform $\NC$, or just $\NC$, to be the class of properties $\Pi$ that can be exactly decided by a logspace-uniform family of circuits $\{C_n\}_{n \in \N}$ of size $\poly(n)$ and depth $\polylog(n)$; exact decision means that $C_n(x) = 1$ if and only if $x \in \Pi_n$, for all $x \in \{0,1\}^n$.

Let us proceed to formally defining interactive and quantum interactive proofs of proximity, the main proof models of this work.


\begin{definition}
    Let $\Pi = (\Pi_n)_{n \in \N}$ be a property, and let $\epsilon = \epsilon(n) \in [0,1)$.
    An interactive (respectively, quantum interactive) protocol $(P,V)$ is an \emph{interactive proof of $\epsilon$-proximity} (respectively, \emph{quantum interactive proof of $\epsilon$-proximity}) for $\Pi$ with completeness $a : \N \to (0,1]$ and soundness $b : \N \to [0,1)$, where $a(n) > b(n)$ for every $n \in \N$, if the following hold for every $n \in \N$ and $x \in \{0,1\}^n$:
    \begin{enumerate}
        \item \emph{Completeness:} If $x \in \Pi_n$, then
        $\Pr[\braket{P(x),V^x(n)} = 1] \ge a(n)$.
        \item \emph{Soundness:} If $x$ is $\epsilon$-far from $\Pi$, then for every (respectively, quantum) prover $P^*$ compatible with $V$,
        $\Pr[\braket{P^*(x),V^x(n)} = 1] \le b(n)$.
    \end{enumerate}
    Unless otherwise specified, interactive proofs of proximity and quantum interactive proofs of proximity have completeness $2/3$ and soundness $1/3$.
\end{definition}

An interactive proof of $\epsilon$-proximity and a quantum interactive proof of $\epsilon$-proximity for $\Pi$ are also called an $\epsilon$-IPP and an $\epsilon$-QIPP for $\Pi$, respectively. The complexity measures and uniformity of a proof of proximity are inherited from the complexity of the underlying protocol.

If an $\epsilon$-QIPP $(P, V)$ for $\Pi$ is a quantum--classical interactive protocol, we say that $(P, V)$ is an \emph{$\epsilon$-QCIPP}.

The set of all properties having $\epsilon$-IPPs (respectively, $\epsilon$-QCIPP and $\epsilon$-QIPP) with completeness $a$, soundness $b$, round complexity $r$, communication complexity $c$, query complexity $q$, and time complexity $t$ is denoted $\mathbf{IPP}_{a,b}(\epsilon, r, c, q, t)$ (respectively, $\mathbf{QCIPP}_{a,b}(\epsilon, r, c, q, t)$ and $\mathbf{QIPP}_{a,b}(\epsilon, r, c, q, t)$). The completeness and soundness $a$ and $b$ may both be omitted to mean completeness $2/3$ and soundness $1/3$. Additionally, $t$ may be omitted, in which case the corresponding class allows for arbitrary nonuniform algorithms or channels.

Note that for $\epsilon = 0$, the soundness condition of proofs of proximity reduces to the standard soundness, holding for every $x \notin \Pi_n$. Thus we have the following definition for exact decision.

\begin{definition}
    An \emph{interactive proof} (respectively, a quantum--classical interactive proof, or quantum interactive proof) for a property $\Pi$ is a $0$-IPP (respectively, $0$-QCIPP, or $0$-QIPP) for $\Pi$.
\end{definition}

An interactive proof for $\Pi$ is called an IP for $\Pi$, and similarly, a quantum--classical interactive proof is abbreviated as a QCIP, and a quantum interactive proof is abbreviated as a QIP.

For a proximity parameter $\epsilon$, a property $\Pi$, and a parameter $c : \N \to \N$ with $1 \le c(n) \le n$, the \emph{$\epsilon$-QIPP query complexity} of $\Pi$ subject to communication $c$ is the minimum number of queries over all $\epsilon$-QIPPs for $\Pi$ with communication at most $c$. This value is denoted by $\sfQIPP_\epsilon(\Pi ; c)$. For QIPs, we denote $\sfQIP(\Pi ; c) := \sfQIPP_0(\Pi ; c)$. The \emph{$\epsilon$-QIPP total complexity} of $\Pi$ is the minimum value of $c + \sfQIPP_\epsilon(\Pi ; c)$ over all $c$, and is denoted $\sfQIPP_\epsilon(\Pi)$. The \emph{QIP total complexity} of $\Pi$ is defined similarly for QIPs, and is denoted by $\sfQIP(\Pi)$.

We recall the following results for IPPs and QIPPs, which will be used several times throughout this work. The first, Lemma~\ref{lem:GS-private-to-public-coin} is the Goldwasser--Sipser transformation for converting private-coin (classical) protocols to public-coin ones.

\begin{lemma}[\cite{goldwasser-sipser:private-public-coins,RVW13}]\label{lem:GS-private-to-public-coin}
    If a property $\Pi$ has an $\epsilon$-IPP with $r(n)$ rounds, $c(n) \ge \log n$ communication complexity, and $q(n)$ query complexity, then $\Pi$ has an $\epsilon$-IPP that is public-coin with deterministic decision having $O(r(n))$ rounds, $c(n) \cdot t(n)$ communication complexity, and $q(n) \cdot t(n)$ query complexity, where $t(n) = O(r(n) \cdot \log (c(n) \cdot r(n)))$.
\end{lemma}

The second establishes soundness reduction by parallel repetition for QIPPs (and hence, QCIPPs as well). The proof follows from perfect AND-parallel repetition~\cite{GutoskiWatrous2007QuantumGames} and the generalized Chernoff bound~\cite{impagliazzo-kabanets:concentration}.

\begin{restatable}[Parallel repetition]{lemma}{LemParallelRepetition}\label{lem:parallel_repetition}
    Suppose a property $\Pi$ admits an $\epsilon$-QIPP with completeness at least $a$, soundness at most $b<a$, and round, communication, and query complexities $r,c,$ and $q$. For $0<\eta<1/2$, it admits an $\epsilon$-QIPP with completeness at least $1-\eta$, soundness at most $\eta$, and respective complexities $r,O(kc),O(kq)$, where
    \[
     k=O\!\left(\frac{\log(1/\eta)}{(a-b)^2}\right).
    \]
    The transformation preserves perfect completeness and public-coin protocols. A verifier with time complexity $t$ yields a verifier with time complexity $O(t\poly(k))$.
\end{restatable}
\begin{proof}
    The proof of this lemma can be found in Appendix~\ref{App:qipp-3-proof}.
\end{proof}

The last lemma we recall is the technique of \emph{amplitude amplification}.

\begin{lemma}[Amplitude amplification,~\cite{BHMT-amplitude-amplification}]\label{lem:ampl-ampl}
    Let $A$ be a randomized algorithm with oracle access to a string $x \in \{0,1\}^n$ that makes $q$ queries, promised that either $A$ outputs $1$ with probability $0$ or outputs $1$ with probability at least $\gamma$ for an input $x$. Then there exists a quantum algorithm $B$ that makes $O(q/\sqrt{\gamma})$ queries to the unitary $U_x \ket{i}\ket{b} = \ket{i}\ket{b \oplus x_i}$, outputs $1$ with probability $0$ when $A(x)$ outputs $1$
    with probability $0$, and outputs $1$ with probability at least
    $2/3$ whenever $A(x)$ outputs $1$ with probability at least $\gamma$.
\end{lemma}

\subsection{Cryptographic definitions}\label{Subsec:crypto}
Our lower bound in Section~\ref{Sec:QIPP_lower_bound} uses cryptographic assumptions. We provide the necessary cryptographic definitions here. For further details, see any standard text on cryptography~\cite{goldreich:cryptography}.

A function $\mu : \N \to \R_{\ge 0}$ is \emph{negligible} if for every constant $c > 0$, we have $\mu(\lambda) = O(1/\lambda^c)$. If a function $\mu(\lambda)$ is negligible, we often write $\mu(\lambda) = \negl(\lambda)$.

\begin{definition}[$T$-secure PRG]\label{def:PRG}
    For a function $T: \N \to \N$, a \emph{$T$-secure pseudorandom generator (PRG)} is a polynomial-time computable function $G : \{0,1\}^\lambda \to \{0,1\}^{\ell(\lambda)}$ for some $\ell(\lambda) > \lambda$ such that for any probabilistic time-$T$ algorithm $D$ (a \emph{distinguisher}), for all $\lambda \in \N$, we have
    \[
    \left| \Pr_{s \gets \{0,1\}^\lambda}[D(G(s)) = 1] - \Pr_{r \gets \{0,1\}^{\ell(\lambda)}}[D(r) = 1] \right| \le \negl(\lambda).
    \]

    Such a function $G$ is a \emph{post-quantum $T$-secure pseudorandom generator} if the security property holds for all time-$T$ quantum algorithms.
\end{definition}

\begin{definition}\label{def:subexp_PRG}
    A $2^{o(\lambda)}$-secure PRG is called a \emph{subexponentially-secure PRG}. Similarly, a post-quantum $2^{o(\lambda)}$-secure PRG is called a \emph{post-quantum subexponentially-secure PRG}.
\end{definition}

\section{Logspace-uniform NC has QCIPPs with \texorpdfstring{$o(\sqrt{n})$}{o(sqrt(n))} complexity}\label{Sec:nc-aa-speedup}


In this section, we show that every property in logspace-uniform $\NC$ admits a QCIPP with subquadratic query and communication complexity, answering in the positive an open question of Dall'Agnol, Gur, Roy Moulik, and Thaler~\cite{dallagnol-gur-roymoulik-thaler:qmap} asking whether every property in logspace-uniform $\NC$ admits QIPPs with both query and communication complexity $o(\sqrt{n})$. Our main theorem is the following.

\begin{theorem}\label{thm:nc-in-qcipp}
    Let $\Pi$ be a property in logspace-uniform $\NC$ and let $\epsilon=\epsilon(n)\in[1/n,1)$ be a proximity parameter. Then $\Pi$ has a public-coin $\epsilon$-QCIPP with perfect completeness, soundness $1/2$, and the following complexities.
    \begin{itemize}
        \item Query complexity
        \(
        \softO\left(\frac{1}{\epsilon^{3/2}\sqrt n}
             +\frac{1}{\sqrt\epsilon}\right).
        \)
        \item Communication complexity $\softO(\epsilon n)$.
        \item Time complexity $\softO(\epsilon n + \frac{1}{\epsilon^{3/2}\sqrt n} +\frac{1}{\sqrt\epsilon})$.
        \item Round complexity $\polylog(n)$.
    \end{itemize}
\end{theorem}

Setting $\epsilon = n^{-3/5}$ in Theorem~\ref{thm:nc-in-qcipp} and simplifying the complexity bounds gives the following corollary.

\begin{corollary}
    Every property in logspace-uniform $\NC$ has a public-coin $n^{-3/5}$-QCIPP with perfect completeness, soundness $1/2$, query complexity $\softO(n^{2/5})$, communication complexity $\softO(n^{2/5})$, time complexity $\softO(n^{2/5})$, and $\polylog(n)$ rounds.
\end{corollary}

Our protocol exploits the linear structure of the queries specified by the Rothblum--Rothblum IPP~\cite{RR20} for logspace-uniform $\NC$. Throughout this section, we use the revised full version of their paper~\cite{RR20-revised} that fixed a mistake in the construction. 
After the interaction, the verifier's remaining task can be viewed as checking a small sample (about $\polylog(n)$) of large linear identities of the form
\[
\sum_{i \in [p]} \alpha_k(i)\cdot x[q_k(i)] = \beta_k,
\]
where the locations $q_k(i) \in [n]$ and coefficients $\alpha_k(i) \in \F$ are computable from a short description in the transcript. Importantly, and as we show, the classical query complexity lies almost entirely in the cost of checking a single one of these identities. Then, instead of asking the prover for all queried values in a sampled identity, the verifier partitions the $p$ terms into $c$ blocks and asks the prover only for the $c$ corresponding partial sums. If the original identity is false but the claimed partial sums add to the claimed total, then at least one block sum must be incorrect; amplitude amplification allows the verifier to find such a block with fewer queries than classical sampling. For $1 \le c \le p$, this reduces the cost of checking an identity from $p$ queries to only $O(\lceil p/c\rceil\sqrt{c}) = O(p/\sqrt{c})$.

This yields a QCIPP for every property in logspace-uniform $\NC$ with query and communication complexity $\softO(n^{2/5})$. In Section~\ref{Sec:advice-aa}, we also construct a more efficient amplitude amplification algorithm that avoids having to load classical data in every amplification iteration. As a consequence, we can reduce the time complexity down to the sum of query and communication complexities (up to polylogarithmic factors), which is $\softO(n^{2/5})$ in the above balanced regime.
Moreover, the protocol is public-coin, so the round-collapse theorem we show later, Theorem~\ref{thm:pub-coin-round-reduction}, can be used to give a three-round QIPP with the same query, communication, and time bounds up to polylogarithmic factors. See Section~\ref{Sec:QIPP_3_rounds} for further details on this round collapse.

\subsection{A structural form of the Rothblum--Rothblum IPP}

We now isolate the part of the Rothblum--Rothblum IPP for logspace-uniform $\NC$~\cite{RR20-revised} that will be used in our quantum speedup. We need the fact that, after the interactive part of the protocol, the verifier is left with a list of linear constraints on the input $x$. These constraints have the form
\[
\sum_{i \in [p]} \alpha_k(i) \cdot x[q_k(i)] = \beta_k,
\]
where $k$ indexes the constraint, the location function $q_k : [p] \to [n]$ tells the verifier which input locations are involved, the weight function $\alpha_k : [p] \to \F$ gives the corresponding coefficients, and $\beta_k \in \F$ is a value claimed by the prover.

The theorem below should be read as a structural restatement of their protocol. Completeness says that, when $x \in \Pi$ and the prover is honest, all of these linear constraints are true. Soundness says that, when $x$ is far from $\Pi$, then for every cheating prover, either the verifier already rejects during the interaction, or a noticeable fraction of the resulting linear constraints are false. All functions and claimed values are fixed before the verifier samples which constraints to check.

\begin{theorem}[Structural form of the revised Rothblum--Rothblum IPP]\label{thm:rr20}
    Let $\Pi$ be a property in logspace-uniform $\NC$ and let $\epsilon = \epsilon(n) \in [1/n,1)$ be a proximity parameter. Then there exists a field $\F$ of characteristic $2$, with $|\F| = \poly(n)$, and a public-coin interactive protocol between a verifier $V$ who gets only $n$ as input and a prover $P$ who gets $x \in \{0,1\}^n$ as input, with $\epsilon$ fixed in the protocol, satisfying the following properties. The verifier makes no queries to $x$ during this protocol.
    \begin{itemize}
        \item \emph{Structure.} 
        There are positive integers $C = C(n,\epsilon)$ and $p = p(n,\epsilon)$, and a parameter $\delta = \delta(n,\epsilon)$, such that
        $p = O(1/\epsilon)$ and $C \le O(n)$, and $\delta \ge 1/\polylog(n)$, along with the following.
        The verifier either rejects or outputs a pair $(\sigma, \beta)$, where $\sigma$ is a substring of length $\polylog(n)$ of the transcript held by the prover and verifier, and $\beta = (\beta_1, \dots, \beta_C) \in \F^C$ consists of values supplied by the prover. Given $\sigma$, a constraint index $k \in [C]$, and some $i \in [p]$, the verifier and prover can compute $q_k(i) \in [n]$ and $\alpha_k(i) \in \F$, defining the $C$ constraints
        \[
        \sum_{i \in [p]} \alpha_k(i)\cdot x[q_k(i)] = \beta_k
        \]
        over all $k \in [C]$. Moreover, $q_k(i)$ and $\alpha_k(i)$ are computable from $\sigma$ in time $\polylog(n)$.

        

        \item \emph{Perfect completeness.} If $x \in \Pi$ and the prover is honest, then with probability $1$ the verifier does not reject and all $C$ constraints are true. That is, for all $k \in [C]$,
        \[
        \sum_{i \in [p]} \alpha_k(i)\cdot x[q_k(i)] = \beta_k.
        \]

        \item \emph{Soundness.} If $x$ is $\epsilon$-far from $\Pi$, then for every prover strategy $P^*$, with at least $1/2$ probability, $V$ rejects when interacting with $P^*$ or at least a $\delta$-fraction of $k\in [C]$ correspond to false constraints:
        \[
        \sum_{i\in[p]}\alpha_k(i)\cdot x[q_k(i)]\ne\beta_k.
        \]

        \item \emph{Complexity.} The communication complexity is $\softO(\epsilon n)$; the verifier's time complexity is $\softO(\epsilon n)$; and the round complexity is $\polylog(n)$.
    \end{itemize}
\end{theorem}

It is helpful to note how Theorem~\ref{thm:rr20} gives an ordinary IPP. The prover and verifier first run the above protocol. If the verifier does not reject, it samples $O(1/\delta)$ constraints independently and checks them directly using its oracle access to $x$. When at least a $\delta$-fraction are false, these checks find a false constraint with constant probability. Together with the theorem's soundness guarantee, this gives a constant rejection probability on far inputs, which can be amplified by a constant number of sequential or parallel repetitions. The query complexity is at most $O(p/\delta) = \polylog(n)/\epsilon$, while the communication and round complexities remain those in the theorem.

Before proving the theorem, let us briefly explain where the linear constraints come from. The Rothblum--Rothblum protocol reduces the original claim to a claim about a smaller derived object, using their distance doubling technique. The verifier receives a claimed version of this reduced object from the prover and needs to check some of its locations. Each such value is obtained by recursively unfolding the reduction back to the original input.

The revised protocol uses \emph{shift permutations}. Viewing the input as $f^{(0)} : \{0,1\}^m \to \F$, each reduction round $r$ defines
\[
f^{(r)}(\mathbf u)
    \coloneqq f^{(r-1)}(0,\mathbf u)
    + c_r\cdot f^{(r-1)}(1,\mathbf u+\mathbf a_r),
\]
where $\mathbf a_r \in \{0,1\}^{m-r}$ and $c_r\in\F$ are chosen by the verifier. 

After $R$ rounds, a queried value of $f^{(R)}$ expands into a linear combination of $2^R$ values of $f^{(0)}$. The leaves give the locations $q_k(i)$, the products of the edge coefficients give the weights $\alpha_k(i)$, and the prover's claimed value gives $\beta_k$.

\begin{proof}
    We use Theorems 5.6 and 6.1 of the recent revision~\cite{RR20-revised} of~\cite{RR20}.
    Assume for simplicity that $n$ is a power of $2$, and let $m = \log n$. View inputs in $\{0,1\}^n$ instead as functions $f : \{0,1\}^m \to \{0,1\}$, adopting their choice of denoting the input as $f$ and using $\mathbf{x}$ to denote indexes; e.g. we may say that $f$ is queried at $\mathbf{x} \in \{0,1\}^m$ as $f(\mathbf{x})$, instead of saying that an input $x \in \{0,1\}^n$ is queried at $i \in [n]$ as $x[i]$. This notation change is only cosmetic and permits consistency with~\cite{RR20-revised} for our proof.

    \paragraph{Padding.}
    We describe the protocol when $n$ is a power of 2. If this is not the case, one may pad the input $x$ of length $n$ with zeros to the next power of two, call it $N$ and run the described protocol on the ``padded'' property 
    \[
    \Lambda_{N} := \{ x 0^{N-n} : x \in \Pi_n\}.
    \]
    Then $\Lambda$ is in logspace-uniform $\NC$. The verifier and prover, on input $x \in \{0,1\}^n$, run the protocol for $\Lambda$ on input $x 0^{N-n}$ and proximity parameter $\epsilon \cdot n/N$. Technically, we need to keep $n$ as an explicit input to accommodate different input lengths that pad to $N$, but this is easily resolved by using the pair-language formalism described in \cite{RR20-revised}. The structural form we present in Theorem~\ref{thm:rr20} readily extends to pair-languages. One could alternatively build the property $\Lambda$ by including a constant-distance, logspace-uniform $\NC$ encoding of the input length $n$ into the last $N$ bits, by using, for example, a Hadamard code. The constant-distance of the code would give the desired soundness.

    \paragraph{Parameters.} For the rest of the paper, we will assume that $n$ is a power of 2. Let $m = \log n$. Choose a sufficiently large fixed constant $A \ge 1$, and define the ``internal'' proximity parameter
    \[
    \eta := \max\left\{\frac{\epsilon}{A m}, \frac{1}{n}\right\}.
    \]
    Thus $\eta\le\epsilon$, so that any proof of $\eta$-proximity is also a proof of $\epsilon$-proximity. Choose the $\pval$ arity to be
    \[
    t :=
       2^{\left\lceil\log\left(\max\{140m^2, \epsilon n/8\}\right)\right\rceil}.
    \]
    For sufficiently large $n$, we have $t\le n$ and
    \[
    t=\Theta(\max\{m^2,\epsilon \cdot n\}).
    \]
    Choose $\F$ of characteristic $2$ with $|\F|=\Theta(n\cdot t^2\cdot m^2)$, taking the constant in this choice sufficiently large. Since $t\le n$, every field element has $O(\log n)$ bits.

    \paragraph{The reduction to PVAL.}
    Run the distance-preserving reduction underlying~\cite[Theorem 5.6]{RR20-revised} with proximity parameter $\eta$, making no input queries. We use the larger field $\F$ just chosen. The same GKR reduction \cite{goldwasser-kalai-rothblum:delegating-computation} works over this field, since field elements may be represented with $\log|\F|=O(\log n)$ bits. For all sufficiently large $n$, the number of evaluation claims (i.e. the $\pval$ arity) is at most $O(\eta \cdot n \cdot \log n) \le \max\{\frac{\epsilon n}{16}, O(\log n)\} \le t$, for a sufficiently large constant $A$ chosen in the definition of $\eta$ and since $t \ge \Omega(m^2) = \omega(\log n)$.

    The verifier either rejects or obtains $\pval(t,\mathbf j,\mathbf v)$.
    If $f\in\Pi$, the honest prover gives an instance containing $f$: for all $k \in [t]$, $f(j_k) = v_k$.
    If $f$ is $\epsilon$-far from $\Pi$, then for every cheating prover,
    with probability at least $1/2$, the verifier either rejects or
    \[
    \Delta(f,\pval(t,\mathbf j,\mathbf v))\ge\eta.
    \]

    The minimum-distance hypothesis of~\cite[Theorem 6.1]{RR20-revised} follows
    from~\cite[Proposition 5.4]{RR20-revised}. Let
    $d=\lceil t/(14m^2)\rceil$ and $\kappa=t/2$.
    Since $\log(n|\F|)=O(m)$ and $t\ge140m^2$, for sufficiently large $n$,
    \[
    d\log(n|\F|)+\kappa\le t.
    \]
    Hence, except with probability less than $2^{-t/2}\le1/100$ over the verifier's coins,
    \[
    \Delta(\pval(t,\mathbf j,\mathbf0))>
       \frac dn\ge\frac{t}{14nm^2}.
    \]

    \paragraph{The protocol.} Recall from the above reduction, the verifier and prover start by sharing a $\pval$ instance $\pval(t, \mathbf{j}, \mathbf{v})$, where the verifier wishes to test that $f: \{0,1\}^m \to \F$ is $\eta$-close to $\pval(t, \mathbf{j}, \mathbf{v})$.

    Initialize $f^{(0)} \coloneqq f$, $\eta^{(0)} \coloneqq \eta$, $\mathbf{j}^{(0)} \coloneqq \mathbf{j}$, and $\mathbf{v}^{(0)} \coloneqq \mathbf{v}$. For round $r = 1, \dots, R$, where $R \coloneqq \max\{0,m-\log(4t)\}$, the verifier and prover execute the following protocol.
    \begin{enumerate}
        \item The prover sends its claimed evaluations of the two halves of $f^{(r-1)}$, and the verifier checks their consistency with the current $\pval$ instance, as in Steps 3--5 of~\cite[Figure 4]{RR20-revised}. The verifier then samples a permutation $\pi^{(r)} : \{0,1\}^{m-r} \to \{0,1\}^{m-r}$ uniformly from the class of shift permutations denoted $\Pi_{m-r}$, and sends it to the prover. Thus, $\pi^{(r)}(\mathbf{x}) = \mathbf{x} + \mathbf{a}^{(r)}$ for a uniform $\mathbf{a}^{(r)} \in \{0,1\}^{m-r}$.
        \item The verifier and prover engage in a certain $\polylog(n)$-round interaction, as in Steps 7--9 of~\cite[Figure 4]{RR20-revised}, during which the prover sends its polynomial claims and the verifier performs all of the consistency and non-triviality checks, or rejects.
        \item Once these polynomial claims are fixed, the verifier samples field elements $c^{(r)}, \xi_1^{(r)}, \dots, \xi_t^{(r)} \gets \F$. Let $\gamma^{(r)}$ denote the shared transcript up to this point, including the initial $\pval$ instance and all preceding rounds. In particular, it contains $\pi^{(r)}$, $c^{(r)}$, and the other information used to compute the reduced instance.
        \item The prover and verifier compute the following reduced $\pval$ instance over $m-r$ dimensions:
        \begin{enumerate}
            \item Let $f^{(r)} : \{0,1\}^{m-r} \to \F$ be defined by, for all $\mathbf{x} \in \{0,1\}^{m-r}$:
            \[
            f^{(r)}(\mathbf{x}) \coloneqq f^{(r-1)}(0, \mathbf{x})
                + c^{(r)} \cdot f^{(r-1)}(1, \pi^{(r)}(\mathbf{x})).
            \]
            \item Let $\mathbf{j}^{(r)} \in (\F^{m-r})^t$ be computed by a deterministic function $\phi_{\mathrm{points}}^{(r)}$ of the transcript $\gamma^{(r)}$.
            \item Let $\mathbf{v}^{(r)} \in \F^t$ be computed by a deterministic function $\phi_{\mathrm{values}}^{(r)}$ of the transcript $\gamma^{(r)}$.
            \item Let
            \[
            \eta^{(r)} \coloneqq \min\left\{
                2\eta^{(r-1)} \cdot (1-1/m),
                \frac{t}{2^{m-r+1}} \cdot \frac{1}{140m^2}
            \right\}.
            \]
        \end{enumerate}
    \end{enumerate}
    
    After all $R$ rounds of the above dimension reductions, the prover sends a claimed table $g^{(R)}$ for $f^{(R)}$ to the verifier. Upon receiving $g^{(R)}$, the verifier tests that $g^{(R)} \in \pval(t, \mathbf{j}^{(R)}, \mathbf{v}^{(R)})$, rejecting if not. The verifier outputs $\sigma := (\mathbf{a}^{(r)}, c^{(r)})_{r \in [R]}$, which is included in the transcript and has length $\polylog(n)$, and the values specified by $g^{(R)}$. Note that each $\mathbf{a}^{(r)}$ is a description of the corresponding shift permutation $\pi^{(r)}$.
    

    \paragraph{Testing $g^{(R)} \in \pval(t, \mathbf{j}^{(R)}, \mathbf{v}^{(R)})$ efficiently.} If one does not care about time complexity, one can perform the test $g^{(R)} \in \pval(t, \mathbf{j}^{(R)}, \mathbf{v}^{(R)})$ using a naive algorithm running in time $\softO(t^2)$, since $g^{(R)}$ has size $\softO(2^{m-R}) \le \softO(t)$ and the verifier can simply evaluate the multilinear extension of $g^{(R)}$ at $t$ points. However, this time complexity is unacceptable for us.

    We resolve this by using a GKR interactive proof~\cite{goldwasser-kalai-rothblum:delegating-computation} for the claim $g^{(R)} \in \pval(t, \mathbf{j}^{(R)}, \mathbf{v}^{(R)})$, which can be decided by a logspace-uniform $\NC$ circuit family given input $(g^{(R)}, t, \mathbf{j}^{(R)}, \mathbf{v}^{(R)})$. The GKR protocol thus allows the prover and verifier to test this exact claim with perfect completeness and soundness $1/100$, with total verifier time and total communication only polylogarithmically more than the length of $(g^{(R)}, t, \mathbf{j}^{(R)}, \mathbf{v}^{(R)})$. Since this GKR-input has length $\softO(t)$, this protocol allows the verifier to test $g^{(R)} \in \pval(t, \mathbf{j}^{(R)}, \mathbf{v}^{(R)})$ in time $\softO(t)$ with perfect completeness and soundness $1/100$, using only $\softO(t)$ communication. The protocol remains public-coin.
    
    \paragraph{Structural properties of the protocol.}
    The formulation of~\cite{RR20-revised} as stated in our theorem uses this protocol, except where the verifier simply ``outputs'' all of the constraints $f^{(R)}(\mathbf{x}) = g^{(R)}(\mathbf{x})$ for all $\mathbf{x} \in \{0,1\}^{m-R}$ without making its queries and deciding. We now prove that the protocol described satisfies the structural properties claimed in the theorem. The constant number of sequential repetitions used for soundness below simply concatenates these lists of constraints.
    
    The protocol is clearly public-coin, and all checks before the omitted queries depend only on the transcript.
    
    For convenience, write $\pi^{(r,0)} \coloneqq \mathrm{id}$, $\pi^{(r,1)} \coloneqq \pi^{(r)}$, $c^{(r,0)} \coloneqq 1$, and $c^{(r,1)} \coloneqq c^{(r)}$. Observe that unfolding the $R$-level recursion, we have the following description of $f^{(R)}$: for all $\mathbf{x} \in \{0,1\}^{m-R}$,
    \[
    \begin{aligned}
    f^{(R)}(\mathbf{x})
        &= \sum_{(b_1, \dots, b_R) \in \{0,1\}^R}
            \left(\prod_{r=1}^R c^{(r,b_r)}\right) \\
        &\qquad\cdot f\Bigl(b_1,
            \pi^{(1,b_1)}\bigl(b_2,
            \pi^{(2,b_2)}\bigl(\cdots\bigl(b_R,
            \pi^{(R,b_R)}(\mathbf{x})\bigr)\cdots\bigr)\bigr)\Bigr),
    \end{aligned}
    \]
    where $(b, s)$ for a bit $b$ and a string $s$ denotes the concatenation of $b$ and $s$.
    The functions $q_{\mathbf{x}} : \{0,1\}^R \to \{0,1\}^m$ and $\alpha_{\mathbf{x}} : \{0,1\}^R \to \F$, and the values $\beta_{\mathbf{x}} \in \F$, for $\mathbf{x} \in \{0,1\}^{m-R}$ as claimed in the theorem, are defined from the transcript as follows. We use the usual bijections between $\{0,1\}^R$ and $[2^R]$, and between $\{0,1\}^m$ and $[n]$, so that we have $p = 2^R$.
    
    For $\mathbf{x} \in \{0,1\}^{m-R}$, and for $\mathbf{b} = (b_1, \dots, b_R) \in \{0,1\}^R$, we define
    \[
    \alpha_{\mathbf{x}}(\mathbf{b}) \coloneqq \prod_{r=1}^R c^{(r,b_r)},
    \]
    which can clearly be computed from $\sigma := (\mathbf{a}^{(r)}, c^{(r)})_{r \in [R]}$.
    
    For the same $\mathbf{x}$ and $\mathbf{b}$, we define
    \[
    q_{\mathbf{x}}(\mathbf{b}) \coloneqq \Bigl(b_1,
        \pi^{(1,b_1)}\bigl(b_2,
        \pi^{(2,b_2)}\bigl(\cdots\bigl(b_R,
        \pi^{(R,b_R)}(\mathbf{x})\bigr)\cdots\bigr)\bigr)\Bigr),
    \]
    which is similarly computable from $\sigma$, since each $\pi^{(r,b)}$ is determined by it. When $R=0$, we interpret $q_{\mathbf{x}}(\emptyset)=\mathbf{x}$ and the empty product defining $\alpha_{\mathbf{x}}$ as $1$.
    
    Finally, let $\beta_{\mathbf{x}} \coloneqq g^{(R)}(\mathbf{x})$, which is indeed sent by the prover. These definitions give exactly the claimed linear constraints.
    
    The number of constraints in one execution is
    \[
    C_0 \coloneqq 2^{m-R} = \min\{n,4t\} = O(n).
    \]
    Now we bound the size of the domain of $q_{\mathbf{x}}$ and $\alpha_{\mathbf{x}}$ as:
    \[
    p = 2^R \le \frac{n}{t}
        \le \frac{8}{\epsilon}
        = O\left(\frac{1}{\epsilon}\right),
    \]
    as claimed.

    \paragraph{Perfect completeness.}
    All input-independent checks pass when $f\in\Pi$ and the prover is
    honest, as in the Rothblum--Rothblum protocol. The honest table is
    $g^{(R)}=f^{(R)}$, so all $C_0$ linear constraints are true. Moving the base
    case up from $t$ to $4t$ does not affect completeness: the honest prover sends the
    corresponding intermediate table, which satisfies its $\pval$ instance.

    \paragraph{Soundness.}
    The distance-doubling analysis in~\cite[Section 6.2]{RR20-revised} uses
    the proximity parameters
    \[
    \eta_r\coloneqq\min\left\{
       2(1-1/m)\eta_{r-1},
       \frac{t}{140m^2\,2^{m-r+1}}
    \right\},\qquad r\in[R].
    \]
    The denominator in the second term is the table size before round $r$.
    Since $t\ge140m^2$ and $\eta\ge1/n$, every $\eta_r\ge1/n$,
    as required by that analysis.

    We first lower bound the final proximity parameter.
    Let $\zeta=1-1/m$, so $\zeta^m\ge1/4$ for $m\ge2$.
    Iterating the above recurrence gives, for $R\ge1$,
    \[
    \eta_R=\min\left\{
       p\zeta^R\eta,
       \min_{1\le r\le R}
       \frac{t}{280m^2 C_0}\,\zeta^{R-r}
    \right\}.
    \]
    Consequently, also when $R=0$,
    \[
    \eta_R\ge\frac14\min\left\{
       p\eta,\frac{t}{280m^2 C_0}
    \right\}=\Omega(1/m^2).
    \]
    Indeed, $C_0\le4t$ gives
    $t/(280m^2 C_0)\ge1/(1120m^2)$, while
    \[
    p\eta\ge
       \frac{\max\{1,\epsilon n/(Am)\}}{4t}
       =\Omega(1/m^2),
    \]
    using $t=\Theta(\max\{m^2,\epsilon n\})$.
    For the latter bound, if $\epsilon n\le m^2$, use the numerator's
    $1$ term; otherwise use its $\epsilon n/(Am)$ term.

    Fix any initial $\pval$ instance satisfying the minimum-distance
    hypothesis and from which $f$ is $\eta$-far.
    The per-round analysis of~\cite[Section 6.2]{RR20-revised}, in particular
    Proposition 6.3 and Claims 6.5--6.7, shows that each folding round,
    conditional on a valid preceding history, either rejects or preserves
    the distance guarantees, except with probability
    at most $1/(2m)$. A union bound over the $R$ folding rounds therefore
    shows that, with probability at least $1-R/(2m)\ge1/2$, the verifier
    either rejects or $f^{(R)}$ is $\eta_R$-far from the final $\pval$ instance.

    By the soundness of the GKR subprotocol, the probability that its
    membership check accepts despite
    $g^{(R)} \notin \pval(t, \mathbf{j}^{(R)}, \mathbf{v}^{(R)})$
    is at most $1/100$. Whenever $g^{(R)}$ belongs to this final instance
    and $f^{(R)}$ is $\eta_R$-far from it, at least an
    $\eta_R$-fraction of $\mathbf u \in \{0,1\}^{m-R}$ satisfy
    $f^{(R)}(\mathbf u)\ne g(\mathbf u)$.
    These are precisely the false linear constraints. Combining the
    folding guarantee with the initial reduction's soundness, the
    $1/100$ bound on a bad initial minimum-distance hypothesis,
    and the additional $1/100$ error of the GKR subprotocol, we obtain
    that the probability that $V$ rejects or at least an
    $\eta_R$-fraction of the constraints are false is at least
    \[
        \frac{1}{2} \left(\frac{1}{2}-\frac{1}{100}\right)-\frac{1}{100}
        = \frac{47}{200} > \frac{1}{5}.
    \]
    
    
    To obtain the stated soundness, run four copies of this interaction
    sequentially, using fresh verifier coins in each. Reject if any copy
    rejects; otherwise concatenate their lists of constraints.
    Thus, $C=4C_0$, and the number $p$ of summands per constraint is
    unchanged. If any copy has at least an $\eta_R$-fraction of false
    constraints, then the concatenated list has at least a
    \[
    \delta\coloneqq\eta_R/4=\Omega(1/m^2)
    \]
    fraction of false constraints. Hence, the probability that $V$ rejects or at least a $\delta$-fraction of constraints are false is at least $1-(4/5)^4 \ge \frac{1}{2}$.

    Perfect completeness is preserved under repetition. The verifier outputs the concatenation of each partial transcript $\sigma$ output in each repetition and the concatenation $\beta$ of the four
    claimed tables, with the corresponding functions $q_k$ and
    $\alpha_k$ defined above.

    \paragraph{Complexities.}
    The reduction to $\pval$ communicates $\softO(t)$ bits and uses $\polylog(n)$ rounds. Each dimension reduction communicates $\softO(t)$ bits and uses $\polylog(n)$ rounds. Since there are $R\le m=\log n$ such reductions, their total communication is
    $\softO(t)$ bits and their total round complexity is
    $\polylog(n)$. The final claimed function-table $g^{(R)}$ has $C_0 \le 4t$ field elements and therefore adds $C_0\log|\F| = O(t\log n)$ bits of communication. The GKR protocol to test $\pval$ membership of $g^{(R)}$ adds only $\softO(t)$ communication. Since
    \[
    t = \Theta(\max\{m^2,\epsilon n\})
      = O(\epsilon \cdot n \cdot m^2),
    \]
    the total communication complexity is $\softO(\epsilon n)$ and the total round complexity is $\polylog(n)$. The constant number of sequential repetitions increases both bounds only by a constant factor.
    The verifier's time complexity is $\softO(\epsilon n)$, inherited from the analysis of \cite{RR20-revised} and noting that the interactive test for $g^{(R)} \in \pval(t, \mathbf{j}^{(R)}, \mathbf{v}^{(R)})$ requires time only $\softO(t) = \softO(\epsilon n)$.
    Lastly, it is easy to see that $q_k(i)$ and $\alpha_k(i)$ are computable from the output partial transcript in time $\polylog(n)$.
\end{proof}

\subsection{Amplitude amplification with classical advice}\label{Sec:advice-aa}

Before proceeding to our main construction in Section~\ref{Sec:nc-aa-speedup}, we design a more efficient amplitude amplification algorithm that will aid our construction to have time complexity only polylogarithmically more than the sum of query and communication complexities.
While the amplitude amplification of Lemma~\ref{lem:ampl-ampl} is sufficient to reduce the sum of query and communication complexity, its time complexity lags behind due to the gate cost of building the unitary corresponding to the classical test.

Consider a test that samples one of $M$ rows of a classical table and evaluates a predicate on that row. A straightforward coherent implementation of this procedure must load the entire table in each amplification iteration, since a unitary that implements the sampling procedure must be able to compute the predicate on every row. We show how to retain the usual quadratic query improvement while paying for the table only once, up to polylogarithmic factors, in the case where a single successful row must be found. The construction combines the parallel-lookup circuit of~\cite{beals} with sample-based reflections of~\cite{kimmel}, and applies them to the fixed-point amplitude amplification of Yoder, Low, and Chuang~\cite{ylc}.

Consider an input $x \in \{0,1\}^n$ of length $n$, an advice string $\sigma \in \{0,1\}^s$ of length $s$, and a table $(d_1, \dots, d_M) \in (\{0,1\}^d)^M$, each row containing $d$ bits. 
Let $f^x(\sigma, r, z)$ be a Boolean predicate, for $r \in [M]$ represented with $\ell = \ceil{\log M}$ bits and $z \in \{0,1\}^d$. 
Note that for all $x \in \{0,1\}^n$, we have $f^x : \{0,1\}^{s+\ell+d} \to \{0,1\}$. A \emph{$(q,t)$ evaluator for $f$} is a reversible query circuit $E^x$ uniformly constructible in time $t$ that makes at most $q$ queries to $x$ and maps $(\sigma, r, z, b, 0) \mapsto (\sigma, r, z, b \oplus f^x(\sigma, r, z), 0)$ for all $\sigma$, $z$ and $r$, and on all $x \in \{0,1\}^n$. The last register is the workspace.

On nonuniform advice $\sigma$ and a dataset $D = (d_1, \dots, d_M)$, the associated randomized test samples $r \in [M]$ uniformly and outputs $f^x(\sigma, r, d_r)$. Its success probability is 
\[
\frac{1}{M} \sum_{r=1}^M f^x(\sigma, r, d_r).
\]
Our task is to amplify the success probability with an efficient quantum algorithm, given input $(\sigma, D)$.
Given such a $(q,t)$-evaluator $E$ for $f$, the usual amplitude amplification algorithm (Lemma~\ref{lem:ampl-ampl}) gives an $O(q \cdot \sqrt{M})$-query quantum algorithm that succeeds with probability $2/3$ whenever there exists $r \in [M]$ such that $f^x(\sigma, r, d_r) = 1$, and succeeds with probability $0$ if such an $r$ does not exist. Unfortunately, the number of gates of this quantum circuit is $O((t+M(d+\ell)) \sqrt{M})$, since each reflection iteration requires loading the entire table $D$. The main lemma of this section, Lemma~\ref{lem:ampl-ampl-better}, shows how to amortize this loading cost, reducing the total number of gates.

For a Boolean predicate $f^x(\sigma, r, z)$, advice $\sigma$ and a dataset $D = (d_1, \dots, d_M)$ of the form specified above, we denote the acceptance probability of the randomized procedure induced by $f$ and $(\sigma, D)$ as
\[
\Acc_f^x(\sigma, D) := \expsub{r \in [M]}{f^x(\sigma, r, d_r)}.
\]

\begin{lemma}[Amplitude amplification with classical advice]\label{lem:ampl-ampl-better}
Let $s, M, d = \poly(n)$.
Let $f^x(\sigma, r, z)$ be a Boolean predicate on an oracle $x \in \{0,1\}^n$, advice $\sigma \in \{0,1\}^s$, selection $r \in [M]$, and data $z \in \{0,1\}^d$. Suppose $f$ has a $(q,t)$ query evaluator, and let $w=d+\ceil{\log(M+1)}$. For every $0<\lambda\leq1$, there is a quantum algorithm with query complexity \[O(q/\sqrt\lambda)\] and time complexity \[\softO((M+\lambda^{-1})(d+\log M) + t/\sqrt{\lambda})\]
that, on input $\sigma \in \{0,1\}^s$ and $D \in (\{0,1\}^d)^M$, returns $1$ or $0$. The quantum algorithm outputs $1$ with probability at least $2/3$ if $\Acc_f^x(\sigma,D)\geq\lambda$, and returns $0$ with certainty if $\Acc_f^x(\sigma,D)=0$.
\end{lemma}

Here an index $r\in[M]$ is successful if $f^x(\sigma,r,d_r)=1$. In particular, taking $\lambda=1/M$ finds a successful row whenever one exists, with $O(q\sqrt M)$ queries and $\softO(M(d+\log M)+t\sqrt M)$ time.

The proof of Lemma~\ref{lem:ampl-ampl-better} is rather technical, so we provide a proof sketch here and delay the full proof to Appendix~\ref{App:ampl-ampl-proof}.

\begin{proof}[Proof sketch]
Write $a=\operatorname{Acc}_f^x(\sigma,D)$, and consider the state
\[
 \ket{\psi_D}=\frac{1}{\sqrt M}\sum_{r=1}^M\ket{r,d_r}.
\]
Measuring this state produces a successful row with probability $a$.
The fixed-point amplitude amplification of \cite{ylc} increases this
probability to at least $9/10$ whenever $a\geq\lambda$, using
$J=O(\lambda^{-1/2})$ applications of each of two kinds of phase
operators. The first multiplies $\ket{\psi_D}$ by a prescribed phase
and fixes its orthogonal complement; the second multiplies successful
computational basis states by a prescribed phase and fixes unsuccessful
ones. The phases depend only on the lower bound $\lambda$, so the
algorithm does not need to know $a$.

The successful-state phase operators are straightforward to implement
from the evaluator $E^x$: compute $f^x(\sigma,r,z)$ into an auxiliary
bit, apply the prescribed phase when that bit is $1$, and uncompute.
Each such operation costs $O(q)$ queries and $O(t)$ gates. To implement the phase operators about
$\ket{\psi_D}$, we use the sample-based simulation of \cite{kimmel}.
For bounded phases, their construction approximates such an
operator to channel error $\eta$ using $O(1/\eta)$ fresh copies of
$\ket{\psi_D}$ and $\softO(w/\eta)$ gates, where
$w=d+\ceil{\log M}$.

Choose $\eta=c/J$ for a sufficiently small absolute constant $c$.
Since errors add under composition, the accumulated error over the
$O(J)$ operations is then a sufficiently small constant. Each
sample-based simulation consumes $O(J)$ copies, so $O(J^2)$ copies
suffice for the entire amplification procedure, including its initial
state. A parallel-lookup circuit of \cite{beals} prepares all these copies
together using
\[
 \softO((M+J^2)w)
\]
gates and no input queries. Its output is a product state, so each
simulation receives fresh, independent copies. This amortizes the
cost of loading the classical table across all amplification steps.
The simulations themselves use $\softO(J^2w)$ gates in total.

Finally, measure the index register, retrieve the corresponding actual
row from $D$, and evaluate the predicate exactly. Accept only if this
check succeeds. The ideal amplification remains supported on actual
table rows, and the accumulated channel error changes its acceptance
probability by only a small constant. Thus the implemented algorithm
accepts with probability at least $2/3$ when $a\geq\lambda$.
When $a=0$, the final exact check always rejects, regardless of
simulation errors.

Including this final check, the total query complexity is
$O(Jq+q)=O(q/\sqrt\lambda)$, and the total gate complexity is
\[
 \softO((M+J^2)w+Jt+Mw+t)
 =\softO\bigl((M+\lambda^{-1})w+t/\sqrt\lambda\bigr).
\]
The same bound holds for circuit construction time. The full proof
in the appendix gives the phase sequence, approximation bounds,
and padding argument for general $M$.
\end{proof}

\subsection{Speeding up IPPs for NC via amplitude amplification}

The structural theorem above reduces the problem to checking a small random sample of linear constraints. A direct application of the previous section would ask the prover to reveal every input value appearing in each sampled constraint and would then use amplitude amplification only to check whether one of these values was falsified. This is wasteful for the logspace-uniform $\NC$ protocol, because the verifier does not need the individual values themselves; it only needs their weighted sum.

The improved protocol therefore asks for more sparse information. For each sampled constraint, the verifier partitions the summation domain $[p]$ into $c$ parts. The prover sends the claimed partial sum over each part. The verifier first checks that these partial sums add to the claimed total $\beta_k$. Then, using amplitude amplification over the $c$ parts, it searches for a part whose claimed partial sum disagrees with the true weighted sum over that part. Checking one part costs at most $\lceil p/c\rceil$ input queries, while amplitude amplification reduces the number of parts that must be checked from $c$ to $O(\sqrt c)$. This is the source of the improved query--communication tradeoff.


\begin{proof}[Proof of Theorem~\ref{thm:nc-in-qcipp}]
    Let $(P',V')$ be the interactive protocol from Theorem~\ref{thm:rr20} for $\Pi$ and proximity parameter $\epsilon$. Let $C,p,\delta$ be its parameters, and set
    $c\coloneqq\min\{p,\ceil{\epsilon n}\}$ and $s \coloneqq \left\lceil\frac{\ln 2}{\delta}\right\rceil$.
    Partition $[p]$ into the $c$ nonempty sets
    \[
    P_j\coloneqq\{i\in[p]:i\equiv j\pmod c\},
    \]
    for $j \in [c]$.
    Every block has size at most $\lceil p/c\rceil$. The choice $c\le p$ will be useful when accounting for this rounding.

    On input $x\in\{0,1\}^n$, the protocol $(P(x),V^x(n))$ proceeds as follows.
    \begin{enumerate}
        \item Emulate $(P'(x),V'(n))$. If $V'$ rejects, output $\mathsf{reject}$. Otherwise, the verifier and prover share a partial transcript $\sigma$ and values $\beta = (\beta_1, \dots, \beta_C) \in \F^C$, and the corresponding constraint-defining functions $q_k : [p] \to [n]$ and $\alpha_k : [p] \to \F$, for $k\in[C]$.

        \item The verifier samples an ordered list $K=(k_1,\dots,k_s)$ of independent uniform indices in $[C]$, with replacement, and sends $K$ to the prover. 

        \item The honest prover responds with the partial sums
        \[
        \left(\sum_{i\in P_j}\alpha_{k_\ell}(i)
                    \cdot x[q_{k_\ell}(i)]\right)_{
                    \ell\in[s],\,j\in[c]}.
        \]
        Denote the received values by $(v_\ell^{(j)})_{\ell\in[s],j\in[c]}$. Reject any incorrectly formatted response.

        \item For every $\ell\in[s]$, check that $\sum_{j\in[c]}v_\ell^{(j)}=\beta_{k_\ell}$. If any check fails, output $\mathsf{reject}$.


        \item For every $\ell\in[s]$, define a Boolean function 
        \[
        f^x(\sigma, r, z) = \mathbf1\left[
            \sum_{i\in P_r}\alpha_{k_\ell}(i)\cdot x[q_{k_\ell}(i)] \ne z \right].
        \]
        It is clear that $f$ has a $(\ceil{p/c}, \softO(\ceil{p/c}))$ query evaluator, so we may apply Lemma~\ref{lem:ampl-ampl-better} to $f$ and $\lambda = 1/c$ to obtain a quantum algorithm $Q$. 
        Then the verifier runs $Q$ on $\sigma$ and $D = (v_\ell^{(1)}, \dots, v_\ell^{(c)})$. If $Q$ outputs $1$ (indicating it found a violated constraint), the verifier rejects.

        \item If no preceding step has rejected, the verifier accepts.
    \end{enumerate}

    The initial protocol is public-coin, and the only additional verifier message is the random list $K$. Thus the resulting protocol is also public-coin.

    \paragraph{Perfect completeness.}
    When $x\in\Pi$ and the prover is honest, the initial verifier does not reject, every linear constraint is true, and every claimed block sum is correct. All checks therefore pass. Combined with the perfect completeness of amplitude amplification of Lemma~\ref{lem:ampl-ampl-better}, we have that this protocol has perfect completeness for $\Pi$.

    \paragraph{Soundness.}
    Suppose $x$ is $\epsilon$-far from $\Pi$. Let $P^*$ be an arbitrary prover strategy, and let all transcript-dependent functions and values ($q_k, \alpha_k, \beta_k$) be with respect to communication with $P^*$.

    Then, with probability at least $1/2$, the verifier either rejects in the initial emulation or at least a $\delta$-fraction of the constraints $\sum_{i \in [p]} \alpha_k(i) \cdot x[q_k(i)] = \beta_k$ for $k \in [C]$ are false, by Theorem~\ref{thm:rr20}. These functions and values are fixed before the verifier samples $K$. Conditional on any initial transcript in which the verifier does not reject and at least a $\delta$-fraction of the constraints are false, the probability that all $s$ independently sampled constraints are true is at most
    \[
    (1-\delta)^s\le e^{-\delta s}\le\frac12,
    \]
    since $s=\lceil\ln2/\delta\rceil$. Thus, with probability at least $1/4$, the verifier has already rejected or some verifier-sampled $k^*=k_{\ell^*} \in K$ satisfies $\sum_{i \in [p]} \alpha_{k^*}(i) \cdot x[q_{k^*}(i)] \ne \beta_{k^*}$.
    Unless the verifier has already rejected, let us condition on the latter event in the below analysis.

    For this $k^*$, the prover $P^*$'s claimed partial sums, denoted $(v_\ell^{(j)})$, must satisfy either
    $\sum_{j\in[c]}v_{\ell^*}^{(j)}\ne\beta_{k^*}$
    or there exists some $j^*\in[c]$ such that
    \[
    \sum_{i\in P_{j^*}}\alpha_{k^*}(i)\cdot x[q_{k^*}(i)]
       \ne v_{\ell^*}^{(j^*)}.
    \]
    In the former case, the verifier rejects with probability $1$. In the latter case, we have 
    \[
    \Acc_f^x(\sigma, (v_{\ell^*}^{(1)}, \dots, v_{\ell^*}^{(c)})) \ge \frac{1}{c},
    \]
    so that Lemma~\ref{lem:ampl-ampl-better} guarantees that the quantum algorithm $Q$ from the lemma outputs $1$ with probability at least $2/3$. Hence, the verifier rejects with probability at least $1/2$.

    This implies that the verifier rejects with probability at least $1/8$. Sequentially repeating the protocol a constant number of times, using fresh verifier randomness and accepting only if every execution accepts, reduces soundness to $1/2$. 

    \paragraph{Query complexity.}
    By the query bound in Lemma~\ref{lem:ampl-ampl-better}, the number of input queries is at most
    \[
    O(s)\cdot O(\sqrt c)\cdot O(\lceil p/c\rceil)
       =O\left(\frac{p}{\delta\sqrt c}\right),
    \]
    where we used $c\le p$, hence $\lceil p/c\rceil\le2p/c$. If $c=\lceil\epsilon n\rceil$, then $p=O(1/\epsilon)$ gives
    \[
    \frac{p}{\sqrt c}=O\left(
       \frac{1}{\epsilon^{3/2}\sqrt n}\right).
    \]
    Otherwise $c=p$, and $p/\sqrt c=\sqrt p=O(1/\sqrt\epsilon)$. Since $1/\delta=\polylog(n)$, these two cases give the stated query complexity. 

    \paragraph{Communication complexity.}
    The initial interaction communicates $\softO(\epsilon n)$ bits. The list $K$ uses $s \cdot \log C=\polylog(n)$ bits, and the prover's block sums use $s\cdot c\cdot \log|\F|$ bits. Since
    \[
    c\le\lceil\epsilon n\rceil\le2\epsilon n,
    \]
    the total communication is $\softO(\epsilon n)$. Constant repetition preserves this bound.

    \paragraph{Time complexity.}
    The initial interaction and classical checks take $\softO(\epsilon n)$ time, by Theorem~\ref{thm:rr20}.
    Lemma~\ref{lem:ampl-ampl-better} gives time
    $\softO(c+\lceil p/c\rceil\sqrt c)$ per search.
    Since $c\le p$, the preceding query and communication analyses imply total time
    \[
    \softO\left(\epsilon n+sc+\frac{sp}{\sqrt c}\right)
    =
    \softO\left(\epsilon n+
    \frac{1}{\epsilon^{3/2}\sqrt n}
    +\frac{1}{\sqrt\epsilon}\right).
    \]

    \paragraph{Round complexity.}
    The initial protocol uses $\polylog(n)$ rounds, and the block-sum check adds only two messages. Constant sequential repetition leaves the round complexity $\polylog(n)$.
\end{proof}

\section{A lower bound on QIPPs for logspace-uniform NC}\label{Sec:QIPP_lower_bound}

Motivated by our subquadratic QCIPP for logspace-uniform $\NC$ properties presented in Section~\ref{Sec:nc-aa-speedup}, we now shift gears to proving lower bounds on the QIPP complexity of logspace-uniform $\NC$ properties. The main result is stated in Theorem~\ref{thm:main_lower_bound}, and restated below. We obtain that, informally, QIPPs for all logspace-uniform $\NC$ properties with query complexity $q$, communication complexity $c$, and polynomial-time verification cannot satisfy $q^2 \cdot c = o(n)$, unless a cryptographic assumption is false. Since Section~\ref{Sec:ipp-aa-speedup} exhibited, for all $\delta \in (0, 1/2]$, a QCIPP for logspace-uniform $\NC$ with query complexity $q = \softO(n^{\delta/2})$ and communication complexity $c = \softO(n^{1-\delta})$, our lower bound is tight up to polylogarithmic factors whenever communication is bounded by a parameter $c \ge \Omega(\sqrt{n})$.


\begin{theorem}\label{thm:main_lower_bound}
Assume that there exists a post-quantum subexponentially secure PRG
$
G:\{0,1\}^{\lambda}\to \{0,1\}^{2\lambda}
$
that is computable in logspace-uniform $\NC$. Let $q : \N \to \N$ be a nondecreasing function computable in space $O(\log n)$ given $n$, such that $q(n)\le n^{1/2-\Omega(1)}$. Then there is a property
$\Pi$ in logspace-uniform $\NC$ and a constant $\epsilon>0$ such that no $\epsilon$-QIPP for $\Pi$
with query complexity $q(n)$, communication complexity $c(n)$, and a polynomial-time verifier can satisfy
\[
    q(n)^2 c(n) = o(n).
\]
\end{theorem}



We view the restriction that the verifier runs in polynomial time to be unimportant with respect to our goal of efficient verification. Indeed, since our primary interest is in \emph{sublinear} time proofs of proximity, a proof of proximity that has sublinear communication and queries but runs in superpolynomial time would have limited utility. The result implies that if the cryptographic assumption is true, then any QIPP for all logspace-uniform $\NC$ properties that achieves $q^2 \cdot c = o(n)$ must require the verifier to run in superpolynomial time.

The rest of this section is dedicated to proving Theorem~\ref{thm:main_lower_bound}. Subsection~\ref{Subsec:lower_bound_construction} gives the construction of the hard property $\Pi$, based directly on the construction of \cite{KR-arguments-of-proximity}. The heart of the reduction is contained in Section~\ref{Subsec:QIPP-to-avg-case-QIP}, where we quantize the reduction of \cite{KR-arguments-of-proximity} using a hybrid method. We conclude Theorem~\ref{thm:main_lower_bound} by combining the result of the previous subsection with the quantum simulation theorem of \cite{BSW-QIP-short-messages} in Proposition~\ref{prop:BSW_small_comm_QIP_simulation}.

\subsection{The construction}\label{Subsec:lower_bound_construction}

For the rest of this section, fix a parameter $q$ from Theorem~\ref{thm:main_lower_bound}: let $q : \N \to \N$ be a nondecreasing function computable in space $O(\log n)$, such that $q(n)\le n^{1/2-\Omega(1)}$.

We recall that constant-rate constant-distance codes that have encoding maps computable in logspace-uniform $\NC$ exist; see Spielman's construction \cite{Spielman1996LinearTime} as well as an explicit discussion of logspace-uniformity in \cite{GuruswamiKabanets2008Hardness}. For a code $C_k: \{0,1\}^k \to \{0,1\}^{\alpha k}$ for a constant $\alpha \ge 1$, denote its absolute distance by $\dist(C_k) = \min \{\dist(C_k(x), C_k(y)) : x, y \in \{0,1\}^k, x \ne y\}$ and its relative distance by $\Delta(C_k) = \frac{\dist(C_k)}{k}$.\footnote{Note that the normalizing factor is the message length $k$, rather than the code length $\alpha k$.}

\begin{theorem}[\cite{Spielman1996LinearTime,GuruswamiKabanets2008Hardness}]
\label{thm:good-nc-codes}
    There exists a logspace-uniform family of circuits $\{C_k : \{0,1\}^k \to \{0,1\}^{\alpha k}\}_{k \in \N}$, for some constant rate $\alpha^{-1} \le 1$ such that:
    \begin{itemize}
        \item each $C_k$ has size $\poly(k)$ and depth $\polylog(k)$;
        \item $\Delta(C_k) = \Omega(1)$.
    \end{itemize}
\end{theorem}

Let $C = \{C_k\}$ denote the ensemble of codes given by Theorem~\ref{thm:good-nc-codes}, and let its rate be $\alpha^{-1}$. Choose a constant $\epsilon > 0$ such that $\Delta(C_k) > \epsilon \cdot \frac{1+2\alpha}{2}$ for all $k$. For a given string $x$, we often denote $C(x) := C_{|x|}(x)$.


Let $G : \{0,1\}^\lambda \to \{0,1\}^{2\lambda}$ be a post-quantum subexponentially-secure PRG with security parameter $\lambda$, such that $G$ is computable in logspace-uniform $\NC$.

We are now ready to define the property $\Pi$ in Theorem~\ref{thm:main_lower_bound}, which we will denote by $\Lambda$.

\begin{definition}
    Let $\Lambda = (\Lambda_n)_{n \in \N}$ be the following property. Let $\Lambda_n$ be the set of all strings $(C(r_1, \dots, r_m), s_1, \dots, s_m)$, where
    \begin{enumerate}
        \item $m \coloneqq 144q^2$,
        \item $\lambda \coloneqq \frac{1}{144(1+2\alpha)} \frac{n}{q^2}$,
        \item every $s_i$ has length $\lambda$, and
        \item $r_i = G(s_i)$ for every $i$.
    \end{enumerate}
\end{definition}
Note that the length of strings in $\Lambda_n$ is indeed $\alpha \cdot m \cdot 2\lambda + m \cdot \lambda = \lambda \cdot m \cdot (2\alpha + 1) = \frac{1}{144(1+2\alpha)}\frac{n}{q^2} \cdot 144q^2 \cdot (2\alpha + 1) = n$, so that $\Lambda_n \subseteq \{0,1\}^n$.

Note that by logspace-uniform $\NC$ computability of $G$ and $C$, and the $O(\log n)$ space computability of $q$, we easily obtain:

\begin{lemma}\label{lem:property_in_NC}
    $\Lambda$ is in logspace-uniform $\NC$.
\end{lemma}


\subsection{Reducing a QIPP to an average-case complete QIP}\label{Subsec:QIPP-to-avg-case-QIP}

As discussed in the technical overview, the first step is to reduce a QIPP to a QIP in the CRS model with average-case completeness. The language we reduce to is the following.
Let $\calL$ be the language that consists of the image of $G$:
\[
\calL \coloneqq \{r : \exists s, r = G(s)\}.
\]

Let us define a QIP in the CRS model formally. 

\begin{definition}\label{def:QIP_with_crs}
    A \emph{quantum interactive protocol with CRS} is described by a quantum interactive protocol $(P,V)$ with $V$ running in quantum polynomial time, together with a probabilistic polynomial-time algorithm $\crs$. For a language $\calL$ and an ensemble of distributions $D = (D_n)_{n \in \N}$ supported on $\calL \cap \{0,1\}^n$, the protocol $(P, V, \crs)$ is an \emph{average-case complete QIP with CRS} for $\calL$ with respect to $D$ with completeness $a$ and soundness $b$ with $0 \le b < a \le 1$ if it satisfies the following.
    \begin{enumerate}
        \item Average-case completeness. For every $n \in \N$,
        \[
        \Pr[x \gets D_n, \crs_0 \gets \crs(1^n) : \braket{P, V}(x, \crs_0) = 1] \ge a.
        \]
        \item Soundness. For every $n \in \N$, for every $x \in \{0,1\}^n \setminus \calL$, and for every $\crs_0$ output by $\crs(1^n)$, any cheating prover $P^*$ compatible with $V$ satisfies
        \[
        \Pr[\braket{P^*, V}(x, \crs_0) = 1] \le b.
        \]
    \end{enumerate}
\end{definition}

In the following lemma, $U_\lambda$ denotes the uniform distribution over $\{0,1\}^\lambda$.

\begin{lemma}\label{lem:lower_bound_reduction}
    Suppose $(P,V)$ is an $\epsilon$-QIPP for $\Lambda$ with query complexity $q(n)$, communication complexity $c(n)$, and a $\poly(n)$-time verifier such that $q(n)^2 \cdot c(n) = o(n)$. For inputs to $\calL$ of length $\tau = 2\lambda$, let $D_{\tau}$ be the distribution given by $G(U_\lambda)$. Then there exists an average-case complete QIP for $\calL$ with CRS with respect to $(D_\tau)_\tau$ and with completeness $1/2$, soundness $1/3$, a polynomial-time verifier, and communication complexity $o(\tau)$ on inputs of length $\tau$.
\end{lemma}
\begin{proof}
    Define the following CRS-model quantum interactive proof for $\calL$, denoted $(P', V', \mathsf{crs})$. On input $r \in \{0,1\}^{\tau}$ for $\tau = 2\lambda$,
    \begin{itemize}
        \item $\crs(1^{2\lambda})$: output $\lambda m + \log m$ uniform bits $(s_1, \dots, s_m, i) \in (\{0,1\}^\lambda)^m \times [m]$.
        \item $V'(r, (s_1, \dots, s_m, i))$:
        \begin{enumerate}
            \item for $j \in [m]$, set $r_j = r$ if $j = i$ and set $r_j = G(s_j)$ if $j \ne i$.
            \item simulate $V$ given query access to the instance 
            $
            (C(r_1, \dots, r_m), s_1, \dots, s_m)
            $
            of $\Lambda$.
        \end{enumerate}
        \item $P'(r, (s_1, \dots, s_m, i))$:
        \begin{enumerate}
            \item for $j \in [m]$, set $r_j = r$ if $j = i$ and set $r_j = G(s_j)$ if $j \ne i$.
            \item let $s$ be a uniform preimage of $r$ under $G$, which is well-defined if $r \in \calL$;
            \item for $j \in [m]$, set $s'_j = s$ if $j = i$ and set $s'_j = s_j$ otherwise.
            \item simulate $P$ on the instance 
            $
            (C(r_1, \dots, r_m), s'_1, \dots, s'_m)
            $
            of $\Lambda$.
        \end{enumerate}
    \end{itemize}

    \paragraph{Running time.} Let $\delta = \Omega(1)$ such that $q = O(n^{\frac{1}{2}-\delta})$. Observe that the length $n$ of the output instance of $\Lambda$ satisfies $\tau = 2\lambda = \Theta(n/q^2) = \Omega(n/n^{1-2\delta}) = \Omega(n^{2\delta})$ since $q^2 = O(n^{1-2\delta})$. This implies that $n = O(\tau^{\frac{1}{2\delta}}) = \poly(\tau)$. Thus, CRS generation is polynomial-time, and $V'$ is a quantum polynomial time algorithm, since $V$ is a quantum polynomial time algorithm.

    \paragraph{Communication complexity.} Observe that the communication complexity of $(P', V', \crs)$ on input length $\tau$ is $c(n) = o(n/q(n)^2)$. Since $n$ satisfies $\tau = \Theta(n/q^2)$, the communication complexity is $o(\tau)$.

    \paragraph{Average-case completeness.}
    Fix an input length $\tau = 2\lambda$, and sample $R$ according to $D_\tau$ in $\calL \cap \{0,1\}^\tau$ by $R = G(U_\lambda)$.

    Consider the distribution of inputs to $P$ and $V$ for the input $R$ and uniform $\crs_0 \gets \crs(1^\tau)$ in an execution of $(P'(R, \crs_0), V'(R, \crs_0))$. Observe from the construction that $P'(R, \crs_0)$ computes an instance $X$ of $\Lambda_n$ and runs $P(X)$, where $n = \poly(\tau)$ is determined precisely by the construction. Let us show that $X$ is uniformly distributed in $\Lambda_n$. Let $S'$ be a uniform preimage of $R$ in $\{0,1\}^\lambda$. Then, for any $s \in \{0,1\}^\lambda$, since $R = G(U_\lambda)$,
    \[
    \Pr[S' = s] = \Pr[S' = s \mid R = G(s)] \cdot \Pr[R = G(s)] = \frac{1}{|G^{-1}(G(s))|} \cdot \frac{|G^{-1}(G(s))|}{2^\lambda} = \frac{1}{2^\lambda}.
    \]
    It follows that $S'$ is uniform in $\{0,1\}^\lambda$.
    Since $S'$ is independent of the CRS, conditional on any $I=i$, the corrected seed tuple consists of independent uniform seeds. Hence $X$ is uniformly distributed in $\Lambda_n$ and independent of $I$.
    
    Additionally, the only difference between the input that $V'(R, \crs_0)$ computes for $V$ and $X$ (from $P'$) is a single uniform $s_i$-block of bits, which is replaced by an independent and uniformly distributed string.

    It follows that the distribution of inputs $(X^1, X^2)$ given to $P$ and $V$ by $P'(R, \crs_0)$ and $V'(R, \crs_0)$, respectively, over $R$ distributed as $D_\tau$ and uniform $\crs_0$ is \emph{identical} to $(X, X(S, I))$, where $X$ is uniformly distributed in $\Lambda_n$, $S$ is uniform in $\{0,1\}^\lambda$, $I$ is uniform in $[m]$, and $X(S,I)$ denotes the string obtained by replacing the $I$th seed block (which corresponds to $s_i$) of $X$ with the seed $S$.
    
    In particular, this implies the following. Let $X$ denote a uniform element of $\Lambda_n$; let $S$ denote a uniform element of $\{0,1\}^\lambda$; let $I$ denote a uniform element of $[m]$, for the rest of this proof. Then
    \[
    \Pr_{R, \crs_0}[\braket{P', V'}(R, \crs_0) = 1] = \Pr_{X, S, I}[\braket{P(X), V^{X(S,I)}(n)} = 1].
    \]
    We make the following observation. If $V$'s oracle $X(S, I)$ (which is not in $\Lambda$ with high probability) is instead replaced with $X$ (which is in $\Lambda$), then the probability that $V$ accepts in the interaction $(P(X), V^X(n))$ is at least $2/3$, by the completeness of $(P,V)$ for $\Lambda$. Thus, it suffices to bound
    \[
    \left|\Pr_{X}[\braket{P(X), V^{X}(n)} = 1] - \Pr_{X, S, I}[\braket{P(X), V^{X(S,I)}(n)} = 1] \right|.
    \]
    We will bound this quantity for all fixed $x \in \Lambda_n$, which is clearly sufficient.
    
    Let $x \in \Lambda_n$. Consider the quantum query algorithm $(P(x), V^{(\cdot)}(n))$ obtained by interacting $V$ on its oracle with the honest prover on input $x$. W.l.o.g. this is a \emph{unitary} algorithm, where an acceptance outcome is decided by a fixed POVM element $\Pi_{acc}$. Note that this is indeed a fixed query algorithm, since, for fixed $x$, the final state only depends on the verifier's oracle. Call this query algorithm $\calA^x(\cdot)$, where $\calA^x(z)$ denotes that the verifier uses the unitary oracle $U_z$. We are interested in comparing the final states of $\calA^x(x)$ and $\calA^x(x(S, I))$. Consider fixed $s \in \{0,1\}^\lambda$ and $i \in [m]$.

    For each $t \in \{0, \dots, q-1\}$, let $\ket{\psi^x_t(z)}$ be the state of $\calA^x(z)$ immediately before the $(t+1)$st oracle call to $U_z$, and let $\ket{\psi^x_q(z)}$ denote the final state of $\calA^x(z)$. By the hybrid method as formulated in~\cite{BBBV-strengths-weaknesses,quantum-query-lower-bounds}, we obtain that for each $t \in \{0, \dots, q-1\}$,
    \[
    \| \ket{\psi^x_{t+1}(x)} - \ket{\psi^x_{t+1}(x(s, i))} \| \le  \| \ket{\psi^x_{t}(x)} - \ket{\psi^x_{t}(x(s, i))} \| + 2 \min_{z \in \{x, x(s,i)\}} \| \Pi_{\calS_i} \ket{\psi^x_t(z)} \|,
    \]
    in the $\ell_2$ norm,
    where $\calS_i$ is the set of query register indexes corresponding to the $i$th seed block, and $\Pi_{\calS_i} = \sum_{j \in \calS_i} \ket{j}\bra{j}$ is the projector onto the subspace spanned by setting the query register to $j$, for $j \in \calS_i$.
    In particular, we have
    \[
    \| \ket{\psi^x_{t+1}(x)} - \ket{\psi^x_{t+1}(x(s, i))} \| \le  \| \ket{\psi^x_{t}(x)} - \ket{\psi^x_{t}(x(s, i))} \| + 2 \| \Pi_{\calS_i} \ket{\psi^x_t(x)} \|.
    \]

    Since the initial states are equal ($\ket{\psi^x_{0}(x)} = \ket{\psi^x_{0}(x(s,i))}$), telescoping these inequalities yields
    \[
    \| \ket{\psi^x_{q}(x)} - \ket{\psi^x_{q}(x(s, i))} \| \le 2 \sum_{t=0}^{q-1} \| \Pi_{\calS_i} \ket{\psi^x_t(x)} \|.
    \]

    Now, consider the \emph{expected $\ell_2$ distance} over uniform $s$, denoted $S$, and uniform $i$, denoted $I$.
    Then, by linearity of expectation and the Cauchy--Schwarz inequality,
    \begin{align}
        \bbE_{S, I}\| \ket{\psi^x_{q}(x)} - \ket{\psi^x_{q}(x(S, I))} \| &\le 2 \sum_{t=0}^{q-1} \bbE_{S, I} \| \Pi_{\calS_I} \ket{\psi^x_t(x)} \|\\
        &= \frac{2}{m} \sum_{t=0}^{q-1} \sum_{i=1}^m \| \Pi_{\calS_i} \ket{\psi^x_t(x)} \|\\
        &\le \frac{2}{m} \sum_{t=0}^{q-1} \sqrt{m \sum_{i=1}^m \| \Pi_{\calS_i} \ket{\psi^x_t(x)} \|^2}\\
        &\le \frac{2}{m} \sum_{t=0}^{q-1} \sqrt{m} = 2\frac{q}{\sqrt{m}},
    \end{align}
    where (2) holds by linearity of expectation, (3) holds by the Cauchy--Schwarz inequality, and (4) holds since $\{\Pi_{\calS_i}\}_{i \in [m]}$ are orthogonal projectors, so that $\sum_{i=1}^m \| \Pi_{\calS_i} \ket{\psi^x_t(x)} \|^2 \le 1$ for all $t$.

    Now we bound the expected difference in acceptance probabilities. Let $p_x = \braket{\psi^x_q(x) \mid \Pi_{acc} \mid \psi^x_q(x)}$ and $p'_{x,s,i} = \braket{\psi^x_q(x(s,i)) \mid \Pi_{acc} \mid \psi^x_q(x(s,i))}$ be the probabilities of acceptance for $\ket{\psi^x_q(x)}$ and $\ket{\psi^x_q(x(s,i))}$, respectively. Note that by definition of these states,
    $p_x = \Pr[\braket{P(x), V^{x}(n)} = 1]$ and $p'_{x,s,i} = \Pr[\braket{P(x), V^{x(s,i)}(n)} = 1]$. Since the trace distance is at most twice the $\ell_2$ distance for pure states, we obtain
    \[
    |p_x - p'_{x,s,i}| \le \frac{1}{2}\| \ket{\psi^x_q(x)}\bra{\psi^x_q(x)} - \ket{\psi^x_q(x(s,i))}\bra{\psi^x_q(x(s,i))}\|_1 \le \| \ket{\psi^x_{q}(x)} - \ket{\psi^x_{q}(x(s, i))} \|.
    \]

    Taking the expectation over uniform $s$ and uniform $i$ of the above inequality yields
    \[
    \bbE_{S,I} |p_x - p'_{x,S,I}| \le \bbE_{S,I} \| \ket{\psi^x_{q}(x)} - \ket{\psi^x_{q}(x(S, I))} \| \le 2 \frac{q}{\sqrt{m}} = \frac{1}{6},
    \]
    since $m = 144q^2$.

    We previously observed that the completeness of $(P,V)$ implies that $p_x = \Pr[\braket{P(x), V^{x}(n)} = 1] \ge 2/3$. Since we have $p'_{x,s,i} \ge p_x - |p_x - p'_{x,s,i}| \ge 2/3 - |p_x - p'_{x,s,i}|$, we obtain
    \[
    \bbE_{S,I} (p'_{x,s,i}) \ge \bbE_{S,I} \left(\frac{2}{3} - |p_x - p'_{x,s,i}|\right) \ge \frac{2}{3}-\frac{1}{6} = \frac{1}{2}.
    \]

    Using $p'_{x,s,i} = \Pr[\braket{P(x), V^{x(s,i)}(n)} = 1]$ and further averaging over uniform $x \in \Lambda_n$ yields the desired inequality
    \[
    \Pr[\braket{P(X), V^{X(S,I)}(n)} = 1] \ge \frac{1}{2}.
    \]
    Average-case completeness follows.

    \paragraph{Soundness.} Let $r \in \{0,1\}^\tau \setminus \calL$, and consider any CRS string $\crs_0 = (s_1, \dots, s_m, i)$. Let $P^*$ be any cheating prover strategy. Notice that $V'(r, \crs_0)$ gives the string $x = (C(r_1, \dots, r_m), s_1, \dots, s_m)$ to $V$, where $r_j = G(s_j)$ if $j \ne i$ and $r_i = r$. Let us bound the Hamming distance between $x$ and $\Lambda_n$. Every string in $\Lambda_n$ has a substring that is a codeword that decodes to a string that is not $(r_1, \cdots, r_m)$, since $r_i \notin \calL$. Since the distance of $C$ on messages of length $2 \lambda m$ is greater than $\epsilon \cdot \frac{1+2\alpha}{2} \cdot 2\lambda m$, the Hamming distance between $x$ and $\Lambda_n$ is greater than $2 \cdot \epsilon \cdot \frac{1+2\alpha}{2} \cdot \lambda m = \epsilon \cdot (\lambda m + \alpha \cdot 2\lambda m) = \epsilon \cdot n$. Thus, the normalized distance is greater than $\epsilon$. By the soundness of $(P,V)$, the strategy $P^*$ cannot make $V^x(n)$ accept with probability greater than $1/3$; thus, $\Pr[\braket{P^*(r, \crs), V'(r, \crs)}=1] \le 1/3$.
\end{proof}

\subsection{Reducing a QIP to a PRG distinguisher}\label{Subsec:QIP-to-PRG-distinguisher}

To complete the proof of Theorem~\ref{thm:main_lower_bound}, we apply the work of Beigi, Shor, and Watrous~\cite{BSW-QIP-short-messages} on quantum interactive proofs with small communication. Recall Proposition~\ref{prop:BSW_small_comm_QIP_simulation} from the technical overview, which shows that QIPs with communication $c$ and time $\poly(n)$ can be simulated by a quantum algorithm in time $2^{O(c)} \poly(n)$ for instances of length $n$.


We will use this result as a black-box to complete the proof of the theorem, noting that a straightforward application is not immediately trivial, since Lemma~\ref{lem:lower_bound_reduction} only guarantees average-case completeness.

\begin{proof}[Proof of Theorem~\ref{thm:main_lower_bound}]
    We use the property $\Lambda$, which is in logspace-uniform $\NC$ by Lemma~\ref{lem:property_in_NC}.
    Towards contradiction, suppose that there exists an $\epsilon$-QIPP for $\Lambda$ with query complexity $q$, communication complexity $c$ and a quantum polynomial time verifier such that $q^2 \cdot c = o(n)$. By Lemma~\ref{lem:lower_bound_reduction}, there exists an average-case complete QIP for $\calL$ in the CRS model with respect to $(D_\tau)_\tau$, denoted $(P, V, \crs)$, with completeness $1/2$, soundness $1/3$, a polynomial-time verifier, and communication complexity $o(\tau)$. 
    
    To invoke Proposition~\ref{prop:BSW_small_comm_QIP_simulation}, define the following auxiliary promise problem.
    The YES instances of $\Pi$ are pairs $(r, \crs_0)$ such that $r \in \calL \cap \{0,1\}^\tau$ and $\crs_0 \gets \crs(1^\tau)$ such that $\Pr[\braket{P, V}(r, \crs_0) = 1] \ge 2/5$; the NO instances of $\Pi$ are pairs $(r, \crs_0)$ such that $r \in \{0,1\}^\tau \setminus \calL$ and $\crs_0 \gets \crs(1^\tau)$. 
    
    It is obvious by construction that $(P, V)$ is a QIP for $\Pi$ with ordinary completeness $2/5$ (by the promise of acceptance probability) and soundness $1/3$ (since soundness holds for \emph{any} CRS string). Thus, we may apply Proposition~\ref{prop:BSW_small_comm_QIP_simulation}.

    Let $A$ be the quantum algorithm for $\Pi$ given by Proposition~\ref{prop:BSW_small_comm_QIP_simulation}, which runs in time $2^{o(\tau)} \poly(\tau) = 2^{o(\tau)}$. By standard amplification techniques, we may assume that $A$ has completeness $1/2$ and soundness $1/100$. Define the following distinguisher $D$ for $G$.

    On input $r \in \{0,1\}^\tau$, do:
    \begin{enumerate}
        \item Sample $\crs_0 \gets \crs(1^\tau)$.
        \item Compute $b \gets A(r, \crs_0)$.
        \item Output $b$.
    \end{enumerate}

    Clearly $D$ runs in time $2^{o(\tau)}$. To show that indeed $D$ distinguishes pseudorandom inputs from truly random inputs, we first bound the probability of $(r, \crs_0)$ satisfying the promise of $\Pi$ in the pseudorandom case.

    Let $R = G(S)$, where $S$ is uniform in $\{0,1\}^\lambda$, for $\tau = 2\lambda$. Then indeed, $R$ is distributed as $D_\tau$. Consider the random variable $Z = \Pr[\braket{P, V}(r, \crs_0) = 1]$, over random $r \gets R$ and $\crs_0 \gets \crs(1^\tau)$. We have the following inequalities, the first by average-case completeness of $(P, V, \crs)$:
    \begin{align*}
        1/2 &\le \Pr_{r \gets R, \crs_0 \gets \crs(1^\tau)}[\braket{P, V}(r, \crs_0) = 1] = \bbE_{r, \crs_0}\Pr[\braket{P, V}(r, \crs_0) = 1] = \bbE [Z]\\
        &= \bbE[Z \mid Z \ge 2/5] \Pr[Z \ge 2/5] + \bbE[Z \mid Z < 2/5] \Pr[Z < 2/5]\\
        &\le 1 \cdot \Pr[Z \ge 2/5] + \frac{2}{5}\cdot \Pr[Z < 2/5] = \Pr[Z \ge 2/5] + \frac{2}{5} \cdot (1-\Pr[Z \ge 2/5])\\
        &= \frac{3}{5}\cdot \Pr[Z \ge 2/5] + \frac{2}{5}.
    \end{align*}
    Thus, 
    \[
    \Pr_{r, \crs_0}[\Pr[\braket{P, V}(r, \crs_0) = 1] \ge 2/5] = \Pr[Z \ge 2/5] \ge \frac{1}{6}.
    \]
    Note that when $r$ is drawn from $D_\tau$, the event that $\Pr[\braket{P, V}(r, \crs_0) = 1] \ge 2/5$ implies that $A$ gets a YES instance of $\Pi$.
    It follows that 
    \[
    \Pr_{s \in \{0,1\}^\lambda}[D(G(s)) = 1] \ge \frac{1}{6} \cdot \frac{1}{2} = \frac{1}{12}.
    \]

    Now consider the case when $D$ is given an input $r$ drawn uniformly from $\{0,1\}^\tau$. With probability at least $1-2^{-\lambda}$, the uniform string $r$ is not in $\calL$, so in this case, $A$ is given an input that is a NO instance of $\Pi$. Thus, 
    \[
    \Pr_{r \in \{0,1\}^\tau} [D(r) = 1] \le 2^{-\lambda} + \frac{1}{100}.
    \]

    It follows that 
    \[
    \left| \Pr_{s \in \{0,1\}^\lambda}[D(G(s)) = 1] - \Pr_{r \in \{0,1\}^\tau} [D(r) = 1] \right| = \Omega(1),
    \]
    contradicting the subexponentially secure pseudorandomness of $G$.
\end{proof}

We conclude this section with the following observation. 
While Theorem~\ref{thm:main_lower_bound} shows a lower bound on the power of QIPPs for logspace-uniform $\NC$ properties that is \emph{tight} up to polylogarithmic factors for a communication bound $c \ge \Omega(\sqrt{n})$, a natural question emerges. For the other regime where communication is bounded by $c \le O(\sqrt{n})$, there is a gap between our algorithmic result for logspace-uniform $\NC$ properties and this lower bound. For example, our QCIPP given in Section~\ref{Sec:nc-aa-speedup} can have query and communication complexity $q+c = \softO(n^{2/5})$, but the lower bound we show in this section only forbids $q + c = o(n^{1/3})$. What is the optimal sum of query and communication complexity for QIPPs for logspace-uniform $\NC$ properties? In the next section, we illustrate a deep, bidirectional connection between this question and the QIP complexity of parity.

\section{QIPPs for logspace-uniform NC and QIPs for Parity}\label{Sec:qipp-nc-equiv-qip-parity}

In this section, we show a bi-directional connection between the complexity of QIPPs for $\NC$ and the QIP complexity of (exact) $\Parity$ in the query model, where $\Parity$ is defined by the property $\Parity_k := \{x \in \{0,1\}^k : |x| \equiv 0 \pmod{2}\}$.
Throughout this section, $k$ denotes the input length for an instance of $\Parity$ and $n$ denotes the input length of whichever property $\Pi$ we are considering.

Recall from the introduction and \cite{dallagnol-gur-roymoulik-thaler:qmap} that $\Parity$ has, for every communication bound $c(k)$, a QCMA proof for $\Parity$ with communication at most $c(k)$ and query complexity $O(\frac{k}{\sqrt{c}})$ for instances of length $k$; setting $c = O(k^{2/3})$ gives a QIP for Parity with total complexity $O(k^{2/3})$. It is unknown whether QIPs in the query model can beat the QCMA construction above, for any communication bound $c$.

We show the following connections:
\begin{enumerate}
    \item If $\Parity$ has a QIP with total complexity $k^{2/3 - \Omega(1)}$ for inputs of length $k$, then every property in logspace-uniform $\NC$ has an $n^{-3/5}$-QIPP with total complexity $n^{2/5-\Omega(1)}$. This is made precise in Theorem~\ref{thm:qip-parity-implies-qipp-nc} and Corollary~\ref{cor:qip-parity-implies-qipp-nc} in Section~\ref{Subsec:qip-parity-implies-qipp-nc}. 
    \begin{enumerate}
        \item As a corollary, we show that Parity is \emph{maximally hard} (up to polylogarithmic factors) in exact-decision QIP complexity over all properties in logspace-uniform $\NC$. See Corollary~\ref{cor:qip-for-all-nc}.
    \end{enumerate}
    
    \item There exists a property $\Pi$ in logspace-uniform $\NC$ such that an $n^{-3/5}$-QIPP for $\Pi$ with total complexity $o(n^{2/5})$ implies a QIP for $\Parity$ with communication complexity $c = o(k^{2/5})$ and query complexity $o(k^{4/5}) = o(\frac{k}{\sqrt{c}})$. See Theorem~\ref{thm:qipp-nc-implies-qip-parity} and Corollary~\ref{cor:qipp-nc-implies-qip-parity} in Section~\ref{Subsec:qipp-nc-implies-qip-parity}.

    \item Toward understanding the QIP complexity of Parity, and thus the QIPP complexity of logspace-uniform $\NC$ properties, we show an $\Omega(\sqrt{k})$ query lower bound on any QIP for Parity. We prove an analogous lower bound for any high-sensitivity property. We refer the reader to Proposition~\ref{prop:qip-sensitivity-lower-bound} and Corollary~\ref{cor:parity-lower-bound} in Section~\ref{Subsec:parity-lower-bound}.
\end{enumerate}

\subsubsection*{A Parity barrier for the $\pval$ approach}

We highlight an important consequence of our reduction to QIPs for Parity. Within the paradigm of \cite{RVW13,RR20}, any improvement to the QIPP for $\pval$ that yields total complexity $o(n^{2/5})$ for all logspace-uniform $\NC$ properties would also improve the known query--communication tradeoff for Parity.
These constructions first use $\Omega(\epsilon n)$ parallel copies of the GKR protocol \cite{goldwasser-kalai-rothblum:delegating-computation} to reduce verification of the original property to an $\epsilon$-proximity claim for $\pval$, and then apply an $\epsilon$-QIPP for $\pval$. Here $\epsilon=\epsilon(n)$ is an internal parameter no larger than the desired proximity parameter. Since the reduction preserves the proximity guarantee, the composed protocol is itself an $\epsilon$-QIPP, even when only constant proximity is required.

To see the consequence, let $\Pi$ be the logspace-uniform $\NC$ property supplied by Corollary~\ref{cor:qipp-nc-implies-qip-parity} for proximity $n^{-3/5}$. Suppose a construction following this paradigm gives $\Pi$ a QIPP with query and communication complexities $q(n),c(n)=o(n^{2/5})$. The initial GKR reduction alone communicates $\Omega(\epsilon n)$ bits, so $\epsilon n=O(c(n))=o(n^{2/5})$, and hence,
\[
\epsilon=o(n^{-3/5}).
\]
Thus the composed protocol is also an $n^{-3/5}$-QIPP for $\Pi$ with total complexity $o(n^{2/5})$. Corollary~\ref{cor:qipp-nc-implies-qip-parity} then yields a QIP for $\Parity_k$ with query complexity $o(k^{4/5})$ and communication complexity $o(k^{2/5})$, improving the known tradeoff for Parity.

\subsection{QIPs for Parity imply QIPPs for logspace-uniform NC}\label{Subsec:qip-parity-implies-qipp-nc}
We begin with a transformation of QIPs for Parity to QIPPs for any property in logspace-uniform $\NC$. The idea is to use the same strategy as in Section~\ref{Sec:nc-aa-speedup}, but instead verify each of the sampled linear constraints with $O(\log n)$ invocations of a QIP for Parity. Specifically, the crucial observation is that sums over a field $\F$ of characteristic $2$ can be computed by adding each index of the sum independently in $\F_2$, where field elements are represented as bit strings. Then, a linear constraint $\sum_{i \in [p]} \alpha_k(i) \cdot x[q_k(i)] = \beta_k$ holds in $\F$ if and only if for each index of representations of elements of $\F$, the linear constraint at that index holds. To verify a linear constraint, one simply needs to verify each index individually, which is an instance of Parity. Hence, one obtains a QIPP for all of logspace-uniform $\NC$ from any QIP for Parity.

\begin{theorem}\label{thm:qip-parity-implies-qipp-nc}
    Suppose that for some constants $0 \le a,b \le 1$, Parity has a QIP with query complexity $O(k^a)$ and communication complexity $O(k^b)$.
    Let $\epsilon = \epsilon(n) \in [1/n, 1)$ be a proximity parameter. Then every property in logspace-uniform $\NC$ has an $\epsilon$-QIPP with:
    \begin{itemize}
        \item Query complexity $\softO(\epsilon^{-a})$.
        \item Communication complexity $\softO(\epsilon n + \epsilon^{-b})$.
    \end{itemize}
\end{theorem}


\begin{proof}
    Define the property
    \[
    \MOD(\F)_{k} = \{(x^1, \dots, x^k) \in \F^k : \sum_{i = 1}^k x^i = 0\},
    \]
    for a field $\F$ of characteristic $2$ with $|\F| = \poly(n)$. A QIP for $\MOD(\F)$ is defined as a QIP for the binary property 
    \[
    \{(x^1, \dots, x^k) \in (\{0,1\}^{\log |\F|})^k : (\braket{x^1}_\F, \cdots, \braket{x^k}_\F) \in \MOD(\F)_k\},
    \]
    where $\braket{x^i}_\F$ denotes the field element that is represented by the bit string $x^i$. 
    
    We first show that any QIP for $\Parity$ with the parameters given in the theorem implies that $\MOD(\F)_k$ has a QIP with query complexity $k^{a} \polylog(n)$ and communication complexity $k^{b} \polylog(n)$. To see this, we first use parallel repetition only $O(\log \log n)$ times to amplify the completeness and soundness errors to at most $\eta \le \frac{1}{3 \log |\F|}$, by Lemma~\ref{lem:parallel_repetition} and using $\log |\F| = O(\log n)$. Observe that addition in a field of characteristic 2 can be reduced to computing bit-wise parity of each of $\log|\F| = O(\log n)$ positions. Indeed, for any $(x^1, \dots, x^k) \in (\{0,1\}^{\log |\F|})^k$, we have $(\braket{x^1}_\F, \cdots, \braket{x^k}_\F) \in \MOD(\F)_k$ if and only if for all $j \in [\log |\F|]$, we have $\bigoplus_{i=1}^k x^i_j = 0$, which happens if and only if for all $j \in [\log |\F|]$, $(x^1_j, \dots, x^k_j) \in \Parity_k$. Thus, we obtain a QIP for $\MOD(\F)_k$ by running the QIP for $\Parity_k$ on each of the $\log |\F| = O(\log n)$ positions. This protocol has completeness at least $(1-\eta)^{\log |\F|} \ge 1-\eta \log |\F| \ge \frac{2}{3}$ and soundness at most $\eta \le \frac{1}{3}$ for $\MOD(\F)_k$. The claimed query complexity and communication complexity of this protocol follow by construction.
    
    Next, for any $t = \polylog(n)$, we may use $t$ invocations of parallel repetition of the QIP for $\MOD(\F)$, outputting the majority decision, to obtain a QIP for $\MOD(\F)$ with completeness and soundness error at most $\zeta(t) = 2^{-\Omega(t)}$; the query and communication complexity get blown up by a factor of at most $t$. Call this $t$-repeated amplified protocol $(P_t, V_t)$.

    Now we show how to apply the established QIP for $\MOD(\F)$ to the protocol given by Theorem~\ref{thm:rr20}, thus obtaining the desired QIPP for $\Pi$. 
    The protocol on input $x \in \{0,1\}^n \subseteq \F^n$ is described as follows. Let $5/6 \le \gamma_s < \gamma_c < 1$ be any constants.

    \begin{enumerate}
        \item Run the interactive protocol of Theorem~\ref{thm:rr20} to $\Pi$, with communication at most $\epsilon \cdot n \cdot \polylog(n)$. Both parties obtain integers $C = O(n)$, $p = O(\frac{1}{\epsilon})$, and $\delta \ge 1/\polylog(n)$, and a pair $(\sigma, \beta)$, for a partial transcript $\sigma$ and $\beta = (\beta_1, \dots, \beta_C) \in \F^C$. Given $\sigma$, the prover and verifier can compute the constraint-defining functions $q_1, \dots, q_C : [p] \to [n]$ and $\alpha_1, \dots, \alpha_C : [p] \to \F$. 

        \item $V$ sets $s = O(1/\delta) = \polylog(n)$ and samples an ordered list $K = (k_1, \dots, k_s)$ of independent uniform indices in $[C]$. Then $V$ sends $K$ to $P$. 

        \item Let $t$ be such that $\zeta(t) \le \min\{\frac{1-\gamma_c}{s}, \frac{1}{2}\}$, so that $t = O(\log s) = O(\log n)$ and $\zeta(t) \le 2^{-\Omega(t)}$. For each $k \in K$, $(P, V)$ in parallel invoke the QIP $(P_t, V_t)$ for $\MOD(\F)_{p+1}$ on instance
        \[
        (\alpha_k(1) \cdot x[q_k(1)], \cdots, \alpha_k(p) \cdot x[q_k(p)], -\beta_k).
        \]
        Note that this instance is in $\MOD(\F)_{p+1}$ if and only if $\sum_{i \in [p]} \alpha_k(i) \cdot x[q_k(i)] = \beta_k$.

        \item If the verifier rejects in any of these $s$ instances, the verifier outputs reject. Otherwise, the verifier outputs accept.
    \end{enumerate}

    By a union bound, this protocol has completeness at least $1-(s \cdot \zeta(t)) \ge \gamma_c$.
    
    For soundness, let $E$ denote the event that the verifier rejects during the initial interaction or at least one sampled constraint is false. Choose $s = O(1/\delta) = \polylog(n)$ so that $(1-\delta)^s \le 1/3$. By Theorem~\ref{thm:rr20}, with probability at least $1/2$, the verifier rejects or at least a $\delta$-fraction of constraints are false. Hence $\Pr[E] \ge \frac12(1-(1-\delta)^s) \ge 1/3$. Conditioned on $E$, the verifier either already rejects or must accept a false constraint, which occurs with probability at most $\zeta(t)$ by soundness of $(P_t,V_t)$. Thus the probability that the verifier accepts is at most
    \[
    \Pr[E]\zeta(t)+1-\Pr[E]
    \le \frac13\zeta(t)+\frac23
    \le \frac56 \le \gamma_s.
    \]
    Hence the protocol has soundness at most $\gamma_s$.
    
    Completeness $\gamma_c$ and soundness $\gamma_s$ can be amplified to error at most $1/3$ with constantly many parallel repetitions.

    Every query to the string of field elements $(\alpha_k(1) \cdot x[q_k(1)], \dots, \alpha_k(p) \cdot x[q_k(p)], -\beta_k)$ for a particular $k \in K$ can be simulated with a single query to $x$. Hence the query complexity of this protocol is
    \begin{align*}
    q(n) &\le p^{a} \polylog(n) \cdot O(1/\delta) \cdot t = \softO(\epsilon^{-a}),
    \end{align*}
    since $p = O(1/\epsilon)$.
    Similarly, the communication complexity is
    \begin{align*}
    c(n) &\le 
    \softO(\epsilon n) + p^{b} \cdot t \cdot \polylog(n) = \softO(\epsilon n + \epsilon^{-b}).
    \end{align*}
\end{proof}

As a consequence of this theorem, we observe the following.

\begin{corollary}\label{cor:qip-parity-implies-qipp-nc}
    Suppose that $\Parity$ has a QIP with query and communication complexity at most $k^{2/3-\Omega(1)}$. Then every property in logspace-uniform $\NC$ has an $n^{-3/5}$-QIPP with query and communication complexity at most $n^{2/5-\Omega(1)}$.
\end{corollary}
\begin{proof}
    Let $\alpha > 0$ be a constant such that Parity has a QIP with query and communication complexity at most $O(k^{2/3-\alpha})$. Let $\epsilon = n^{-(3/5+\alpha/2)}$, so that $\epsilon \le n^{-3/5}$.
    Applying Theorem~\ref{thm:qip-parity-implies-qipp-nc} with proximity parameter $\epsilon$ gives an $\epsilon$-QIPP, and thus an $n^{-3/5}$-QIPP, for all logspace-uniform $\NC$ properties with query complexity
    \[
    \softO((n^{3/5+\alpha/2})^{2/3-\alpha}) = \softO(n^{2/5+\alpha/3 -3\alpha/5 - \alpha^2/2}) = n^{2/5-\Omega(1)}
    \]
    and communication complexity
    \[
    \softO(n^{-(3/5 + \alpha/2)} \cdot n + (n^{3/5 + \alpha/2})^{2/3 - \alpha}) = n^{2/5-\Omega(1)}.
    \]
\end{proof}

As a further consequence of Theorem~\ref{thm:qip-parity-implies-qipp-nc}, we show that Parity is \emph{maximally hard} in QIP complexity over every property in logspace-uniform $\NC$, up to polylogarithmic factors. This result is striking, as it implies that the entire class of logspace-uniform $\NC$ properties admits \emph{exact-decision} QIPs with sublinear queries and communication. Classically, this is not possible, as Parity is a logspace-uniform $\NC$ property that requires a linear number of queries in any exact-decision interactive proof, regardless of communication \cite{dallagnol-gur-roymoulik-thaler:qmap}.

\begin{corollary}\label{cor:qip-for-all-nc}
    Suppose that for some constants $0 \le a,b \le 1$, $\Parity$ has a QIP with query complexity $O(k^a)$ and communication complexity $O(k^b)$. Then every property in logspace-uniform $\NC$ has a QIP with query complexity $\softO(k^a)$ and communication complexity $\softO(k^b)$.
\end{corollary}
\begin{proof}
    Let $\Pi$ be such a property. Let $D(\Pi)$ denote the property $D(\Pi)_{2k} = \{(x,x) : x \in \Pi_k\}$. This property is in logspace-uniform $\NC$, so we may apply Theorem~\ref{thm:qip-parity-implies-qipp-nc} to $D(\Pi)$ with proximity parameter $\epsilon = 1/(2k)$ for instances of length $2k$. This yields an $\epsilon$-QIPP for $D(\Pi)$ with query complexity $\softO(k^a)$ and communication complexity $\softO(k^b)$. We can obtain a QIP for $\Pi$ as follows: given input $x \in \{0,1\}^k$, simulate the QIPP on input $(x, x)$, simulating one query with the corresponding query to $x$. Completeness is preserved. For soundness, if $x \notin \Pi_k$, then $(x,x)$ differs from $D(\Pi)$ in at least two positions, so that $(x,x)$ is indeed $\epsilon$-far from $D(\Pi)$; the soundness of the QIPP for $D(\Pi)$ thus implies soundness for this QIP for $\Pi$. Lastly, the query and communication complexities are preserved, proving the claim.
\end{proof}

\subsection{QIPPs for logspace-uniform NC imply QIPs for Parity}\label{Subsec:qipp-nc-implies-qip-parity}
Now we prove the converse, for small proximity parameters. Our construction uses the same ensemble of codes of~\cite{Spielman1996LinearTime,GuruswamiKabanets2008Hardness} used in the construction of Section~\ref{Sec:QIPP_lower_bound}. Given such a code $C_m : \{0,1\}^m \to \{0,1\}^{\alpha m}$ of constant rate and constant distance, one can construct a ``locally encodable'' code (still computable in logspace-uniform $\NC$) $C_m^\ell : \{0,1\}^{m\ell} \to \{0,1\}^{\alpha m \ell}$, for a parameter $\ell$, by using an independent copy of $C_m$ to encode each consecutive substring of length $m$ of the message. Then, the property $\Pi$ we construct is an encoded Parity:
\[
\Pi_n = \{C_m^\ell(x) : x \in \Parity_{m \ell}\},
\]
for a specific tradeoff of $m$ and $\ell$ determined by the proximity parameter.
The reduction to Parity goes as follows: on an input $x \in \{0,1\}^k$, simulate a QIPP for $\Pi$ on input $C_m^\ell(x)$, where queries to $C_m^\ell(x)$ are simulated by querying the substring of $x$ that encodes to the needed query location. As long as the proximity parameter is set sufficiently small, an instance $x$ not in $\Parity$ encodes to an instance that is sufficiently far from $\Pi$, since the distance of $C_m^\ell$ is exactly the (constant) distance of $C_m$ divided by $\ell$.

The code construction clearly loses in relative distance, but gains in being locally computable, which is needed to reduce a QIPP for $\Pi$ to a QIP for Parity. This is formalized in Theorem~\ref{thm:qipp-nc-implies-qip-parity} below.

\begin{theorem}\label{thm:qipp-nc-implies-qip-parity}
    Let $\epsilon = \epsilon(n) \in [1/n,\epsilon_0)$ be a proximity parameter computable in space $O(\log n)$ satisfying $\epsilon(O(n)) = O(\epsilon(n))$, where $\epsilon_0 \in (0,1)$ is a universal constant. Then there exists a property $\Pi$ in logspace-uniform $\NC$ such that the following holds. If there exists an $\epsilon$-QIPP for $\Pi$ with query complexity at most $q(n)$ and communication complexity at most $c(n)$ satisfying $q(O(n)) = O(q(n))$ and $c(O(n)) = O(c(n))$, then there exists a QIP for $\Parity$ with query complexity $O(\epsilon(k) \cdot k \cdot q(k))$ and communication complexity $O(c(k))$ on inputs of length $k$. 
\end{theorem}
\begin{proof}
    Let $\epsilon$ satisfy the theorem conditions.
    Let $\{C_m : \{0,1\}^m \to \{0,1\}^{\alpha m}\}_{m \in \N}$ be the code family from Theorem \ref{thm:good-nc-codes}, noting that we use $m$ to denote the message length instead of $k$. Without loss of generality, we assume that each $C_m$ is \emph{systematic}: for all $x$, the first $m$ encoded bits of $C_m(x)$ are precisely $x$. Let $\Delta \in (0,1)$ be a constant such that $\Delta(C_m)/\alpha > \Delta$ for all $m$. For $\ell \in \N$, define the code $C_m^\ell : \{0,1\}^{m\ell} \to \{0,1\}^{\alpha m \ell}$ as
    \[
    C_m^\ell(x_1, \dots, x_\ell) = C_m(x_1) || C_m(x_2) || \cdots || C_m(x_\ell),
    \]
    for $|x_i| = m$, and $||$ denotes concatenation.
    It is easy to see that $\Delta(C_m^\ell) \ge \frac{\Delta}{\ell}$.

    Now we construct the property $\Pi = (\Pi_n)$. For each $n$, let $m = m(n) = \frac{\epsilon n}{\alpha \Delta}$ and $\ell = \ell(n) = \frac{\Delta}{\epsilon}$. Define
    \[
    \Pi_n = \{C_m^\ell(x) : x \in \Parity_{m \ell}\}.
    \]
    We show that $\Pi$ is a logspace-uniform $\NC$ property by providing such an algorithm that decides $\Pi$:
    \begin{enumerate}
        \item Compute $m$ and $\ell$ from $n$ and $\epsilon$, using the fact that $\epsilon(n)$ is computable in logarithmic space.
        \item Given a string $z \in \{0,1\}^{\alpha m\ell}$ claimed to be an encoding of a string in $\Parity_{m\ell}$, recover the underlying message $x$ using that $C_m^\ell$ is systematic.
        \item Compute the encoding $C_m^\ell(x)$ in logspace-uniform $\NC$ (recalling that $\{C_m\}$ is computable in logspace-uniform $\NC$ by Theorem~\ref{thm:good-nc-codes}), and check that $C_m^\ell(x) = z$.
        \item If the check passes, compute the parity of $x$ as $y = \bigoplus_{i=1}^{m\ell} x_i$ and accept if and only if $y=0$.
    \end{enumerate}
    The circuit family described by this algorithm clearly has polynomial size and polylogarithmic depth, and is logspace-uniform. Hence $\Pi$ is a logspace-uniform $\NC$ property.
    
    To reduce an $\epsilon$-QIPP for $\Pi$ to a QIP for Parity, we use the distance bound $\Delta(C_m^\ell)/\alpha > \frac{\Delta}{\ell} = \epsilon$.
    
    Suppose that $\Pi$ has an $\epsilon$-QIPP $(P, V)$ with query complexity $q(n)$ and communication complexity $c(n)$. We will reduce this QIPP to an exact-decision QIP for $\Parity$. Let $k = n/\alpha = \Theta(n)$; so $k = m\ell$. 
    
    Given $x \in \{0,1\}^k$, the verifier and prover simulate $(P, V)$ on input $C_m^\ell(x)$. When the simulated verifier $V$ queries index $i$, the real verifier computes the corresponding block $b \in [\ell]$ containing $i$ as $b = \ceil{i/(\alpha m)}$; then the verifier computes $C_m(x[(b-1)m+1], \dots, x[bm])$ with $m$ queries to $x$ and returns the bit at index $1 + ((i-1) \bmod (\alpha m))$. This procedure extends to arbitrary superposition states in the usual way, given that the verifier has access to the unitary $U_x \ket{i}\ket{b} = \ket{i}\ket{b\oplus x_i}$.

    Completeness follows from the completeness of $(P, V)$ since $x \in \Parity_k$ implies $C_m^\ell(x) \in \Pi_n$. Soundness holds as follows: for $x \notin \Parity_k$, we have that $\Delta(C_m^\ell(x), \Pi) \ge \dist(C_m^\ell)/(\alpha m \ell) = \Delta(C_m^\ell)/\alpha > \epsilon$, so that we get soundness from the soundness of $(P, V)$.

    The query complexity of the QIP for $\Parity$ is $q(n) \cdot m(n) = O(\epsilon(n) \cdot n \cdot q(n)) = O(\epsilon(k) \cdot k \cdot q(k))$. The communication complexity is just $c(n) = c(\alpha k) = O(c(k))$.
\end{proof}

As a consequence of Theorem~\ref{thm:qipp-nc-implies-qip-parity}, we obtain the following by setting $\epsilon = n^{-3/5}$.
\begin{corollary}\label{cor:qipp-nc-implies-qip-parity}
    There exists a property $\Pi$ in logspace-uniform $\NC$ such that
    if $\Pi$ has an $n^{-3/5}$-QIPP with total complexity $o(n^{2/5})$,
    then $\Parity_k$ has a QIP with query complexity $o(k^{4/5})$
    and communication complexity $o(k^{2/5})$.
\end{corollary}

\subsection{A query lower bound for Parity}\label{Subsec:parity-lower-bound}

Motivated by the apparently deep connection between QIPPs for all logspace-uniform $\NC$ and QIPs for Parity, better understanding the QIP complexity of Parity is of significant interest. In this section, we show that any QIP for Parity requires $\Omega(\sqrt{k})$ queries, regardless of communication. In fact, we show an analogous lower bound for all properties $\Pi$ with high one-sided sensitivity, defined below.

For a property $\Pi \subseteq \{0,1\}^k$, define its (one-sided) sensitivity $s_1(\Pi)$ to be the maximum over $x \in \Pi$ of $|\{i \in [k] : x^i \notin \Pi\}|$, where $x^i$ denotes the string obtained by flipping the $i$th bit of $x$. This is a one-sided analogue of the usual sensitivity definition for Boolean functions.

Our main result of this section, formalized in Proposition~\ref{prop:qip-sensitivity-lower-bound}, is that every QIP for any property $\Pi \subseteq \{0,1\}^k$ requires at least $\Omega(\sqrt{s_1(\Pi)})$ queries. The proof is based on a standard hybrid method \cite{BBBV-strengths-weaknesses}, where we fix an honest prover execution on some $x \in \Pi$ and consider changing the verifier's oracle to a random neighboring instance that is not in $\Pi$. The induced query algorithm must be able to distinguish these cases with constant probability, but a single random bit change can only change the expected distance in states by a factor proportional to the square of the query complexity. The formal proof using this idea is written below.

\begin{proposition}\label{prop:qip-sensitivity-lower-bound}
    Let $\Pi \subseteq \{0,1\}^k$. Then every QIP for $\Pi$ has query complexity at least $\Omega(\sqrt{s_1(\Pi)})$.
\end{proposition}
\begin{proof}
    Let $V$ be the verifier of an arbitrary QIP for $\Pi$. Let $x \in \Pi$ such that $|\{i \in [k] : x^i \notin \Pi\}| = s_1(\Pi)$. Let $P$ be a prover strategy such that $\Pr[\braket{P, V^x(k)} = 1] \ge 2/3$, which exists by completeness. On the other hand, soundness implies that $\Pr[\braket{P, V^{x^i}(k)} = 1] \le 1/3$ for every $i$ such that $x^i \notin \Pi$. Let $A$ denote the quantum query algorithm given by $(P, V^x(k))$ on input oracle $x$. Then $A$ has the same query complexity as $V$, which we call $q$.

    Without loss of generality, we may purify the channels of $P$ and delay any measurements of $V$, so that $A$ is unitary prior to the measurement of the output bit. 

    Immediately before query $t$ of $A$ on input $x$, denote the state by $\ket{\psi_t}$; at the same point in $A$ but on input $x^i$, denote the corresponding state by $\ket{\psi_t^i}$. Let $\ket{\psi}$ denote the final state of $A$ on $x$, and let $\ket{\psi^i}$ denote the final state of $A$ on $x^i$. Let $\Pi_i$ denote the projector onto the subspace of states whose query index register is set to $\ket{i}$.

    A standard hybrid bound and Cauchy--Schwarz shows that
    \[
    \|\psi - \psi^i\| \le 2 \sum_{t=1}^q \|\Pi_i \ket{\psi_t}\| \le 2 \sqrt{q \sum_{t=1}^q \|\Pi_i \ket{\psi_t}\|^2 }
    \]
    Denote by $S$ the set $\{i \in [k] : x^i \notin \Pi\}$. Then,
    \begin{align}
        \sum_{i \in S} \|\psi - \psi^i\|^2 &\le \sum_{i \in S} 4 q \sum_{t=1}^q \|\Pi_i \ket{\psi_t}\|^2\\
        &\le 4q \sum_{t=1}^q \sum_{i \in S} \|\Pi_i \ket{\psi_t}\|^2\\
        &\le 4q^2.
    \end{align}

    No measurement can distinguish two states with advantage greater than their trace distance; for pure states, trace distance is at most the $L^2$ distance. Hence, $\|\psi - \psi^i\| \ge 1/3$ for all $i \in S$, since indeed $A$ distinguishes $x$ from $x^i$ with advantage $1/3$. It follows that
    \[
    \frac{|S|}{9} \le \sum_{i \in S} \|\psi - \psi^i\|^2 \le 4q^2.
    \]
    Thus, $q \ge \frac{\sqrt{s_1(\Pi)}}{6} = \Omega(\sqrt{s_1(\Pi)})$ since $|S| = s_1(\Pi)$.
\end{proof}

As a corollary, we obtain the following, noting that $s_1(\Parity_k) = k$.

\begin{corollary}\label{cor:parity-lower-bound}
    Every QIP for $\Parity$ has query complexity at least $\Omega(\sqrt{k})$.
\end{corollary}

\section{Random properties require linear QIPP complexity}\label{Sec:QIPP-random-lower-bound}

In this paper so far, we have proven many positive results on what QIPPs \emph{can} do, and limitations under either computational assumptions (Section~\ref{Sec:QIPP_lower_bound}) or with a subconstant proximity parameter (Section~\ref{Subsec:qipp-nc-implies-qip-parity}). Moreover, both of these lower bounds were for specific constructions designed for hardness.

In this section, we ask: for which properties $\Pi$ can we prove an \emph{unconditional} lower bound on the total complexity (i.e. sum of query and communication complexity) of any QIPP for $\Pi$? We prove that \emph{almost every} property $\Pi \subseteq \{0,1\}^n$ with cardinality $2^{\beta n}$ for a constant $\beta \in (0,1)$ requires \emph{linear} total complexity for any QIPP for $\Pi$. Furthermore, this linear lower bound is \emph{unconditional} and requires no assumptions on running time. Note that every property can be trivially tested with $n$ queries and no communication by reading the entire input; hence our lower bound is tight up to constant factors.

More formally, we prove the following. Let $\beta \in (0,1)$ be any constant. Then there exist constants $\epsilon, \alpha > 0$ such that the following holds. If $\Pi \subseteq \{0,1\}^n$ is sampled uniformly conditioned on $|\Pi| = 2^{\beta n}$, then the probability that $\Pi$ has an $\epsilon$-QIPP with total complexity at most $\alpha n$ is inverse doubly exponentially small in $n$. See Theorem~\ref{thm:qipp-random-lower-bound} for the complete statement, including the bounds on $\epsilon$ and $\alpha$ in terms of $\beta$.


We use the following notation throughout this section.
For a property $\Pi\subseteq \{0,1\}^n$ and some $\epsilon > 0$, recall that $\sfQIPP_\epsilon(\Pi)$ denotes the minimum value of $q(V) + c(V)$ over all $\epsilon$-QIPP verifiers $V$ for $\Pi$ with completeness and soundness errors at most $1/3$, where $q(V)$ denotes the query complexity of $V$ and $c(V)$ denotes the communication complexity of $V$. For $p \in [0,1]$, let $H(p) := - p \log p - (1-p) \log(1-p)$ denote the binary entropy function.

\begin{theorem}\label{thm:qipp-random-lower-bound}
    Let $0<\alpha<1/4$, $0<\beta<1$, and $0<\epsilon<1/2$ be constants such that $\beta + H(\epsilon) < 1$ and $\beta > 2\alpha + H(2\alpha)$. Let $m = 2^{\beta n}$. Then, for all sufficiently large $n$,
    \[
    \Pr_{|\Pi| = m} [\sfQIPP_\epsilon(\Pi) \le \alpha n]
    \le 2^{-2^{\beta n - o(1)}},
    \]
    where $\Pi$ is a uniform subset of $\{0,1\}^n$ conditioned on $|\Pi| = m$.
\end{theorem}

Our main technical contribution is to extend counting-based lower bounds
from quantum testers to arbitrary quantum interactive proofs, with bounds
depending only on query and communication complexity. We build on the
random-property argument of Buhrman, Fortnow, Newman, and
R\"ohrig~\cite{BFNR08}, which discretizes low-query testers and counts the
resulting descriptions.

The challenge is that optimization over the prover need not preserve low
polynomial degree, and the verifier's private workspace is unrestricted.
Using the strategy formalism of Gutoski and
Watrous~\cite{GutoskiWatrous2007QuantumGames}, in the verifier-channel form used by Beigi,
Shor, and Watrous~\cite{BSW-QIP-short-messages}, we encode a verifier making
$q$ queries and exchanging $c$ qubits by an accepting operator $\rho_1^x$
on $c$ qubits. The polynomial method of~\cite{BealsEtAlPolynomials} implies that every matrix entry has
degree at most $2q$ in the oracle input $x$. Our key step is to construct
a single finite net for these input-dependent operators by rounding their
Fourier coefficients. We control the approximation error simultaneously
for every input and every compatible prover strategy, obtaining a bound
on the number of verifier behaviors that is independent of running time
and private workspace.

Finally, each rounded description defines an acceptance set containing
$\Pi$ and contained in its $\epsilon$-neighborhood. A Hamming-ball estimate
bounds the number of properties compatible with each such set. Union
bounding over the rounded descriptions then proves the theorem.

\subsection{Proof of Theorem~\ref{thm:qipp-random-lower-bound}}

For brevity, let $\Sigma=\{0,1\}$.

\begin{lemma}\label{lem:rounded-qipp-verifiers}
    Let $0<\alpha<1/4$. There exists a collection $\calS$ of subsets of $\Sigma^n$ such that
    \[
    |\calS|
    \le 2^{2^{(2\alpha + H(2\alpha))n + O(\log n)}}
    \]
    and a map $S$ on quantum interactive verifier strategies $V$ of total complexity at most $\alpha \cdot n$ to sets $S(V) \in \calS$ such that for all $x \in \Sigma^n$:
    \begin{enumerate}
        \item $\omega_V(x) \ge 2/3$ implies $x \in S(V)$; and
        \item $\omega_V(x) \le 1/3$ implies $x \notin S(V)$.
    \end{enumerate}
\end{lemma}

\begin{proof}
    Let $V$ be a quantum interactive verifier with total complexity at most $\alpha \cdot n$. Let $q$ denote its query complexity, and let $c$ denote its communication complexity, so that $q + c \le \alpha \cdot n$. On some input $x \in \Sigma^n$, the verifier $V$ is defined by quantum channels $V_0^x, \dots, V_r^x$ that can be implemented by circuits using at most $q$ total queries to the oracle $U_x$. Without loss of generality, we assume that $V$ sends the first state. Let $c_i^{(0)}$ denote the number of qubits sent by $V_{i-1}^x$, for $i \in \{1, \dots, r\}$, and let $c_i^{(1)}$ denote the number of qubits received by $V_i^x$, for $i \in \{1, \dots, r\}$. For all $i$, let $\calQ_i = \C(\Sigma^{c_i^{(0)}})$ denote the Hilbert space for the verifier's $i$th send (i.e. ``question'') register, and let $\calR_i = \C(\Sigma^{c_i^{(1)}})$ denote the Hilbert space for the verifier's $i$th received register; let $\calA = \C(\Sigma)$ denote the Hilbert space for the verifier's single output qubit. Here, $\C(A)$ for a set $A$ denotes the space whose basis states are $\ket{a}$ for $a \in A$. Note that $\dim \calQ_i = 2^{c_i^{(0)}}$ and $\dim \calR_i = 2^{c_i^{(1)}}$ and that $\sum_{i=1}^r (c_i^{(0)} + c_i^{(1)}) = c$. Also, for $b \in \{0,1\}$, let $c^{(b)} = \sum_{i=1}^r c_i^{(b)}$.

    Then, the verifier circuits are quantum channels that must take the following form. There must exist Hilbert spaces $\calK_1, \dots, \calK_r$ of arbitrary dimension such that:
    \begin{itemize}
        \item $V_0^x : \sfL(\C) \to \sfL(\calQ_1 \otimes \calK_1)$;
        \item for $1 \le i \le r-1$, we have
        \[
        V_i^x : \sfL(\calR_i \otimes \calK_i)
        \to \sfL(\calQ_{i+1} \otimes \calK_{i+1});
        \]
        and
        \item $V_r^x : \sfL(\calR_r \otimes \calK_r) \to \sfL(\calA)$.
    \end{itemize}
    Furthermore, each $V_i^x$ is a quantum channel, i.e. a completely positive trace-preserving map. These $r+1$ channels can be equivalently described by a single channel
    \[
    \Phi^x :
    \sfL(\calR_1 \otimes \cdots \otimes \calR_r)
    \to
    \sfL(\calA \otimes \calQ_1 \otimes \cdots \otimes \calQ_r),
    \]
    given by Gutoski and Watrous \cite{GutoskiWatrous2007QuantumGames}. See Figure~\ref{Fig:channel-description} for a depiction of this channel, adapted from \cite{BSW-QIP-short-messages}.

    \begin{figure}[t]
        \centering
        \resizebox{\linewidth}{!}{%
        \begin{tikzpicture}[
            channel/.style={draw, minimum width=10mm,
                minimum height=14mm, inner sep=0pt},
            wire/.style={-{Latex[length=1.8mm]}, line width=.5pt},
            every node/.style={font=\small}
        ]
            \foreach \j in {0,...,4} {
                \node[channel] (v\j) at ({2.2*\j},0)
                    {$V_{\j}^{x}$};
            }

            \foreach \i [evaluate=\i as \j using int(\i-1)]
                in {1,...,4} {
                \draw[wire] (v\j.east)
                    -- node[below=3pt,inner sep=1pt]
                    {$\mathcal K_{\i}$} (v\i.west);
            }

            \foreach \i [evaluate=\i as \j using int(\i-1)]
                in {1,...,4} {
                \pgfmathsetmacro{\yin}{-1.6-.7*(\i-1)}
                \pgfmathsetmacro{\yout}{3.5-.7*(\i-1)}

                \draw[wire] (-1.2,\yin)
                    -- ({2.2*\i-1.35},\yin)
                    -- ([xshift=-2mm,yshift=-4mm]v\i.west)
                    -- ([yshift=-4mm]v\i.west);
                \node[above=3pt,inner sep=1pt]
                    at (-.65,\yin) {$\mathcal R_{\i}$};

                \draw[wire] ([yshift=4mm]v\j.east)
                    -- ([xshift=3mm,yshift=4mm]v\j.east)
                    -- ({2.2*\j+1.1},\yout)
                    -- (10,\yout);
                \node[above=3pt,inner sep=1pt]
                    at (9.55,\yout) {$\mathcal Q_{\i}$};
            }

            \draw[wire] (v4.east)
                -- node[above=3pt,inner sep=1pt]
                {$\mathcal A$} (10,0);

            \draw[decorate,decoration={brace,amplitude=5pt}]
                (-1.4,-3.7) -- (-1.4,-1.6)
                node[midway,left=7pt] {$\sigma$};
            \draw[decorate,decoration={brace,amplitude=5pt}]
                (10.15,3.5) -- (10.15,0)
                node[midway,right=7pt] {$\Phi^x(\sigma)$};
        \end{tikzpicture}%
        }
        \caption{The channel $\Phi^x$ for $r=4$. Each wire
        represents an entire register. The input registers are
        $\mathcal R_1,\ldots,\mathcal R_4$, the output registers
        are $\mathcal Q_1,\ldots,\mathcal Q_4,\mathcal A$, and
        $\mathcal K_i$ is the verifier's private memory.
        Adapted from \cite{BSW-QIP-short-messages}.}
        \label{Fig:channel-description}
    \end{figure}
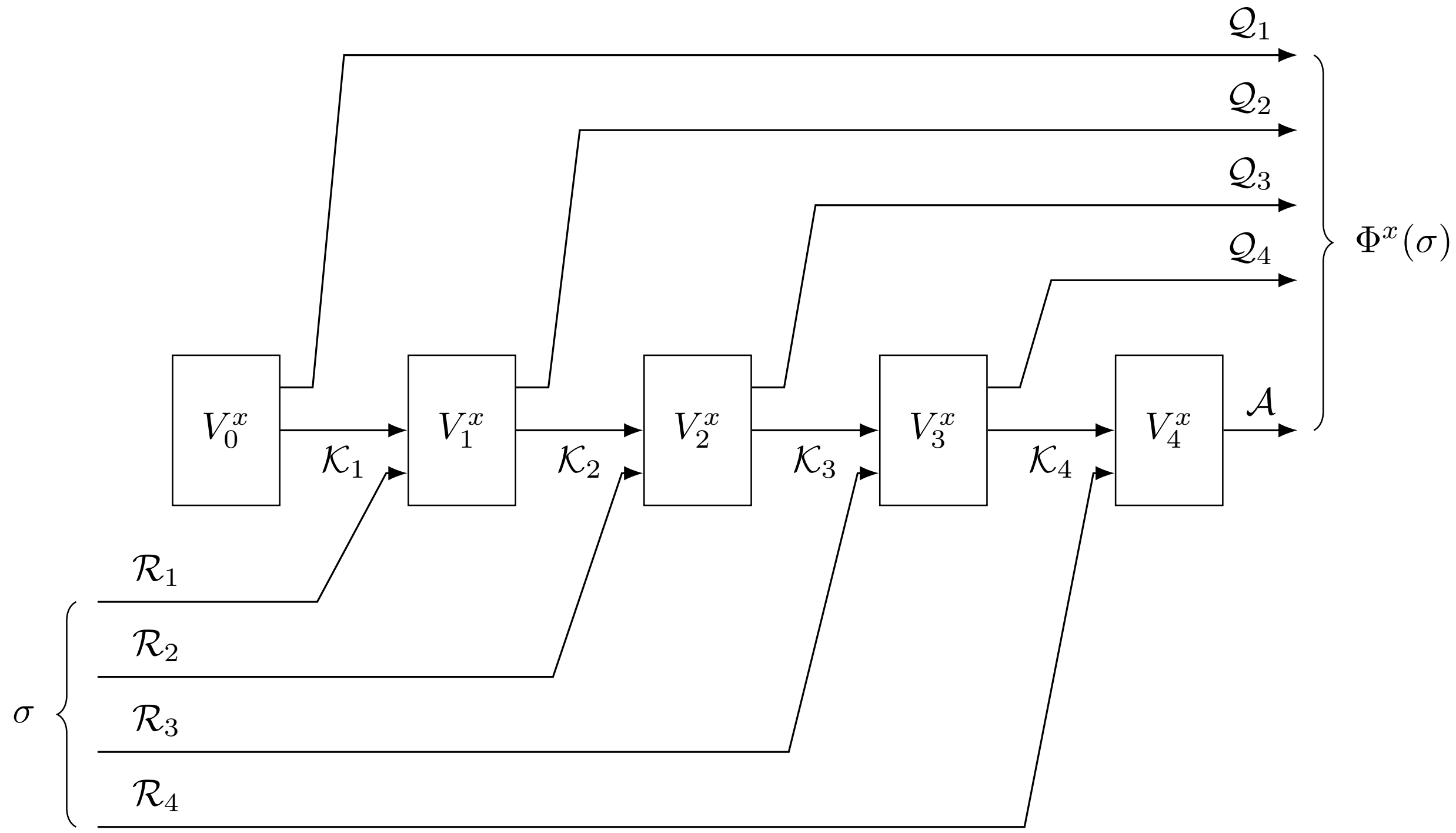

    Now let
    \[
    \rho^x =
    \frac{1}{2^{c^{(1)}}}
    \sum_{y,z \in \Sigma^{c^{(1)}}}
    \Phi^x(\ket{y}\bra{z}) \otimes \ket{y}\bra{z}
    \in
    \sfL(\calA \otimes \calQ_1 \otimes \cdots \otimes \calQ_r
    \otimes \calR_1 \otimes \cdots \otimes \calR_r)
    \]
    denote the normalized Choi-Jamiołkowski representation of $\Phi^x$. We will need the fact that this state can be prepared by preparing the maximally entangled state
    $\frac{1}{2^{c^{(1)}/2}} \sum_{y \in \Sigma^{c^{(1)}}} \ket{y} \otimes \ket{y}$
    and feeding the first $c^{(1)}$ qubits of this state through $\Phi^x$ and keeping the state of the other $c^{(1)}$ qubits unmodified. In particular, since $\Phi^x$ can be represented with a $q$-query quantum circuit, the state $\rho^x$ can be prepared by a $q$-query quantum circuit.

    Let
    \[
    \rho^x_1 =
    (\bra{1} \otimes \bbI)\rho^x(\ket{1} \otimes \bbI)
    \in
    \sfL(\calQ_1 \otimes \cdots \otimes \calQ_r
    \otimes \calR_1 \otimes \cdots \otimes \calR_r)
    \]
    be the projection of $\rho^x$ onto the subspace on which the output register $\calA$ has been set to $\ket{1}$. Using the terminology of Gutoski and Watrous \cite{GutoskiWatrous2007QuantumGames}, $\rho^x_1$ is, up to normalization and transposition, an accepting \emph{co-strategy representation}. They show that for a specific compact set of positive semidefinite operators $\calP$ in the space
    $\sfL(\calQ_1 \otimes \cdots \otimes \calQ_r \otimes \calR_1 \otimes \cdots \otimes \calR_r)$,
    consisting of transposed prover \emph{strategy representations}\footnote{We absorb the transpose in the co-strategy convention of Gutoski and Watrous into the prover's representation.}, we have
    \[
    \omega_V(x) =
    \max_{P \in \calP} 2^{c^{(1)}}\braket{\rho^x_1,P},
    \]
    where $\braket{A,B}=\tr(A^*B)$ denotes the Hilbert--Schmidt inner product. The set $\calP$ depends only on the ordered message lengths. The only other property we will need about $\calP$ is that every $P \in \calP$ has $\tr(P)=2^{c^{(0)}}$.

    Let us establish that every matrix entry of $\rho^x_1$ with respect to the standard basis can be written as a multilinear polynomial in $x$ of degree at most $2q$. Recall that $\rho^x$ is the outcome of some $q$-query quantum circuit. Every such circuit can be purified so that intermediate computation is unitary. Let this purified algorithm be on the Hilbert space
    $\calA \otimes \calQ_1 \otimes \cdots \otimes \calQ_r \otimes \calR_1 \otimes \cdots \otimes \calR_r \otimes \calH$,
    for some auxiliary Hilbert space $\calH$, and let $\ket{\psi^x}$ be the pure output state of the algorithm.

    Beals, Buhrman, Cleve, Mosca, and de Wolf \cite{BealsEtAlPolynomials} establish that for any basis state $\ket{i}$, we can write $\braket{i \mid \psi^x}=p_i(x)$ for some multilinear polynomial $p_i$ of degree at most $q$; furthermore, such polynomials $p_i$ always exist for every state $\ket{i}$, by linearity. Define $\theta=\ket{\psi^x}\bra{\psi^x}$. Then every entry satisfies
    $\braket{i \mid \theta \mid j}
    =\braket{i \mid \psi^x}\braket{\psi^x \mid j}
    =p_i(x)p_j(x)^*$,
    which, after multilinearizing using $x_k^2=x_k$ on $\Sigma^n$, has degree at most $2q$. Since $\rho^x=\tr_\calH(\theta)$ and partial tracing only takes linear combinations of matrix entries, every matrix entry $\braket{i \mid \rho^x \mid j}$ of $\rho^x$ has degree at most $2q$ as well. Lastly, taking projections preserves degree of matrix entries, so every matrix entry $\braket{i \mid \rho^x_1 \mid j}$ of $\rho^x_1$ must be a polynomial in $x$ of degree at most $2q$.

    Before proceeding to the main netting argument, we need to establish a bound on the Fourier coefficients of the matrix entries of $\rho^x_1$ given by the polynomial representation we have just established. Note that $\rho_1$ is represented by a $2^c$-by-$2^c$ matrix. The polynomial representation we just proved shows that for all $i,j \in [2^c]$, we can write
    \[
    \braket{i \mid \rho^x_1 \mid j}
    =
    \sum_{T \subseteq [n]: |T| \le 2q}
    \widehat{\rho_{i,j}}(T)\chi_T(x),
    \]
    where $\widehat{\rho_{i,j}}(T)\in\C$ for every $T$ and
    $\chi_T(x)=\prod_{k\in T}(-1)^{x_k}$.

    Let us prove that for all $i,j$ and for all $T$, we have
    $|\widehat{\rho_{i,j}}(T)|\le1$. First, recall that
    \[
    \widehat{\rho_{i,j}}(T)
    =
    2^{-n}\sum_{x\in\Sigma^n}
    \braket{i \mid \rho^x_1 \mid j}\cdot\chi_T(x).
    \]
    Since $\rho^x$ is positive semidefinite, so is $\rho^x_1$. Together with $\tr(\rho^x_1)\le\tr(\rho^x)=1$, this implies that
    \[
    |\braket{i \mid \rho^x_1 \mid j}|
    \le \|\rho^x_1\|_{\mathrm{op}}
    \le \tr(\rho^x_1)
    \le 1.
    \]
    Using the expression for $\widehat{\rho_{i,j}}(T)$ noted above, we obtain
    \[
    |\widehat{\rho_{i,j}}(T)|
    \le
    2^{-n}\sum_{x\in\Sigma^n}
    |\braket{i \mid \rho^x_1 \mid j}|
    \cdot|\chi_T(x)|
    \le1.
    \]

    Now we are ready to apply the netting method to the operator $\rho^x_1$. Let
    \[
    M=
    |\{T\subseteq[n]:|T|\le2q\}|
    =
    \sum_{k=0}^{2q}\binom nk.
    \]
    Let $\delta=\frac{1}{12\cdot2^{2c}\cdot M}$, and let
    \[
    \calN=
    \left\{
    \frac{\delta}{2}\cdot(j+ki)\in\C:
    j,k\in\Z,\;
    -\frac2\delta\le j,k\le\frac2\delta
    \right\}.
    \]
    Then, every $z\in\C$ such that $|z|\le1$ is at most $\delta$-close to some point in $\calN$. Then, every $\widehat{\rho_{i,j}}(T)$ is $\delta$-close to a point in $\calN$. For each $i\le j$ and every $T$, choose such a point $\widehat{\theta_{i,j}}(T)\in\calN$, choosing it in $\calN\cap\R$ when $i=j$. To preserve that $\rho^x_1$ is Hermitian in the rounded operator, we keep conjugate symmetry by setting
    $\widehat{\theta_{i,j}}(T)=\widehat{\theta_{j,i}}(T)^*$
    for $i>j$. We define the unique linear operator $\theta^x$ such that for all $i,j\in[2^c]$,
    \[
    \braket{i \mid \theta^x \mid j}
    =
    \sum_{T\subseteq[n]:|T|\le2q}
    \widehat{\theta_{i,j}}(T)\chi_T(x).
    \]
    By construction, $\theta^x$ is Hermitian, although it may not be an actual state. Recall that
    $\omega_V(x)=\max_{P\in\calP}2^{c^{(1)}}\braket{\rho^x_1,P}$;
    define the value induced by this rounded operator as
    \[
    \widetilde{\omega}_V(x)
    =
    \max_{P\in\calP}2^{c^{(1)}}\braket{\theta^x,P},
    \]
    noting that indeed $\braket{\theta^x,P}\in\R$ since both $\theta^x$ and $P$ are Hermitian.

    Our task is to show that $\widetilde{\omega}_V(x)$ is close to $\omega_V(x)$. To this end, select any $P\in\calP$. Recall that $P$ is positive semidefinite and $\tr(P)=2^{c^{(0)}}$, so that $\|P\|_1=\tr(P)=2^{c^{(0)}}$. Then, using H\"older's inequality for Schatten norms,
    \begin{align*}
        |\braket{\rho^x_1,P}-\braket{\theta^x,P}|
        &=
        |\braket{\rho^x_1-\theta^x,P}|\\
        &\le
        \|\rho^x_1-\theta^x\|_{\mathrm{op}}\cdot\|P\|_1\\
        &=
        \|\rho^x_1-\theta^x\|_{\mathrm{op}}
        \cdot2^{c^{(0)}}\\
        &\le
        2^{c^{(0)}}\cdot\|\rho^x_1-\theta^x\|_F.
    \end{align*}
    It remains to bound
    $\|\rho^x_1-\theta^x\|_F
    =\sqrt{\sum_{i,j}
    |\braket{i \mid (\rho^x_1-\theta^x) \mid j}|^2}$.
    Fortunately, this follows simply from our netting construction of $\theta^x$. In particular, for all $i,j$, by the triangle inequality,
    \[
    |\braket{i \mid (\rho^x_1-\theta^x) \mid j}|
    \le
    \sum_{T\subseteq[n]:|T|\le2q}
    |\widehat{\rho_{i,j}}(T)-\widehat{\theta_{i,j}}(T)|
    \le\delta\cdot M,
    \]
    where
    $M=|\{T\subseteq[n]:|T|\le2q\}|
    =\sum_{k=0}^{2q}\binom nk$.
    It follows that
    \[
    \|\rho^x_1-\theta^x\|_F
    \le
    \sqrt{(2^c)^2\cdot(\delta\cdot M)^2}
    =
    2^c\cdot\delta\cdot M.
    \]
    Hence,
    $|\braket{\rho^x_1-\theta^x,P}|
    \le2^{c^{(0)}}\cdot2^c\cdot\delta\cdot M$.
    This implies the desired bound
    \begin{align*}
        |\omega_V(x)-\widetilde{\omega}_V(x)|
        &\le
        \max_{P\in\calP}
        2^{c^{(1)}}|\braket{\rho^x_1-\theta^x,P}|\\
        &\le
        2^{c^{(1)}+c^{(0)}+c}\delta M
        =
        2^{2c}\delta M
        =
        \frac1{12}.
    \end{align*}

    We are now ready to define the set $S(V)\subseteq\Sigma^n$ from the theorem statement. Let
    \[
    S(V)=
    \left\{
    x\in\Sigma^n:
    \widetilde{\omega}_V(x)\ge\frac12
    \right\}.
    \]
    The structural properties of $S$ are now immediate from the bound
    $|\omega_V(x)-\widetilde{\omega}_V(x)|\le\frac1{12}$.
    Indeed, if $\omega_V(x)\ge2/3$, then
    $\widetilde{\omega}_V(x)\ge\frac7{12}>\frac12$,
    so $x\in S(V)$; and if $\omega_V(x)\le1/3$, then
    $\widetilde{\omega}_V(x)\le\frac5{12}<\frac12$,
    so $x\notin S(V)$.

    Let $\calS$ be the set of all $S(V)$ over all of the given verifier strategies $V$. It remains only to bound $|\calS|$ to complete the proof. Verifiers making no queries give only the two possible sets $\emptyset$ and $\Sigma^n$, so it suffices to count verifiers with $q\ge1$. If $c=0$, we may pad the protocol with a dummy message containing a single bit; this additive increase can be absorbed into the $O(\log n)$ term in the claimed bound, so it suffices to assume $c \ge 1$ as well. For positive $q$ and $c$ such that $q+c\le\alpha n$, let $\calS(q,c)$ be the class of such sets corresponding to verifiers with query complexity $q$ and communication complexity $c$. Then
    \[
    |\calS|
    \le
    2+\sum_{\substack{q,c\ge1\\q+c\le\alpha n}}
    |\calS(q,c)|,
    \]
    so it suffices to bound $|\calS(q,c)|$.

    To this end, first fix the message layout
    \[
    \lambda=
    (c_1^{(0)},c_1^{(1)},\ldots,c_r^{(0)},c_r^{(1)}).
    \]
    This fixes $\calP$, so the sets $S(V)$ for this layout are determined by the collections of Hermitian operators $\{\theta^x\}_{x\in\Sigma^n}$. These collections are defined only by the complex numbers $\widehat{\theta_{i,j}}(T)$, each belonging to $\calN$, since $\calN$ is closed under conjugation. There are $2^{2c}$ pairs $(i,j)$ and $M$ choices of $T$, so there are at most
    \[
    |\calN|^{2^{2c}\cdot M}
    \]
    such collections for each fixed layout.

    By deleting exchanges in which both messages are empty and composing adjacent local operations, we may assume that $r\le c$. The number of layouts with total communication $c$ is therefore at most
    \[
    \sum_{r=1}^c
    \binom{c+2r-1}{2r-1}
    =
    2^{O(c)}.
    \]
    It follows that
    \[
    |\calS(q,c)|
    \le
    2^{O(c)}|\calN|^{2^{2c}\cdot M}.
    \]
    We have $|\calN|=O(\delta^{-2})=O(2^{4c}M^2)$, so
    \[
    \log|\calS(q,c)|
    \le
    O(c)+2^{2c}M\cdot O(c\log M)
    =
    O(2^{2c}M c\log M).
    \]
    The standard entropy bound, together with $2q\le2\alpha n$, gives
    $M=\sum_{k=0}^{2q}\binom nk\le2^{H(2\alpha)n}$.
    Also, $c\le\alpha n$, so
    \[
    \log|\calS(q,c)|
    \le
    2^{2\alpha n}\cdot2^{H(2\alpha)n}\cdot O(n^2)
    =
    2^{(2\alpha+H(2\alpha))n+O(\log n)}.
    \]
    After union bounding over all $O(n^2)$ possible pairs of $(q,c)$ with $q+c\le\alpha n$, we obtain
    \[
    |\calS|
    \le
    2+O(n^2)\cdot
    2^{2^{(2\alpha+H(2\alpha))n+O(\log n)}}
    =
    2^{2^{(2\alpha+H(2\alpha))n+O(\log n)}}.
    \]
\end{proof}

\begin{proof}[Proof of Theorem~\ref{thm:qipp-random-lower-bound}]
    Let $\calS$ be the family of subsets of $\Sigma^n$ given by Lemma~\ref{lem:rounded-qipp-verifiers}. Let $\Pi$ be any property with $|\Pi|=m$ such that $\sfQIPP_\epsilon(\Pi)\le\alpha\cdot n$, and let $V$ be the verifier of some $\epsilon$-QIPP for $\Pi$ with $q(V)+c(V)\le\alpha\cdot n$. Completeness implies that $\omega_V(x)\ge\frac23$ for all $x\in\Pi$ and soundness implies that $\omega_V(x)\le\frac13$ for all $x$ that are $\epsilon$-far from $\Pi$. Hence, there must exist some $S\in\calS$ (i.e. $S(V)$) such that
    \[
    \Pi\subseteq S\subseteq N_\epsilon(\Pi),
    \]
    where
    $N_\epsilon(\Pi)
    =\{x\in\Sigma^n:\Delta(x,\Pi)\le\epsilon\}$.

    Let us bound, for a fixed $S\in\calS$, the number of properties $\Pi$ with $|\Pi|=m$ that satisfy $\Pi\subseteq S\subseteq N_\epsilon(\Pi)$. To this end, observe that a union bound implies
    \[
    |S|
    \le
    |N_\epsilon(\Pi)|
    \le
    m\cdot\sum_{k=0}^{\floor{\epsilon n}}\binom nk
    \le
    m\cdot2^{H(\epsilon)n}
    =:K.
    \]
    Thus, every property $\Pi$ we are counting is an $m$-element subset of a set of size $K$. Hence the number of such $\Pi$ is at most $\binom Km$.

    Union bounding over all $S\in\calS$ shows that the total number of $\Pi$ we need to count is at most
    \[
    \binom Km
    \cdot
    2^{2^{(2\alpha+H(2\alpha))n+O(\log n)}}.
    \]
    The desired probability is then at most
    \[
    \Pr_{|\Pi|=m}
    [\sfQIPP_\epsilon(\Pi)\le\alpha n]
    \le
    \frac{
        \binom Km\cdot
        2^{2^{(2\alpha+H(2\alpha))n+O(\log n)}}
    }{
        \binom{2^n}m
    }.
    \]
    Note that the standard bounds of
    $(n/k)^k\le\binom nk\le(ne/k)^k$ imply
    \[
    \frac{\binom Km}{\binom{2^n}m}
    \le
    \frac{(Ke/m)^m}{(2^n/m)^m}
    \le
    \left(
        \frac{e\cdot m\cdot2^{H(\epsilon)n}}{2^n}
    \right)^m
    =
    e^m\cdot2^{(\beta+H(\epsilon)-1)nm}
    =
    2^{(\beta+H(\epsilon)-1+o(1))nm},
    \]
    using $m=2^{\beta n}$. Since $\beta+H(\epsilon)<1$ and $\beta>2\alpha+H(2\alpha)$, the desired probability is bounded by
    \begin{align*}
        &2^{(\beta+H(\epsilon)-1+o(1))n2^{\beta n}}
        \cdot
        2^{2^{(2\alpha+H(2\alpha))n+O(\log n)}}\\
        &\qquad\le
        2^{-\Omega(n2^{\beta n})
        +2^{(2\alpha+H(2\alpha))n+O(\log n)}}
        \le
        2^{-2^{\beta n-o(1)}}.
    \end{align*}
\end{proof}

\section{A quadratic quantum speedup for all IPPs}\label{Sec:ipp-aa-speedup}

We have now studied QIPPs for logspace-uniform $\NC$ extensively, showing a quantum speedup over classical IPPs, a lower bound that matches our construction in a wide range of parameter tradeoffs, and a bidirectional connection to Parity. Additionally, we have shown an unconditional limitation for unstructured random properties. But what about QIPPs for properties that already admit classical IPPs? Can quantum queries or quantum communication provide any advantage over classical IPPs?

We answer this question in the affirmative. In this section, we show that every property that admits an $\epsilon$-IPP having communication complexity $c(n) \ge \log n$, query complexity $q(n)$, and $\polylog(n)$ rounds also admits an $\epsilon$-QCIPP (a QIPP with classical communication) with communication complexity $(c(n)+q(n)) \polylog(n)$, query complexity $\sqrt{q(n)} \polylog(n)$, and $\polylog(n)$ rounds. Formally, we prove the following theorem.

\begin{theorem}\label{thm:ipp-to-qcipp}
    Let $\Pi$ be any property and let $\epsilon = \epsilon(n) \in (0,1)$. If there exists an $\epsilon$-IPP for $\Pi$ with query complexity $q(n)$, communication complexity $c(n) \ge \log n$, and round complexity $r(n)$, then there exists a public-coin $\epsilon$-QCIPP for $\Pi$ with query complexity $O(\sqrt{q(n) \cdot t(n)})$, communication complexity $O(c(n)+q(n)) \cdot t(n)$, and round complexity $O(r(n))$, where $t(n) = O(r(n) \cdot \log (c(n) \cdot r(n)))$.
\end{theorem}

The idea for this transformation is the following. First, we use the Goldwasser--Sipser transformation~\cite{goldwasser-sipser:private-public-coins} to render the IPP to be public-coin with deterministic decision. This transformation is made formal in Lemma~\ref{lem:GS-private-to-public-coin}. Without loss of generality, all oracle queries in a public-coin protocol occur after all communication. This step increases the query and communication complexities by a factor of at most $O(r\log(cr))$. Let $q'$ and $c'$ be the new query and communication complexities after this step.
Importantly for the next step, the verifier's queries and decision are deterministic given the transcript. 


Our main contribution of this section is the following transformation on the established public-coin with deterministic decision IPP, yielding the following QCIPP.
The verifier and prover execute the interaction of the specified IPP, after which all oracle queries occur. Appended to its last message, the prover simulates the verifier's deterministic decision algorithm and sends its $q'$ oracle answers in query order. Using these supplied answers, the verifier reconstructs the query sequence and its decision locally, even if the queries are adaptive. Finally, amplitude amplification~\cite{BHMT-amplitude-amplification} checks that every supplied answer used in the simulation agrees with the actual input, taking only $O(\sqrt q')$ oracle queries. The total communication is $c'+q'$, accounting for the original interaction and the $q'$ answer bits returned by the prover.

One can see that this protocol is sound as follows. A cheating prover only gains a small constant advantage over the optimal IPP cheating prover by potentially cheating in amplitude amplification. Even if a cheating prover chooses whether to send correct answers or lie based on the received randomness, the probability of accepting with a false supplied answer is at most a small constant, so the soundness is at most the original IPP's soundness plus that constant. Nonetheless, the protocol is still sound.



\subsection{Proof of Theorem~\ref{thm:ipp-to-qcipp}}

Before proving the theorem, we briefly recall the technique of quantum \emph{amplitude amplification}, which intuitively accomplishes the following. Let $A$ be a randomized algorithm that finds a preimage of $1$ of a Boolean function $f : \{0,1\}^n \to \{0,1\}$ with probability $\gamma > 0$. Amplitude amplification allows us to repeat this algorithm only $O(1/\sqrt{\gamma})$ times to find a preimage of 1 with constant probability, improving upon the $O(1/\gamma)$ repetitions required classically. Recall Lemma~\ref{lem:ampl-ampl}.

Let us proceed to proving Theorem~\ref{thm:ipp-to-qcipp}.
Our first step is to reduce the initial IPP for $\Pi$ into one that is public-coin with deterministic decision. This follows from the Goldwasser--Sipser transformation for interactive proofs \cite{goldwasser-sipser:private-public-coins} adapted to the property testing setting \cite{RVW13}. This step follows from Lemma~\ref{lem:GS-private-to-public-coin}.

Now we prove the main technical contribution of this Section, captured in the following lemma.

\begin{lemma}\label{lem:ipp-to-qcipp-adaptive}
    If a property $\Pi$ admits a public-coin with deterministic decision $\epsilon$-IPP having query complexity $q(n)$, communication complexity $c(n)$, and round complexity $r(n)$, then $\Pi$ has a public-coin $\epsilon$-QCIPP with query complexity $O(\sqrt{q(n)})$, communication complexity $O(c(n) + q(n))$, and round complexity $r(n)$.
\end{lemma}
\begin{proof}
    Denote the given $\epsilon$-IPP by $(P,V)$, which has completeness at least $2/3$ and soundness at most $1/3$. Since $(P,V)$ is public-coin with deterministic decision, we assume without loss of generality that all oracle queries occur after communication. Let $A$ denote the decision algorithm of $V$, which takes the transcript and oracle access to $x$. We construct the following $\epsilon$-QCIPP, denoted $(P_1,V_1)$, for $\Pi$. On input $x\in\{0,1\}^n$, do the following:
    \begin{enumerate}
        \item $(P_1(x),V_1^x(n))$ communicate according to $(P(x),V^x(n))$, obtaining a transcript $\gamma$.
        \item Appended to its last message in the previous step, $P_1(x)$ simulates $A^x(\gamma)$ and sends its oracle answers $y_1,\ldots,y_q$ in query order, padding unused entries if necessary.
        \item $V_1^x(n)$ simulates $A(\gamma)$, answering its $j$th query with $y_j$ and recording the corresponding query location $i_j$. Let $\ell\le q$ be the number of queries used in this simulation. The verifier rejects if the simulated decision is $\mathsf{reject}$.
        \item Otherwise, $V_1^x(n)$ applies Lemma~\ref{lem:ampl-ampl} to the following one-query procedure:
        \begin{enumerate}
            \item Sample $j\in[q]$ uniformly.
            \item Return $1$ if $j\le\ell$ and $x_{i_j}\ne y_j$, and return $0$ otherwise.
        \end{enumerate}
        Use constant error reduction so that any false supplied answer among the used query slots is detected with probability at least $11/12$. Output $\mathsf{reject}$ if a false answer is detected, and output $\mathsf{accept}$ otherwise.
    \end{enumerate}

    The simulation in step 3 uses no queries to $x$, since all answers are supplied by the prover. If any used answer is false, the procedure in step 4 returns $1$ with probability at least $1/q$. Hence amplitude amplification uses $O(\sqrt q)$ input queries. The string of the prover's values appended to its last message has length $q$, so the total communication is at most $c+q$. Thus the protocol remains public-coin and has the stated query, communication, and round complexities.

    We prove that the completeness of $(P_1,V_1)$ is identical to that of $(P,V)$ before amplification. For every transcript $\gamma$, the honest prover supplies the correct answers in query order. Thus the simulation in step 3 follows exactly the query path of $A^x(\gamma)$ and has the same decision. Since all used supplied answers are correct, the one-sided guarantee in Lemma~\ref{lem:ampl-ampl} ensures that step 4 never rejects. Therefore $(P_1,V_1)$ has completeness at least $2/3$; if $(P,V)$ has perfect completeness, then $(P_1,V_1)$ has perfect completeness as well.

    For soundness, suppose that $x$ is $\epsilon$-far from $\Pi$, and fix any cheating prover. Let $\gamma$ be the transcript from step 1. The soundness of $(P,V)$ implies
    \[
    \Pr_{\gamma}[A^x(\gamma)=1]\le\frac{1}{3}.
    \]
    Let $E$ be the event that some supplied answer used in the simulation is false, namely $y_j\ne x_{i_j}$ for some $j\le\ell$. On $E^c$, the simulated computation agrees with $A^x(\gamma)$. On $E$, the consistency check misses all false answers with probability at most $1/12$. This holds for every response, even if it is chosen adaptively after seeing $\gamma$. Therefore
    \begin{align*}
        \Pr[V_1^x(n)\text{ accepts}]
        &=\Pr[V_1^x(n)\text{ accepts}\wedge E^c]
          +\Pr[V_1^x(n)\text{ accepts}\wedge E]\\
        &\le \Pr_{\gamma}[A^x(\gamma)=1]
          +\frac1{12}\Pr[E]\\
        &\le \frac13+\frac1{12}=\frac5{12}.
    \end{align*}

    Repeating in parallel $O(1)$ times forces the completeness and soundness errors to at most $1/3$, increasing each desired complexity measure by only a constant factor and preserving the number of rounds.
\end{proof}

\begin{proof}[Proof of Theorem~\ref{thm:ipp-to-qcipp}]
    Write $t(n)=O(r(n)\log(c(n)r(n)))$. By Lemma~\ref{lem:GS-private-to-public-coin}, there is a public-coin with deterministic decision IPP for $\Pi$ with query complexity $O(q(n)t(n))$, communication complexity $O(c(n)t(n))$, and $O(r(n))$ rounds. Applying Lemma~\ref{lem:ipp-to-qcipp-adaptive} then gives query complexity $O(\sqrt{q(n)t(n)})$, communication complexity $O((c(n)+q(n))t(n))$, and $O(r(n))$ rounds.
\end{proof}

\section{A round collapse theorem for QIPPs}\label{Sec:QIPP_3_rounds}

In this section, we show that any property that admits a \emph{public-coin} QIPP having query, communication, and round complexities $q, c$, and $r$, respectively, also admits a 3-round QIPP with query and communication complexities $q \poly(r)$ and $c \poly(r)$, respectively; moreover, if the initial QIPP has time complexity $t$, then the reduced QIPP has time complexity $t \poly(r)$.
The reduction applies the transformation of Kempe, Kobayashi, Matsumoto, and Vidick \cite{KKMV}, which they use to reduce any multiprover quantum interactive proof to one with only 3-rounds. The main theorem of this section is stated in Theorem~\ref{thm:pub-coin-round-reduction} below. This theorem implies a round collapse of all of the constructions of Section~\ref{Sec:ipp-aa-speedup} and Section~\ref{Sec:nc-aa-speedup}, since they are public-coin. These consequences are contained in Section~\ref{Sec:QIPP-3-corollaries}.



\begin{theorem}
    \label{thm:pub-coin-round-reduction}
    Suppose a property $\Pi$ admits a public-coin $\epsilon$-QIPP
    with query complexity $q$,
    communication complexity $c$, and round complexity $r$. Then $\Pi$ admits a three-round
    $\epsilon$-QIPP with query complexity $q\poly(r)$ and
    communication complexity $c\poly(r)$.
    If the initial QIPP has time complexity $t$, then the
    three-round QIPP has time complexity $t\poly(r)$.
    Perfect completeness is preserved.
\end{theorem}

We proceed directly to the heart of the reduction: a reduction to reduce the number of messages to three. The transformation is the KKMV transformation~\cite{KKMV} applied to QIPPs. After this, we will show how to amplify the completeness and soundness as needed via parallel repetition. 

\begin{restatable}[Round reduction]{lemma}{LemRoundReduction}\label{lem:general_round_reduction}
Suppose a property $\Pi$ admits a public-coin $\epsilon$-QIPP with completeness at least $1-\delta_a$, soundness at most $1-\delta_b$, and query, communication, and round complexities $q, c$, and $r$, respectively. Assume that $\delta_b > 2(r-1)\delta_a$.
Then $\Pi$ admits a 3-round $\epsilon$-QIPP with completeness at least $1-\frac{2\delta_a}{r-1}$, soundness at most $1-\frac{\delta_b}{(r-1)^2}$, query complexity $q\poly(r)$, and communication complexity $c\poly(r)$.

Moreover, if the initial verifier has time complexity $t$, the reduced verifier has time complexity $t\poly(r)$. Perfect completeness is preserved.
\end{restatable}

\begin{proof}[Proof of Lemma~\ref{lem:general_round_reduction}]
    To prove Lemma~\ref{lem:general_round_reduction}, we first establish the structural properties needed to apply the KKMV transformation to public-coin QIPPs, and then describe the KKMV transformation and analyze it in the property testing setting.
    
    \paragraph{Structural properties.}
    Without loss of generality, we assume that the initial QIPP, denoted $(P, V)$, takes the form of public-coin protocols described in Section~\ref{Sec:preliminaries}. Specifically, we assume that in every verifier-receiving round, the verifier receives states and holds them in a private register, while every verifier-sending round has the verifier generate its classical public-coins, copy them into a private register, and send the coins to the prover. The verifier's decision is obtained by applying some quantum channel to its terminating state, and measuring a dedicated output single-qubit register as its decision. 

    We assume that $t \ge c$ without loss of generality. Indeed, if $c > t$, then at least $c-t$ qubits received by the verifier are not ever used, and can be removed.
    
    Furthermore, we assume that $(P, V)$ purifies the coin-toss and copy operations in each verifier-send round. This ensures that $V$'s actions prior to its decision channel are unitary circuits. It is easy to see that the verifier's private state at each round during the protocol (prior to the decision), consists of at most $O(c)$ qubits, while the communication complexity remains at most $c$, the query complexity remains at most $q$, and the time complexity becomes $O(t)$ after purifying.

    Lastly, we pad the number of rounds to be $2^{k+1}+1$ for the least $k$ such that $2^{k+1}+1 \ge r$.

    \paragraph{The reduction.} Let $P^{(0)} := P$ and let $V^{(0)} := V$. For each iteration $i = 0, \dots, k-1$, we describe a reduction from the QIPP $(P^{(i)}, V^{(i)})$ using $2^{k-i+1}+1$ rounds to a QIPP $(P^{(i+1)}, V^{(i+1)})$ with $2^{k-i}+1$ rounds.

    Fix an input string $x \in \{0,1\}^n$, which we omit from the notation of verifier and prover actions below.
    Write the (purified) prover actions on input $x$ as $P^{(i)}[1], \dots, P^{(i)}[2^{k-i}+1]$, which are all unitary operators.
    Write the verifier actions of $V^{(i)}$ on input $x$ by $V^{(i)}[1], \dots, V^{(i)}[2^{k-i}+1]$, where $V^{(i)}[j]$ is unitary for $1 \le j \le 2^{k-i}$ by the purification assumption, but $V^{(i)}[2^{k-i}+1]$ may be a non-unitary quantum channel. The prover sends the first message in $(P^{(i)}, V^{(i)})$ since there are an odd number of rounds.\footnote{In particular, the verifier's initial operator $V_0$ given by Section~\ref{Sec:preliminaries} is the unique (trivial) channel $V_0 : \sfL(\C) \to \sfL(\C)$.} The QIPP $(P^{(i+1)}, V^{(i+1)})$ is described below.

    \begin{enumerate}
        \item The prover sends the internal state of $V^{(i)}$ right before it would apply $V^{(i)}[2^{k-i-1}+1]$, including both the verifier's private register and the incoming register sent by the prover in that round. The verifier sets the received state into its private register and the received message register for that round.
        \item The verifier samples a uniform $d \in \{0,1\}$.
        \begin{enumerate}
            \item If $d=0$, the verifier applies $V^{(i)}[2^{k-i-1}+1]$, sends its outgoing message together with $d$, and follows the remaining interaction \emph{forward in time}. It uses the original acceptance test.
            \item If $d=1$, the verifier returns the incoming message register together with $d$. It then follows the interaction \emph{backward in time}: for $j=2^{k-i-1},\ldots,1$, receive the outgoing register of $V^{(i)}[j]$, apply $(V^{(i)}[j])^\dagger$, and, if $j>1$, return its incoming register. One can apply the inverse $U_x^\dagger$ of the query oracle with just $U_x$, since $U_x^\dagger = U_x$. Accept exactly when the initial private register of $V^{(i)}$ is zero.
        \end{enumerate}
    \end{enumerate}

    Then, we must purify this protocol to obtain $(P^{(i+1)}, V^{(i+1)})$, since the next iteration assumes purification. This requires implementing both branches at once, which can be done assuming controlled channels or controlled gates, and a controlled query $U_x$. Controlled channels and controlled gates are trivial to implement, and implementing a controlled query $U_x$ can be done with two ordinary queries as follows: on basis input $\ket{c} \otimes \ket{i, b}$ for a control qubit $\ket{c}$, compute $x_i$ into an ancilla, add it to $\ket{b}$ conditioned on $\ket{c}$, and uncompute the ancilla.

    \paragraph{Round complexity.}
    The number of rounds of $(P^{(i+1)}, V^{(i+1)})$ is indeed $2^{k-i}+1$, for either branch. In particular, $(P^{(k)}, V^{(k)})$ uses only 3 rounds.

    \paragraph{Completeness and soundness.} Completeness for $x \in \Pi$ and soundness for $\Delta(x, \Pi) > \epsilon$ follow from the analysis of \cite[Lemma~12]{KKMV}. Perfect completeness is clearly preserved.

    \paragraph{Query complexity.} Similarly to communication, the query complexity increases by a universal constant factor in every iteration (including the cost of controlled queries from purification), so the query complexity of $(P^{(k)}, V^{(k)})$ is $q \poly(r)$.

    \paragraph{Communication complexity.} By the public-coin hypothesis on the initial protocol, one can easily observe that there exists a constant $C_0$ such that the communication complexity of $(P^{(i+1)}, V^{(i+1)})$ is at most $C_0$ times the communication complexity of $(P^{(i)}, V^{(i)})$. Hence, the communication complexity of $(P^{(k)}, V^{(k)})$ is at most $O(c) \cdot C_0^k = c \poly(r)$, since $k = O(\log r)$. 

    \paragraph{Time complexity.} If a circuit description of the entire verifier $V$ can be generated uniformly in time $t$, then one can generate a circuit description of each verifier $V^{(i)}$ by generating $V$ first and applying the transformation described at most $i$ times.    
    This transformation multiplies the total time complexity by a universal constant in each iteration, using $c \le t$. Since there are $k = O(\log r)$ iterations, a circuit description of $V^{(k)}$ can be generated in time $t \poly(r)$, as desired.
\end{proof}

Clearly this reduction does not preserve the completeness--soundness gap and requires a strict initial gap. We amplify the gap by parallel repetition, the soundness and complexity of which is established in Lemma~\ref{lem:parallel_repetition}. 


Putting the above results together, we prove Theorem~\ref{thm:pub-coin-round-reduction}.
\begin{proof}[Proof of Theorem~\ref{thm:pub-coin-round-reduction}]
    First, use Lemma~\ref{lem:parallel_repetition} to reduce the initial completeness and soundness errors to $\delta = \frac{1}{6(r-1)}$ with only $O(\log r)$ copies of parallel repetition. Note that the obtained completeness $1-\delta_a = 1-\frac{1}{6(r-1)}$ and soundness $1-\delta_b = 1-(1-\frac{1}{6(r-1)})$ satisfy 
    \[
    \delta_b = 1-\frac{1}{6(r-1)} > \frac{1}{3} = 2(r-1)\cdot \frac{1}{6(r-1)} = 2(r-1)\delta_a,
    \]
    meeting the gap condition of Lemma~\ref{lem:general_round_reduction}. Then, apply Lemma~\ref{lem:general_round_reduction} to obtain completeness at least 
    \[
    1-\frac{2\delta}{r-1} \ge 1-\frac{1}{3(r-1)^2}
    \]
    and soundness at most 
    \[
    1-\frac{1-\delta}{(r-1)^2} = 1-\frac{1}{(r-1)^2}+\frac{\delta}{(r-1)^2} \le 1-\frac{5}{6}\frac{1}{(r-1)^2}.
    \]
    Then, the completeness--soundness gap is at least $\frac{1}{2(r-1)^2}$. Running another $O((2(r-1)^2)^2) = O(r^4)$ copies of parallel repetition allows us to obtain completeness $2/3$ and soundness $1/3$, as desired.

    It only remains to note that every step of this reduction preserves perfect completeness, and all of the complexity parameters only increase by a factor of $\poly(r)$.
\end{proof}

\subsection{Corollaries}\label{Sec:QIPP-3-corollaries}

We conclude this section with several interesting applications of Theorem~\ref{thm:pub-coin-round-reduction} and our previous results.

Combining Theorem~\ref{thm:nc-in-qcipp} with Theorem~\ref{thm:pub-coin-round-reduction} implies the following.

\begin{corollary}\label{cor:nc-in-qipp-3}
    Let $\Pi$ be a property in logspace-uniform $\NC$ and let $\epsilon=\epsilon(n)\in[1/n,1)$ be a proximity parameter. Then $\Pi$ has a 3-round $\epsilon$-QIPP with perfect completeness, and the following complexities.
    \begin{itemize}
        \item Query complexity
        \(
        \softO\left(\frac{1}{\epsilon^{3/2}\sqrt n}
             +\frac{1}{\sqrt\epsilon}\right).
        \)
        \item Communication complexity $\softO(\epsilon n)$.
        \item Time complexity $\softO(\epsilon n + \frac{1}{\epsilon^{3/2}\sqrt n}
             +\frac{1}{\sqrt\epsilon})$.
    \end{itemize}
\end{corollary}


We conclude this section by the following corollary, showing that \emph{every} IPP collapses to a 3-round QIPP with the improved query--communication tradeoff achieved by Theorem~\ref{thm:ipp-to-qcipp}.

\begin{corollary}\label{cor:ipp-to-3-round-qipp}
    Let $\Pi$ be any property and let $\epsilon = \epsilon(n) \in (0,1)$. If there exists an $\epsilon$-IPP for $\Pi$ with query complexity $q(n)$, communication complexity $c(n) \ge \log n$, and round complexity $r(n)$, then there exists a 3-round $\epsilon$-QIPP for $\Pi$ with query complexity $\sqrt{q(n)} \cdot t(n)$ and communication complexity $(c(n)+q(n)) \cdot t(n)$, where $t(n) = \poly(r(n) \cdot \log n)$.
\end{corollary}

\section{Separating QIPPs from IPPs and QMAPs}\label{Sec:qipp-ipp-qmap-separation}

In this section, we exhibit a property based on Aaronson and Ambainis's \emph{Forrelation} problem~\cite{aaronson-ambainis:forrelation} that exponentially separates quantum property testers from classical IPPs. Our proof is based on the recent result of Bouland, Huang, Natarajan, Shalit, and Tal~\cite{BQP-IP-separation-26} that shows $\BQP \subseteq \IP$ does not relativize.
As a consequence, when combined with the known exponential separation of IPPs from QMAPs~\cite{GLR-MA-AM-proximity,sherstov-thaler:vanishing-error-qma}, we obtain an exponential separation of QIPPs from both IPPs and QMAPs.

Formally, we show the following theorem, noting that $\mathbf{QPT}(\epsilon, q)$ denotes the properties that can be $\epsilon$-tested with at most $q$ quantum queries.

\begin{theorem}\label{thm:separation}
    There exists a property $\Pi\subseteq\{0,1\}^n$ such that the
    following holds.
    \begin{enumerate}
        \item For every sufficiently small $\epsilon>0$, we have
        $\Pi\in\QPT(\epsilon,O(1/\epsilon))$.
        \item There exists a constant $\epsilon>0$ such that if
        $\Pi\in\IPP(\epsilon,r,c,q)$ with $c(n)\ge\log n$, then
        \[
            c(n)r(n)^3(q(n)+c(n))^2
            \log^3(c(n)r(n))\log^{3/2}n
            \ge\Omega(n^{1/8}).
        \]
        In particular, every classical $\epsilon$-IPP for $\Pi$,
        with any number of rounds, communication $c(n)\ge1$ and query complexity $q(n)$,
        satisfies
        \[
            q(n)+c(n)
            \ge\Omega\!\left(\frac{n^{1/48}}{\log^{3/4}n}\right).
        \]
    \end{enumerate}
\end{theorem}

As a corollary, we exhibit a property that exponentially separates 2-round QCIPPs from both arbitrary-round IPPs and QMAPs, based on known results that the property containing all bijective functions $f : [k] \to [k]$ exponentially separates 2-round IPPs from QMAPs~\cite{GLR-MA-AM-proximity,sherstov-thaler:vanishing-error-qma}.

\begin{corollary}\label{cor:qipp-vs-ipp-cup-qmap}
    There exists a property $\Pi$ such that the following holds.
    \begin{enumerate}
        \item For all sufficiently small $\epsilon>0$,
        \[
            \Pi\in\QCIPP\bigl(\epsilon,2,
                O(\log n/\epsilon),O(\log n/\epsilon)\bigr).
        \]
        \item For some sufficiently small constant $\epsilon>0$,
        \[
            \Pi\notin
            \IPP\bigl(\epsilon,r,c(n),q_1(n)\bigr)
            \cup\QMAP\bigl(\epsilon,p(n),q_2(n)\bigr)
        \]
        whenever
        \[
            q_1(n)+c(n)
            =o\!\left(\frac{n^{1/48}}{\log^{3/4}n}\right),
            \qquad
            p(n)q_2(n)^3=o(n/\log n).
        \]
        There is no restriction on the number of rounds $r$.
    \end{enumerate}
\end{corollary}

The rest of this section is dedicated to proving Theorem~\ref{thm:separation}. The proof of Corollary~\ref{cor:qipp-vs-ipp-cup-qmap} can be found in Appendix~\ref{Sec:proof-of-qipp-vs-ipp-qmap}, since it is fairly straightforward but tedious.

We begin by recalling the definition of Forrelation, and the property testing property it defines. Let $n \ge 4$ be a power of 2.

\begin{definition}
    Given two functions $f, g: \{0,1\}^{\log (n/2)} \to \{1,-1\}$, their \emph{Forrelation} is defined as
    \[
    \Phi(f,g) \coloneqq (n/2)^{-3/2} \sum_{a,b \in \{0,1\}^{\log(n/2)}} f(a) (-1)^{\langle a, b\rangle} g(b).
    \]
    Define $\Pi_F \subseteq \{0,1\}^n$ as follows. Given $x \in \{0,1\}^n$, the first $n/2$ and the last $n/2$ bits of $x$ define functions $f,g : \{0,1\}^{\log(n/2)} \to \{1, -1\}$ via the natural correspondences $\{0,1\}^{\log(n/2)} \cong [n/2]$ and $\{0,1\} \cong \{1,-1\}$; we define $\Phi(x) \coloneqq \Phi(f,g)$. Let $x \in \Pi_F$ if and only if $\Phi(x) \le \frac{1}{16}$.
\end{definition}

Aaronson and Ambainis~\cite{aaronson-ambainis:forrelation} show the following.

\begin{lemma}\label{lem:forr-has-qpt}
    For all $\epsilon \in (0,1)$, we have $\Pi_F \in \mathbf{QPT}(\epsilon, O(1/\epsilon))$.
\end{lemma}

The following consequence of~\cite{BQP-IP-separation-26}
encapsulates the distributional lower bound we need. For a classical
interactive verifier $V$, let $\omega_V(x)$ denote its maximum acceptance
probability on input $x$, over all prover strategies.

\begin{theorem}[\cite{BQP-IP-separation-26}]
\label{thm:dist-lower-bound}
    There exist distributions $D_0,D_1$ over $\{0,1\}^n$ such that the following holds.
    \begin{enumerate}
        \item $\Pr_{x\sim D_0}[\Phi(x)\le 1/16]
        \ge 1-O(1/n)$.

        \item $\Pr_{x\sim D_1}[\Phi(x)\ge 1/4]\ge 1-O(1/n)$.

        \item For every classical verifier $V$ that is public-coin with deterministic decision, with query complexity $q$ and communication complexity $c \ge 1$, we have
        \[
            \left|
                \bbE_{x\sim D_0}[\omega_V(x)]
                -\bbE_{x\sim D_1}[\omega_V(x)]
            \right|
            \le O\left(\frac{c(q+c)^2\log^{3/2}n}{n^{1/8}}\right).
        \]
    \end{enumerate}
\end{theorem}
\begin{proof}
    Set $N=n/2$, let $H$ be the normalized $N\times N$ Hadamard
    matrix, and identify bits $\{0,1\}$ with signs $\{+1, -1\}$. Note that the magnitude of every entry of $H$ is $N^{-1/2}$. Let $D_0$ be uniform and,
    following~\cite[Section~5]{BQP-IP-separation-26}, let $D_1$ be the distribution obtained by sampling $z$ from $\calN(0, \Sigma)$ and outputting $\operatorname{sgn}(z)$, where $\Sigma$ is the covariance matrix
    \[
    \Sigma = \begin{pmatrix}I_N&H\\H&I_N\end{pmatrix}.
    \]
    Condition (1) follows from \cite[Lemma~A.2]{BQP-IP-separation-26}, while (2) is~\cite[Lemma~A.1]{BQP-IP-separation-26}.

    Applying~\cite[Lemma 4.2]{BQP-IP-separation-26} and~\cite[Corollary~7.1]{BQP-IP-separation-26} to the distribution $D_1$ gives, for every public-coin $r$-round classical verifier with query complexity $q$ and communication complexity $c\ge 1$ with $q+c \le n$,
    \begin{align*}
        \left| \bbE_{x\sim D_0}[\omega_V(x)] -\bbE_{x\sim D_1}[\omega_V(x)] \right|
        &\le O\!\left(n^{-1/8}(r+1)(q+c)^2\log^{3/2}(n(q+c))\right)
        +(2n(q+c))^{-4}\\
        &\le O\left( \frac{r(q+c)^2\log^{3/2}n}{n^{1/8}} \right) + O(n^{-4})\\
        &\le O\left( \frac{c(q+c)^2\log^{3/2}n}{n^{1/8}} \right),
    \end{align*}
    where the last inequality follows from $r \le c$. If $q + c > n$, the claimed indistinguishability bound follows trivially from $0 \le \omega_V(x) \le 1$. Thus (3) follows.
\end{proof}

\subsection{IPP lower bound}

Now we prove the lower bound (2) given in Theorem~\ref{thm:separation}. The proof will proceed as follows. First, using the Goldwasser--Sipser transformation stated in Lemma~\ref{lem:GS-private-to-public-coin} from~\cite{goldwasser-sipser:private-public-coins,RVW13}, we reduce any initial IPP to a public-coin IPP with small overhead.

To apply Theorem~\ref{thm:dist-lower-bound}, we show that $D_0$
produces members of $\Pi_F$ with probability $1-o(1)$, while $D_1$
produces inputs that are $\epsilon$-far from $\Pi_F$ with constant
probability, for some constant $\epsilon>0$.

The first of these claims is explicit in Theorem~\ref{thm:dist-lower-bound}, implying that
$
\Pr_{z \sim D_0} [z \in \Pi_F] \ge 1 - O(1/n).
$
The second claim is slightly more involved, but can be shown with a simple hybrid method, based on the method of~\cite{aaronson-ambainis:forrelation} for a classical property testing lower bound on Forrelation.

\begin{lemma}\label{lem:d-far-from-property}
    There exists a \emph{constant} $\epsilon \in (0,1)$ such that $\Pr_{z \sim D_1} [\Delta(z, \Pi_F) > \epsilon] \ge 1-O(1/n)$.
\end{lemma}
\begin{proof}
    Let $0 < \epsilon < (3/64)^2$.
    Recall from Theorem~\ref{thm:dist-lower-bound} that $\Pr_{z \sim D_1} [\Phi(z) \ge 1/4] \ge 1-O(1/n)$. It suffices to show that every $z \in \{0,1\}^n$ satisfying $\Phi(z) \ge 1/4$ is $\Omega(1)$-far from $\Pi_F$.

    Suppose $z \in \{0,1\}^n$ satisfies $\Phi(z) \ge 1/4$. Let $y \in \{0,1\}^n$ such that $\Delta(y,z) \le \epsilon$; we wish to show that $\Phi(y) > 1/16$. 
    
    Note that~\cite{aaronson-ambainis:forrelation} gives a quantum query algorithm $Q$ that makes a single query to the phase oracle $O_z \ket{i} = (-1)^{z_i} \ket{i}$ (for $i \in [n]$) and accepts with probability precisely $\frac{1 + \Phi(z)}{2}$. Furthermore, the pre-query state of $Q$ is
    $
    \ket{\psi} = n^{-1/2} \sum_{i=1}^n \ket{i}.
    $
    Observe that for all $i \in [n]$, we have $|\braket{i \mid \psi}|^2 = \frac{1}{n}$. Let $D = \{i \in [n]: z_i \ne y_i\}$, so that $|D| \le \epsilon n$ by $\Delta(y, z) \le \epsilon$. Since for all $i \in D$, we have the set equality $\{(-1)^{z_i}, (-1)^{y_i}\} = \{1, -1\}$, this implies
    \begin{align*}
        \|O_z \ket{\psi} - O_y \ket{\psi}\|_2^2 &= \sum_{i = 1}^n |\braket{i \mid O_z \mid \psi} - \braket{i \mid O_y \mid \psi} |^2\\
        &=\sum_{i \in D} |\braket{i \mid O_z \mid \psi} - \braket{i \mid O_y \mid \psi} |^2\\
        &= \sum_{i \in D} |2\braket{i \mid \psi}|^2 = 4 \cdot \frac{|D|}{n} \le 4 \epsilon.
    \end{align*}
    Thus, the difference in acceptance probabilities of $Q$ on $y$ and on $z$ is
    \[
    \left| \frac{1+\Phi(z)}{2} - \frac{1+\Phi(y)}{2} \right| \le \|O_z \ket{\psi} - O_y \ket{\psi}\|_2 \le 2 \sqrt{\epsilon},
    \]
    implying that $ \left| \Phi(z)- \Phi(y) \right| \le 4 \sqrt{\epsilon}$. Since $\epsilon < (3/64)^2$, we obtain $y \notin \Pi_F$ from the inequality
    \[
    \Phi(y) \ge \Phi(z) - |\Phi(z)-\Phi(y)| \ge \frac{1}{4} - 4\sqrt{\epsilon} > \frac{1}{16}.
    \]
\end{proof}

Now we prove the lower bound of Theorem~\ref{thm:separation}, using Lemma~\ref{lem:GS-private-to-public-coin}, Lemma~\ref{lem:d-far-from-property}, and Theorem~\ref{thm:dist-lower-bound}.

\begin{lemma}\label{lem:forr-no-ipp}
    There exists a constant $\epsilon>0$ such that if $\Pi_F$ has
    an $\epsilon$-IPP with $r(n)$ rounds, communication complexity
    $c(n)\ge\log n$, and query complexity $q(n)$, then
    \[
        c(n)r(n)^3(q(n)+c(n))^2
        \log^3(c(n)r(n))\log^{3/2}n
        \ge\Omega(n^{1/8}).
    \]
\end{lemma}
\begin{proof}
    Let $\epsilon$ be the constant from
    Lemma~\ref{lem:d-far-from-property}, and abbreviate
    $q=q(n)$, $c=c(n)$, and $r=r(n)$.
    By Lemma~\ref{lem:GS-private-to-public-coin}, there is a public-coin
    $\epsilon$-IPP with verifier $V$, query complexity $q'=O(qt)$,
    and communication complexity $c'=O(ct)$, where
    $t=O(r\log(cr))$.

    Completeness and Theorem~\ref{thm:dist-lower-bound}(1) give
    \[
        \bbE_{x\sim D_0}[\omega_V(x)]\ge\frac23-O(1/n).
    \]
    Soundness and Lemma~\ref{lem:d-far-from-property} give
    \[
        \bbE_{x\sim D_1}[\omega_V(x)]\le\frac13+O(1/n).
    \]
    Consequently,
    \begin{equation}\label{eq:acc-diff}
        \bbE_{x\sim D_0}[\omega_V(x)]
        -\bbE_{x\sim D_1}[\omega_V(x)]\ge\Omega(1).
    \end{equation}
    Applying Theorem~\ref{thm:dist-lower-bound}(3), we obtain
    \[
        \Omega(1)
        \le O\!\left(\frac{c'(q'+c')^2\log^{3/2}n}{n^{1/8}}\right)
        \le O\!\left(\frac{c(q+c)^2t^3\log^{3/2}n}{n^{1/8}}\right).
    \]
    Substituting $t=O(r\log(cr))$ proves the claim.
\end{proof}

\begin{proof}[Proof of Theorem~\ref{thm:separation}]
    Take $\Pi=\Pi_F$. The quantum upper bound follows from
    Lemma~\ref{lem:forr-has-qpt}, and the IPP tradeoff follows from
    Lemma~\ref{lem:forr-no-ipp}.

    For the round-independent consequence, write $s=q(n)+c(n)$.
    If $s>n$, the claimed bound is immediate. Otherwise, using
    $r(n)\le c(n)\le s\le n$ in the tradeoff gives
    \[
        s^6\log^{9/2}n\ge\Omega(n^{1/8}),
    \]
    and hence $s\ge\Omega(n^{1/48}/\log^{3/4}n)$.
    If initially $c(n)<\log n$, append an irrelevant uniform verifier
    message of length $\lceil\log n\rceil$. The padded protocol has
    total complexity $s+O(\log n)$ and communication at least
    $\log n$, so the same bound applies. Since
    $n^{1/48}/\log^{3/4}n=\omega(\log n)$, it also holds for $s$.
\end{proof}

\section*{Acknowledgments}

The author thanks Tom Gur for many helpful discussions and feedback
on the manuscript, as well as Marcel Dall’Agnol for discussions on the construction of Section~\ref{Sec:nc-aa-speedup}.

Most of this work was completed while the author was at the
University of Cambridge, supported by a Churchill Scholarship
from the Winston Churchill Foundation of the United States.

This material is based on work supported by the National Science Foundation Graduate Research
Fellowship Program under Grant No. DGE-2545526. Any opinions, findings, conclusions, or
recommendations expressed in this material are those of the author(s) and do not necessarily reflect
the views of the National Science Foundation.

\section*{AI Disclosure}

AI tools contributed to the development of two results in this
paper, as detailed below.

OpenAI's GPT 6.1 Sol identified an error in the original
running-time bound for Theorem~\ref{thm:nc-in-qcipp}.
The author developed the quantum amplitude amplification
algorithm with classical advice through subsequent interaction
with this model. 
The author then independently verified the
algorithm's correctness and simplified and clarified the proof.

OpenAI's GPT 5.6 initially proposed the linear lower bound
for QIPPs on random properties,
Theorem~\ref{thm:qipp-random-lower-bound}, together with an
argument containing errors. The author corrected and simplified
the argument and wrote the proof presented here without
AI assistance.

The main ideas underlying the remaining results were developed
without AI assistance. 
AI tools were also used in a limited capacity to assist with literature
review, checking mathematical arguments, and exposition.
The author takes full responsibility
for the correctness and exposition of all results in this paper.









\bibliographystyle{alphaurl}
\bibliography{references}

\appendix

\section{Proofs from Section~\ref{Sec:preliminaries}}\label{App:qipp-3-proof}



In this section, we prove the parallel repetition lemma of Section~\ref{Sec:preliminaries}. We also restate the lemma for convenience.

\LemParallelRepetition*

\begin{proof}[Proof of Lemma~\ref{lem:parallel_repetition}]
    Run $k$ independent copies in parallel and accept if at least $\lceil k(a+b)/2\rceil$ accept. Independent honest strategies give completeness by a Chernoff bound. For soundness, let $X_i$ indicate acceptance of copy $i$ under an arbitrary, possibly entangled prover strategy. Perfect AND-parallel repetition, which follows from~\cite[Theorem~9]{GutoskiWatrous2007QuantumGames}, gives
    \[
     \Pr\left[\bigwedge_{i \in S} X_i=1\right]\leq b^{|S|}
    \]
    for every $S\subseteq[k]$.
    The generalized Chernoff bound~\cite[Theorem~1.1]{impagliazzo-kabanets:concentration} therefore bounds the acceptance probability by $2^{-\Omega(k(a-b)^2)}$, which can be made at most $\eta$ for $k=O\!\left(\frac{\log(1/\eta)}{(a-b)^2}\right).$ The structural and complexity guarantees claimed in the lemma are straightforward.
\end{proof}



\section{Proofs from Section~\ref{Sec:nc-aa-speedup}}\label{App:ampl-ampl-proof}

In this section, we provide the full proof of the amplitude amplification algorithm with classical advice of Lemma~\ref{lem:ampl-ampl-better}.

\begin{proof}[Proof of Lemma~\ref{lem:ampl-ampl-better}]
Assume for simplicity that $M$ is a power of two. Let $w = d + \ceil{\log M}$. Encode the classical data $D = (d_1, \dots, d_M)$ by the $w$-qubit state
\[
 \ket{\psi_D}=\frac1{\sqrt M}\sum_{r=1}^M\ket{r,d_r}.
\]
Let $L = O(\lambda^{-1})$, with the constant determined later.
Then the parallel-lookup circuit of \cite[Theorem~5]{beals} easily allows us to prepare $L$ copies of this state together using $\softO((M+L)w)$ gates and no input queries, and this circuit can be constructed in time $\softO((M+L)w)$. Indeed, initialize $L$ independent uniform registers each holding $M^{-1/2}\sum_{r = 1}^M \ket{r}$ and load the desired $d_r$ values simultaneously using \cite[Eq.~(14)]{beals}. The lookup circuit leaves $D$ unchanged, so its requested outputs are exactly the product state $\ket{\psi_D}^{\otimes L}$. Their theorem states width and depth bounds; multiplying them gives the total number of bounded-arity gates used here. The time needed to construct the circuit is at most a constant factor larger than the number of gates, by inspection.

The evaluator supplies a phase oracle
\[
 P_f\ket{r,z}=(-1)^{f^x(\sigma,r,z)}\ket{r,z}
\]
at cost $O(q)$ queries and $O(t)$ gates per call. Ideal amplitude amplification alternates applying this phase oracle and reflections about $\ket{\psi_D}$. Fortunately, as \cite[Theorem~1]{kimmel} shows, many copies of $\ket{\psi_D}$ are sufficient to simulate the reflection channel $\mathbf{1} - 2\ket{\psi_D}\bra{\psi_D}$. 
We use the sample-based search construction in the proof of \cite[Theorem~7]{kimmel}, with the fixed-point amplification of \cite{ylc}. Fix $x,\sigma,D$, and set
\[
 a=\Acc_f^x(\sigma,D),\qquad
 \lambda_0=\max\{\lambda,1/M\}.
\]
Since $a$ is an integer multiple of $1/M$, $a\geq\lambda$ implies $a\geq\lambda_0$. Let $J$ be the least odd integer at least $\ln(8)/\sqrt{\lambda_0}$. Thus $J=O(\lambda^{-1/2})$ and $J=O(\sqrt M)$.

Define the phase operators
\[
 S_D(\alpha)
 =I-(1-e^{-i\alpha})\ket{\psi_D}\bra{\psi_D}
 \quad\text{and}\quad
 S_f(\beta)\ket{r,z}
 =e^{i\beta f^x(\sigma,r,z)}\ket{r,z}.
\]
Thus $S_D(\alpha)$ multiplies $\ket{\psi_D}$ by $e^{-i\alpha}$
and fixes every vector orthogonal to $\ket{\psi_D}$, while
$S_f(\beta)$ multiplies successful computational basis states by
$e^{i\beta}$ and fixes unsuccessful ones. When
$\alpha=\beta=\pi$, they reduce to the usual reflections used in
amplitude amplification.

To identify the subspace on which amplification takes place, suppose
first that $0<a<1$, and define
\[
 \ket{\mathsf{good}}=\frac{1}{\sqrt{Ma}}
 \sum_{\substack{r\in[M]\\ f^x(\sigma,r,d_r)=1}}\ket{r,d_r},
 \qquad
 \ket{\mathsf{bad}}=\frac{1}{\sqrt{M(1-a)}}
 \sum_{\substack{r\in[M]\\ f^x(\sigma,r,d_r)=0}}\ket{r,d_r}.
\]
These are the normalized superpositions of successful and unsuccessful
rows, respectively, and
\[
 \ket{\psi_D}=\sqrt{1-a}\ket{\mathsf{bad}}+\sqrt a\ket{\mathsf{good}}.
\]
Both phase operators preserve the two-dimensional subspace
$\mathcal T=\operatorname{span}\{\ket{\mathsf{bad}},\ket{\mathsf{good}}\}$.
With respect to its ordered orthonormal basis $(\ket{\mathsf{bad}},\ket{\mathsf{good}})$,
their restrictions have matrices
\[
 \left.S_D(\alpha)\right|_{\mathcal T}
 =I_2-(1-e^{-i\alpha})
 \begin{pmatrix}
  1-a&\sqrt{a(1-a)}\\
  \sqrt{a(1-a)}&a
 \end{pmatrix},
 \qquad
 \left.S_f(\beta)\right|_{\mathcal T}
 =\begin{pmatrix}1&0\\0&e^{i\beta}\end{pmatrix}.
\]
These are precisely the matrices in \cite[Eqs.~(6) and (7)]{ylc},
with their initial state $\ket s$ corresponding to $\ket{\psi_D}$,
their target direction $\ket t$ corresponding to $\ket{\mathsf{good}}$, and
their initial success probability $\lambda$ corresponding to $a$.

Let $\ell=(J-1)/2$. Choose the phases
$\alpha_1,\ldots,\alpha_\ell$ and
$\beta_1,\ldots,\beta_\ell$ from \cite[Eq.~(11)]{ylc}, with their
length parameter $L$ set to $J$ and their error parameter set to
$\delta=1/4$. Their length parameter is unrelated to our number $L$
of prepared copies. The phases depend only on $J$ and $\delta$,
not on the unknown value of $a$, and may be chosen with
$|\alpha_j|,|\beta_j|\leq2\pi$.
Following \cite[Eq.~(10)]{ylc}, define
\[
 G_j=-S_D(\alpha_j)S_f(\beta_j),
 \qquad
 \ket\phi=G_\ell\cdots G_1\ket{\psi_D}.
\]
Each $G_j$ applies the successful-state phase operator first and
the initial-state phase operator second; its overall minus sign
does not affect measurement probabilities. The sequence uses
$\ell\leq J$ applications of each phase operator.
Since $a\geq\lambda_0$ implies
\[
 J\geq\frac{\ln8}{\sqrt{\lambda_0}}
 \geq\frac{\ln8}{\sqrt a},
\]
\cite[Eqs.~(1) and (3)]{ylc} imply that measuring $\ket\phi$
produces a successful row with probability at least
$1-\delta^2=15/16$. If $a=1$, both phase operators preserve
$\ket{\psi_D}$ up to phase, so the same success guarantee holds.

We next implement this ideal sequence approximately. For channels
$\mathcal E$ and $\mathcal F$, measure their distance by
$\tfrac12\|\mathcal E-\mathcal F\|_\diamond$.
This distance bounds the change in the probability of any subsequent
measurement outcome, even in the presence of other, entangled
registers. Set $\eta=1/(100J)$.
Applied to the pure state $\ket{\psi_D}$,
\cite[Theorem~1 and Remark~2, including footnote~1]{kimmel}
allows us to simulate the phase operator $S_D(\alpha)$ using
$O(\alpha^2/\eta)$ fresh copies of $\ket{\psi_D}$.
As $|\alpha|\leq2\pi$, each simulation can achieve channel
error at most $\eta$ using
\[
 N=\ceil{CJ}=O(J)
\]
copies and $\softO(Jw)$ gates, for a sufficiently large fixed
constant $C$. This gate bound includes approximation by a fixed
finite gate set. Reserve a separate set of $N$ copies for each
simulation and one copy for the initial state. It therefore
suffices to prepare
\[
 L=1+JN=O(J^2)=O(\lambda^{-1})
\]
copies in total. The parallel-lookup construction above prepares these copies simultaneously as a product state. 

The operator $S_f(\beta)$ is implemented directly using the evaluator.
For a computational basis input, write $v=f^x(\sigma,r,z)$.
Suppressing the fixed advice and clean workspace, computing $v$,
applying $\operatorname{diag}(1,e^{i\beta})$ to the answer bit,
and uncomputing gives
\[
 \ket{r,z}\ket0
 \xrightarrow{E^x}
 \ket{r,z}\ket v
 \xrightarrow{\operatorname{diag}(1,e^{i\beta})}
 e^{i\beta v}\ket{r,z}\ket v
 \xrightarrow{(E^x)^\dagger}
 e^{i\beta v}\ket{r,z}\ket0.
\]
This is exactly the defining action of $S_f(\beta)$, and by
linearity it also holds on superpositions. Each call costs
$O(q)$ queries and $O(t)$ gates before synthesizing the one-qubit
phase gate. Approximate that gate so that the implemented
$S_f(\beta)$ has channel error at most $\eta$.
The sequence contains at most $J$ calls of each kind. By a
telescoping bound for compositions of channels, its total channel
error is at most
\[
 J\eta+J\eta=2J\eta=1/50.
\]

Finally, measure the index register to obtain $r$, load its actual
row $d_r$ from the original classical table $D$, and evaluate
$f^x(\sigma,r,d_r)$ exactly. Output $1$ if this value is $1$,
and output $0$ otherwise; in the accepting case, $r$ is also
a successful row. This check costs $O(Mw+t)$ gates and $O(q)$
queries. To analyze it, observe that both ideal phase operators
preserve
\[
 \mathcal V_D=\operatorname{span}\{\ket{r,d_r}:r\in[M]\}.
\]
Indeed, $S_f(\beta)$ only changes the phases of these basis
states, while $S_D(\alpha)$ changes a vector in this subspace
only by adding a scalar multiple of $\ket{\psi_D}$.
Thus the ideal state remains supported on actual table rows,
and its successful outcomes pass the final check. Applying
the same check to the ideal and implemented outputs changes
the acceptance probability by at most $1/50$. Consequently,
the implemented algorithm accepts with probability at least
\[
 15/16-1/50>2/3
\]
whenever $a\geq\lambda$, since then $a\geq\lambda_0$.
If $a=0$, every actual row fails the exact check, so the
algorithm rejects with certainty regardless of simulation
errors. 

There are $O(J)$ evaluator calls during amplification and
one final evaluation, giving total query complexity
\[
 O(Jq+q)=O(q/\sqrt\lambda).
\]
Preparing the copies costs $\softO((M+L)w)$ gates, the
$S_D$ simulations cost $\softO(J^2w)$ gates in total, and
the $S_f$ calls cost $\softO(Jt)$ gates, including phase-gate
synthesis. Together with the final check, the total is
\[
 \softO((M+L)w+J^2w+Jt+Mw+t)
 =\softO\bigl((M+\lambda^{-1})w+t/\sqrt\lambda\bigr).
\]
For circuit construction, the phases in \cite[Eq.~(11)]{ylc}
are explicit functions of $J$ and the fixed constant $\delta$.
Computing each amplification phase to additive error $O(1/J)$
suffices to keep its contribution within the per-call error
budget, after adjusting the fixed constants. Since
$J=O(\sqrt M)$, this requires only $O(\log(M+1))$ bits of
precision, and each phase is computable in $\polylog(n)$ time.
The construction bounds in \cite[Theorem~1 and Remark~2]{kimmel},
constructive gate synthesis \cite{dn}, and the assumed
construction times for $E^x$ and the parallel-lookup circuit
therefore give the same bound on circuit construction time.

For general $M$, pad the table to $M'=2^{\ceil{\log M}}$ rows,
define the padded predicate to be zero on indices $r>M$, and
apply the construction with threshold $\lambda/2$.
Its initial success probability is $(M/M')a\geq a/2$, so
$a\geq\lambda$ implies that the padded instance meets this
threshold. The range test adds $O(\log M)$ gates per evaluation
and no queries; invalid indices are also rejected by the final
check. As $M'<2M$, padding changes the stated complexity bounds
only by constant factors, apart from the already suppressed
logarithmic factors. Both guarantees and both complexity bounds
therefore hold for every $M$.
\end{proof}

\section{Proofs from Section~\ref{Sec:qipp-ipp-qmap-separation}}
\label{Sec:proof-of-qipp-vs-ipp-qmap}

We prove Corollary~\ref{cor:qipp-vs-ipp-cup-qmap} by combining the
Forrelation property $\Pi_F$ from Theorem~\ref{thm:separation} with a
binary encoding of the permutation property. The former separates
quantum testers from classical IPPs with any number of rounds, while
the latter separates
2-round IPPs from QMAPs.

Permutation testing concerns functions $f:[k]\to[k]$, rather than
binary strings. The natural binary representation can lose a factor
of $\log k$ in relative distance. To preserve constant proximity,
we instead encode each function value using a constant-rate,
constant-distance code.

Let $\Pi\subseteq\{f:[k]\to[k]\}$ be the property of bijections.
Let $C:[k]\to\{0,1\}^{\ell}$ be an injective code with
$\ell=\Theta(\log k)$ and constant relative distance $\delta>0$.
Thus
\[
    \dist(C(a),C(b))\ge\delta\ell
\]
for all distinct $a,b\in[k]$. Set
$n=k\ell=\Theta(k\log k)$ and define
\[
    \Pi_P
    \coloneqq
    \{(C(f(1)),\ldots,C(f(k))):f\in\Pi\}
    \subseteq\{0,1\}^n.
\]

We use the following results for $\Pi$. In the function model,
a query to $f$ at $i\in[k]$ returns the value $f(i)$, and distance
is the fraction of inputs on which two functions disagree.

\begin{lemma}[{\cite[Lemma~4.2]{GLR-MA-AM-proximity}}]
\label{lem:perm-ipp}
    For every $\epsilon\in(0,1)$, there exists a 2-round IPP for
    $\epsilon$-testing $\Pi$ with perfect completeness, soundness
    $1/2$, query complexity $O(1/\epsilon)$, and communication
    complexity $O(\log k/\epsilon)$.
\end{lemma}

\begin{lemma}[{\cite[Theorem~1.2]{sherstov-thaler:vanishing-error-qma}}]
\label{lem:perm-qmap}
    For every fixed $\epsilon\in(0,1)$, any QMAP for
    $\epsilon$-testing $\Pi$ with proof length $p\ge1$ and query
    complexity $q$ satisfies
    \[
        p\cdot q^3=\Omega(k).
    \]
\end{lemma}

We now transfer these results to $\Pi_P$.

\begin{lemma}
    For every $\epsilon\in(0,1)$, there exists a 2-round IPP for
    $\epsilon$-testing $\Pi_P$ with perfect completeness, soundness
    $1/2$, query complexity $O(\log n/\epsilon)$, and communication
    complexity $O(\log n/\epsilon)$.
\end{lemma}

\begin{proof}
    Write the input as $z=(z_1,\ldots,z_k)$, where
    $z_i\in\{0,1\}^{\ell}$. Use the range-sampling protocol underlying
    Lemma~\ref{lem:perm-ipp}: the verifier samples
    $s=O(1/\epsilon)$ independent uniform labels
    $a_1,\ldots,a_s\in[k]$ and sends them to the prover.
    The prover responds with indices $i_1,\ldots,i_s\in[k]$.
    The verifier reads the corresponding blocks and accepts if
    and only if
    $z_{i_j}=C(a_j)$ for every $j\in[s]$.
    Choose $s$ sufficiently large that $(1-\epsilon)^s\le1/2$.

    Perfect completeness follows by having the prover return the
    preimages of the sampled labels under the encoded permutation.

    For soundness, define
    \[
        R=\{a\in[k]:z_i=C(a)\text{ for some }i\in[k]\}.
    \]
    By retaining one occurrence of each label in $R$ and changing
    the remaining $k-|R|$ blocks, one obtains an encoding of a
    permutation using at most $(k-|R|)\ell$ bit changes.
    Therefore, if $z$ is $\epsilon$-far from $\Pi_P$, then
    $|R|<(1-\epsilon)k$.
    A sampled label outside $R$ cannot pass the verifier's check,
    including when the prover supplies an invalid codeword block.
    Hence the acceptance probability is at most
    \[
        \left(\frac{|R|}{k}\right)^s
        \le(1-\epsilon)^s
        \le\frac12.
    \]

    The verifier reads $s$ blocks of length
    $\ell=\Theta(\log k)=\Theta(\log n)$.
    Both messages contain $s$ indices from $[k]$, giving the
    claimed query and communication complexities.
\end{proof}

\begin{lemma}
    For every fixed $\epsilon\in(0,\delta)$, any QMAP for
    $\epsilon$-testing $\Pi_P$ with proof length $p\ge1$ and
    bit-query complexity $q$ satisfies
    \[
        p\cdot q^3
        =\Omega\!\left(\frac{n}{\log n}\right).
    \]
\end{lemma}

\begin{proof}
    Given oracle access to $f:[k]\to[k]$, simulate the QMAP on
    \[
        x=(C(f(1)),\ldots,C(f(k))).
    \]
    Each bit-query to $x$ can be simulated coherently using
    $O(1)$ queries to $f$: compute the relevant value $f(i)$,
    XOR the requested bit of $C(f(i))$ into the answer register,
    and uncompute. The resulting QMAP for $\Pi$ has the same
    proof length and query complexity $O(q)$.
    Completeness is immediate.

    If $f$ is $\epsilon/\delta$-far from $\Pi$, then for every
    permutation $\pi$,
    \[
    \begin{split}
        \dist\bigl(x,(C(\pi(1)),\ldots,C(\pi(k)))\bigr)
        &\ge
        \delta\ell\,
        |\{i\in[k]:f(i)\ne\pi(i)\}|\\
        &>\epsilon k\ell
        =\epsilon n.
    \end{split}
    \]
    Thus $x$ is $\epsilon$-far from $\Pi_P$, proving soundness.
    Since $\epsilon/\delta\in(0,1)$ is constant,
    Lemma~\ref{lem:perm-qmap} gives
    $p\cdot O(q)^3=\Omega(k)$.
    Finally, $n=\Theta(k\log k)$ implies
    $k=\Theta(n/\log n)$, yielding the claim.
\end{proof}

We can now prove Corollary~\ref{cor:qipp-vs-ipp-cup-qmap}.

\begin{proof}[Proof of Corollary~\ref{cor:qipp-vs-ipp-cup-qmap}]
    Let $\Pi_F$ be the property from Theorem~\ref{thm:separation},
    and let $\Pi_P$ be the binary permutation property above.
    Define
    \[
        \Pi_n
        \coloneqq
        \{(x,y):
          x\in(\Pi_F)_{n/2},\
          y\in(\Pi_P)_{n/2}\}.
    \]
    As usual, compatible input lengths may be assumed by padding.

    First, fix any sufficiently small $\epsilon>0$.
    Run the quantum $\epsilon$-tester for $\Pi_F$ on $x$ and two
    independent copies of the 2-round $\epsilon$-IPP for $\Pi_P$
    on $y$, accepting if and only if all three accept.
    The two IPP copies run in parallel, so the resulting protocol
    still has two messages. Their conjunction has perfect
    completeness and soundness at most $1/4$.

    If $(x,y)\in\Pi_n$, the resulting verifier accepts with
    probability at least $2/3$. If $(x,y)$ is $\epsilon$-far
    from $\Pi_n$, then at least one component is $\epsilon$-far
    from its property, since
    \[
        \Delta((x,y),\Pi_n)
        =
        \frac12\Delta(x,\Pi_F)
        +\frac12\Delta(y,\Pi_P).
    \]
    If $x$ is far, acceptance has probability at most $1/3$;
    if $y$ is far, it has probability at most $1/4$.
    Thus this is a 2-round $\sfQCIPP$ with query and communication
    complexities $O(\log n/\epsilon)$.

    For the lower bounds, let $\epsilon_F>0$ be the constant from
    Theorem~\ref{thm:separation}, and fix
    \[
        0<\epsilon<\frac12\min\{\epsilon_F,\delta\}.
    \]

    Suppose $\Pi_n$ has an $\epsilon$-IPP with any number of rounds,
    query complexity $q_1(n)$, and communication complexity $c(n)$
    satisfying
    \[
        q_1(n)+c(n)
        =o\!\left(\frac{n^{1/48}}{\log^{3/4}n}\right).
    \]

    Fix any $y_0\in(\Pi_P)_{n/2}$ and simulate the protocol on
    $(x,y_0)$, answering queries to the second half using $y_0$.
    This gives an IPP for $\Pi_F$ with the same communication
    complexity and no larger query complexity.
    If $x$ is $2\epsilon$-far from $\Pi_F$, then $(x,y_0)$ is
    $\epsilon$-far from $\Pi_n$.
    Since $2\epsilon<\epsilon_F$, the simulated protocol is also
    an $\epsilon_F$-IPP.

    Applying the round-independent lower bound of
    Theorem~\ref{thm:separation} at input length $n/2$ gives
    \[
        q_1(n)+c(n)
        \ge\Omega\!\left(
            \frac{(n/2)^{1/48}}{\log^{3/4}(n/2)}
        \right)
        =\Omega\!\left(\frac{n^{1/48}}{\log^{3/4}n}\right),
    \]
    contradicting the assumed bound. 

    Finally, suppose $\Pi_n$ has an $\epsilon$-QMAP with proof
    length $p(n)$ and query complexity $q_2(n)$.
    Fix any $x_0\in(\Pi_F)_{n/2}$ and simulate the protocol on
    $(x_0,y)$, answering queries to the first half using $x_0$.
    This gives a QMAP for $\Pi_P$ with the same proof length
    and query complexity $O(q_2(n))$.
    If $y$ is $2\epsilon$-far from $\Pi_P$, then $(x_0,y)$ is
    $\epsilon$-far from $\Pi_n$, so the simulated protocol is sound.
    Since $2\epsilon<\delta$, the binary permutation lower bound
    applies and gives
    \[
        p(n)\cdot q_2(n)^3
        =\Omega\!\left(\frac{n}{\log n}\right).
    \]
    This excludes the claimed QMAP parameter range and completes
    the proof.
\end{proof}

\label{lastpage}

\end{document}